\documentclass[11pt,reqno,english,american]{article}

\usepackage[T1]{fontenc}
\usepackage[latin9]{inputenc}
\usepackage{xcolor}
\usepackage{babel}
\usepackage{mathtools}
\usepackage{enumitem}
\usepackage{amsmath}
\usepackage{amsthm}
\usepackage{amssymb}
\usepackage{cancel}
\usepackage{mathdots}
\usepackage{stmaryrd}
\usepackage{stackrel}
\usepackage{geometry}
\usepackage{setspace}
\PassOptionsToPackage{version=3}{mhchem}
\usepackage{mhchem}
\usepackage{esint}
\usepackage[authoryear,comma]{natbib}
\usepackage[pdfusetitle,
 bookmarks=true,bookmarksnumbered=true,bookmarksopen=false,
 breaklinks=true,pdfborder={0 0 0},pdfborderstyle={},backref=false,colorlinks=true]
 {hyperref}
\hypersetup{
 linkcolor=blue,citecolor=blue,urlcolor=blue}

\makeatletter

\global\long\def\c{\mathcal{C}}%
\global\long\def\var{\mathrm{Var}}%
\global\long\def\eig{\mathrm{eig}}%
\usepackage{authblk}  
\author[a]{Wen Jiang\thanks{Email address: \protect\href{mailto:jiangwen@mju.edu.cn}{jiangwen@mju.edu.cn}.}}

\author[b]{Yachen Wang\thanks{Email address: \protect\href{mailto:wangyachen@stu.xmu.edu.cn}{wangyachen@stu.xmu.edu.cn}.}}
\author[c]{Zeqi Wu\thanks{Email address: \protect\href{mailto:wuzeqi@ruc.edu.cn}{wuzeqi@ruc.edu.cn}.}}
\author[d]{Xingbai Xu\thanks{Email address: \protect\href{mailto:xuxingbai@xmu.edu.cn}{xuxingbai@xmu.edu.cn}.}}
\affil[a]{Newhuadu Business School, Minjiang University}

\affil[b]{Paula and Gregory Chow Institute for Studies in Economics, Xiamen University}
\affil[c]{Institute of Statistics and Big Data, Renmin University of China}
\affil[d]{MOE Key Lab of Econometrics and Fujian Key Lab of Statistics, Wang Yanan Institute for Studies in Economics (WISE), Department of Statistics and Data Science, School of Economics, Xiamen University}

\numberwithin{equation}{section}
\newlist{casenv}{enumerate}{4}
\setlist[casenv]{leftmargin=*,align=left,widest={iiii}}
\setlist[casenv,1]{label={{\itshape\ \casename} \arabic*.},ref=\arabic*}
\setlist[casenv,2]{label={{\itshape\ \casename} \roman*.},ref=\roman*}
\setlist[casenv,3]{label={{\itshape\ \casename\ \alph*.}},ref=\alph*}
\setlist[casenv,4]{label={{\itshape\ \casename} \arabic*.},ref=\arabic*}

\@ifundefined{date}{}{\date{}}
\allowdisplaybreaks[4]
\usepackage{babel}
\usepackage{appendix}\usepackage{titlesec}
\titlelabel{\thetitle.\quad}
\usepackage{booktabs}
\usepackage{multirow}

\makeatother

\theoremstyle{definition}
\newtheorem{defn}{\protect\definitionname}[section]
\theoremstyle{remark}
\newtheorem{rem}{\protect\remarkname}[section]
\theoremstyle{plain}
\newtheorem{lem}{\protect\lemmaname}[section]
\newtheorem{thm}{\protect\theoremname}[section]
\newtheorem{assumption}{\protect\assumptionname}
\newtheorem{prop}{\protect\propositionname}[section]
\theoremstyle{definition}
\newtheorem{example}{\protect\examplename}[section]
\addto\captionsamerican{\renewcommand{\assumptionname}{Assumption}}
\addto\captionsamerican{\renewcommand{\casename}{Case}}
\addto\captionsamerican{\renewcommand{\definitionname}{Definition}}
\addto\captionsamerican{\renewcommand{\examplename}{Example}}
\addto\captionsamerican{\renewcommand{\lemmaname}{Lemma}}
\addto\captionsamerican{\renewcommand{\propositionname}{Proposition}}
\addto\captionsamerican{\renewcommand{\remarkname}{Remark}}
\addto\captionsamerican{\renewcommand{\theoremname}{Theorem}}
\addto\captionsenglish{\renewcommand{\assumptionname}{Assumption}}
\addto\captionsenglish{\renewcommand{\casename}{Case}}
\addto\captionsenglish{\renewcommand{\definitionname}{Definition}}
\addto\captionsenglish{\renewcommand{\examplename}{Example}}
\addto\captionsenglish{\renewcommand{\lemmaname}{Lemma}}
\addto\captionsenglish{\renewcommand{\propositionname}{Proposition}}
\addto\captionsenglish{\renewcommand{\remarkname}{Remark}}
\addto\captionsenglish{\renewcommand{\theoremname}{Theorem}}
\providecommand{\assumptionname}{Assumption}
\providecommand{\casename}{Case}
\providecommand{\definitionname}{Definition}
\providecommand{\examplename}{Example}
\providecommand{\lemmaname}{Lemma}
\providecommand{\propositionname}{Proposition}
\providecommand{\remarkname}{Remark}
\providecommand{\theoremname}{Theorem}

\begin{document}
\title{Limit Theorems for Network Data without Metric Structure\thanks{The authors are listed in alphabetical order and contributed equally.
Zeqi Wu and Xingbai Xu are corresponding authors.}}

\maketitle
\global\long\def\E{\mathbb{E}}%
\global\long\def\p{\mathbb{P}}%
\global\long\def\R{\mathbb{R}}%
\global\long\def\N{\mathbb{N}}%
\global\long\def\Q{\mathbb{Q}}%
\global\long\def\Z{\mathbb{Z}}%
\global\long\def\F{\mathcal{F}}%
\foreignlanguage{english}{
\global\long\def\tr{\text{tr}}%
}
\vspace{-0.9em}

\begin{abstract}
This paper develops limit theorems for random variables with network
dependence, without requiring the individuals in the network to be
located in a Euclidean or metric space. This distinguishes our approach
from most existing limit theorems in network statistics and econometrics,
which are based on weak dependence concepts such as strong mixing,
near-epoch dependence, or $\psi$-dependence. All these weak dependence
concepts presuppose an underlying metric. By relaxing the assumption
of an underlying metric space, our theorems can be applied to a broader
range of network data, including financial and social networks. To
derive the limit theorems, we generalize the concept of functional
dependence (also known as physical dependence) from time series to
random variables with network dependence. Using this framework, we
establish several inequalities, a law of large numbers, and central
limit theorems. Furthermore, we demonstrate the verifiability of our
high-level conditions by deriving primitive sufficient conditions
for spatial autoregressive models, which are widely used in network
data analysis.
\end{abstract}
\textit{Keywords:} functional dependence, network data, law of large
numbers, central limit theorem, concentration inequality, spatial
autoregressive model

\newpage{}

\section{Introduction}

Network-indexed dependence arises in a broad class of statistical
and econometric models. To study estimators and test statistics in
such settings, one needs probabilistic tools, such as moment inequalities,
concentration inequalities, laws of large numbers, and central limit
theorems, to accommodate dependence structures that are neither sequentially
ordered nor naturally organized by a metric. This challenge is especially
acute in nonlinear network models, where martingale methods available
for linear specifications are typically not applicable.

A substantial body of literature has studied weak dependence concepts
for spatial and network-indexed data. For linear spatial autoregressive
(SAR) models, martingale difference array arguments play a central
role \citep{h._kelejian_asymptotic_2001,lee2004asymptotic,lee_gmm_2007,lee2022qml}.
For more general dependent arrays, the literature develops strong-mixing
and $\phi$-mixing spatial random fields \citep{jenish2009central},
spatial near-epoch dependence (NED) \citep{jenish2012spatial}, spatial
functional dependence \citep{wu2023application}, spatial mixingales
based on model-dependent random metrics \citep{kuersteiner2019limit},
and network adaptations of $\psi$-dependence \citep{kojevnikov_limit_2021};
see also \citet{kuersteiner_limit_2013,KuersteinerPrucha2020,leung_weak_2019,leung_normal_2019}.
These contributions have substantially advanced asymptotic theory
for dependent data and have also been applied in a variety of settings;
see, for example, \citet{xu2015maximum,gao2025Causal,kojevnikov2021bootstrap,leung2022Causal,XU201896,qu_estimating_2015,liu2022robust}.
However, most of them rely on a metric, a geodesic distance, or a
model-dependent random metric. Accordingly, dependence is characterized
through distance-based shells and decay with distance.

This reliance on metric structure constitutes a major bottleneck for
network data. Many network models employed in statistics and econometrics
do not possess a natural embedding in Euclidean space. Even when a
metric can be imposed, it may not accurately reflect the actual propagation
of dependence. The problem is especially acute in networks with short
graph distances, high connectivity, or moderate diameter. For example,
consider an Erd\H{o}s-R\'{e}nyi model with mean degree $D_{n}\propto n^{\alpha}$
($0<\alpha<1)$. Classical results on the diameter of Erd\H{o}s-R\'{e}nyi
graphs imply that graph diameter is of constant order with probability
approaching one in this setting \citep{klee1981diameters}. In such
settings, there are only a few relevant distance shells, so weak dependence
cannot be naturally organized via gradual decay over expanding metric
neighborhoods. What is needed, therefore, is a notion of weak dependence
that does not require any metric structure, and can be directly verified
from primitive conditions.

This restriction is substantive, not merely technical. In many network
applications, it is unclear whether the nodes admit any meaningful
Euclidean embedding. For instance, \citet{xu2022dynamic} applied
spatial NED theory to quantile regression in a dynamic network model
for stocks traded on the NYSE and NASDAQ, even though it is not obvious
that such assets should be viewed as points in a Euclidean space.
Likewise, latent-space representations for social networks \citep{hoff2002Latent,athreya2018Statistical,smith2019Geometry}
provide useful modeling devices, but the underlying metric is typically
imposed for tractability rather than being dictated by the dependence
mechanism itself. More generally, in online and social networks, nodes
that are far apart in any Euclidean sense may still interact strongly.
These examples indicate that metric-based weak dependence conditions
can impose a substantive modeling restriction in applications where
no natural metric structure is available.

There is also a methodological mismatch between estimation and asymptotic
theory. Many standard procedures, including quasi-maximum likelihood
estimation and the generalized method of moments, can be formulated
for network data without any metric structure. However, available
asymptotic justifications often rely on metric-based weak dependence
conditions. For instance, the robust estimator for SAR models proposed
by \citet{liu2022robust} ultimately invokes the spatial NED framework
of \citet{jenish2012spatial}. A partial step in this direction is
\citet{kuersteiner2019limit}, who replaces a fixed metric with a
model-dependent random metric. This underscores the need for a notion
of weak dependence that is both metric-free and directly verifiable
from the model. 

We address this problem by extending the concept of functional dependence
measure (FDM) to network data. This measure is defined by perturbing
the underlying innovations and quantifying the resulting effect on
each component of the system. The construction is entirely node-based
and does not rely on Euclidean distance, geodesic distance, or any
other metric. It extends the functional dependence concept of \citet{wu2005nonlinear},
\citet{ELMACHKOURI20131}, and \citet{wu2023application} from time
series and spatial processes to general network dependence. Unlike
strong mixing and related notions, the proposed FDM avoids suprema
over sub-$\sigma$-fields and is therefore substantially easier to
verify in specific network models.

The contributions of this paper are fourfold. First, we introduce
a metric-free functional dependence framework for network data. This
extends the functional dependence paradigm to dependence structures
that are intrinsic to networks and not organized by any metric. Second,
under this framework we develop a general probabilistic toolkit for
network data. In particular, we establish a moment inequality, a concentration
inequality, a weak law of large numbers, and central limit theorems.
Third, for the CLT theory, we introduce a second-order functional
dependence measure and derive sufficient conditions directly in terms
of first- and second-order dependence measures. These conditions are
invariant under relabeling of the nodes and are designed for nonlinear
systems, thereby going beyond martingale-based arguments available
for linear models. Fourth, we show that the proposed dependence concept
is verifiable and useful in concrete models. We compute the dependence
measures for nonlinear SAR models, derive primitive sufficient conditions
for LLNs and CLTs, and verify these conditions under several network
designs, including dominant-unit structures and random graph models.
In addition, we establish the consistency of the maximum likelihood
estimator for the SAR Tobit model using the tools developed in this
paper.

The remainder of the paper is organized as follows. Section~\ref{sec:Func depend}
introduces the functional dependence measure and the associated notion
of $(L^{p},q)$-functional dependence. Section~\ref{sec:Property}
develops the general theoretical tools, including a moment inequality,
a concentration inequality, a weak law of large numbers, and central
limit theorems; it also introduces a second-order functional dependence
measure for the CLT analysis. Section~\ref{Sec: Example} applies
the framework to nonlinear SAR models, derives primitive sufficient
conditions for the LLN and CLTs, and verifies these conditions under
several network designs; it also studies the SAR Tobit model. Section~\ref{sec: Transformation}
investigates how the functional dependence measure behaves under common
transformations. Section~\ref{sec:Conclusion} concludes. Proofs
of the main results are collected in Appendix~\ref{sec:Proofs-for-Section},
and additional proofs are deferred to the supplementary appendix.

\textbf{Notation}. The set of positive integers is $\mathbb{N}\equiv\{1,2,\cdots\}$,
$\mathbb{Z}$ denotes the set of integers, and $\mathbb{Z}^{d}$ is
the $d$-dimensional integer lattice. For any column vector $x=\left(x_{1},x_{2},\ldots,x_{d}\right)'\in\mathbb{R}^{d}$,
where $\mathbb{R}^{d}$ is the $d$-dimensional Euclidean space, $\left\Vert x\right\Vert =(x'x)^{1/2}$
denotes its Euclidean norm, $\left\Vert x\right\Vert _{\infty}=\max_{1\leq k\leq d}\left|x_{k}\right|$
represents its infinity norm, and $\left\Vert x\right\Vert _{1}=\sum^{d}_{k=1}\left|x_{k}\right|$
is its 1-norm. For any matrix $A=\left(a_{ij}\right)_{n\times m}$,
its Frobenius norm is $\left\Vert A\right\Vert \equiv\sqrt{\sum_{ij}a^{2}_{ij}}$,
its 1-norm (also called column sum norm) is defined as $\left\Vert A\right\Vert _{1}=\max_{1\leq j\leq m}\sum^{n}_{i=1}\left|a_{ij}\right|$,
and its infinity norm (also called row sum norm) is defined as $\left\Vert A\right\Vert _{\infty}=\max_{1\leq i\leq n}\sum^{m}_{j=1}\left|a_{ij}\right|$.
For any symmetric matrix $A$, $\min\eig(A)$ denotes its minimum
eigenvalue. For any two sequences $a_{n}$ and $b_{n}$, $a_{n}\sim b_{n}$
if and only if $a_{n}/b_{n}\to1$ as $n\to\infty$. Let $\left(\Omega,\mathcal{F},\p\right)$
be a probability space. Denote probability and expectation by $\p$
and $\E$, respectively. For any random vector (matrix) $X$, its
$L^{p}$ norm is defined as $\left\Vert X\right\Vert _{L^{p}}\equiv\left[\E\left(\left\Vert X\right\Vert ^{p}\right)\right]^{1/p}$
for any constant $p\geq1$. The symbols ``$\xrightarrow{\p}$'' and
``$\ensuremath{\xrightarrow{d}}$'' denote convergence in probability
and convergence in distribution, respectively. For two positive non-random
sequences $a_{n},b_{n}$ and random vector sequence $X_{n}$, $X_{n}=o_{\p}(a_{n})$
means $\p\left(\left\Vert X_{n}\right\Vert >\epsilon a_{n}\right)\to0$
as $n\to\infty$ for any $\epsilon>0$ and $X_{n}=O_{\p}(a_{n})$
means for any $\epsilon>0$, there exists a constant $M>0$ such that
$\limsup_{n\to\infty}\p\left(\left\Vert X_{n}\right\Vert \geq Ma_{n}\right)<\epsilon$.
For any sub-$\sigma$-field $\c$ of $\F$, denote the conditional
probability, conditional expectation, and conditional variance by
$\p_{\c}\left(\cdot\right)\equiv\p\left(\cdot\mid\c\right)$, $\E_{\c}\left(\cdot\right)\equiv\E\left(\cdot\mid\c\right)$,
and $\var_{\c}\left(\cdot\right)\equiv\var\left(\cdot|\c\right)$,
respectively. Besides, for a sub-$\sigma$-field $\mathcal{G}$ of
$\F$, we denote $\E_{\c}\left(\cdot|\mathcal{G}\right)\equiv\E\left(\cdot\mid\mathcal{G}\bigvee\c\right)$,
where $\mathcal{G}\bigvee\c$ denotes the $\sigma$-field generated
by $\mathcal{G}$ and $\c$. For a random vector (matrix) $X$, let
$\left\Vert X\right\Vert _{L^{p},\c}\equiv\left[\E_{\c}\left(\left\Vert X\right\Vert ^{p}\right)\right]^{1/p}$. 

\section{\label{sec:Func depend}Definition of Functional Dependence Measure}

Consider a network with $n$ nodes (also referred to as individuals
or units). For simplicity, name these $n$ nodes as $1,2,\cdots,n$,
and denote $[n]=\{1,2,\cdots,n\}$. Although we use numeric labels,
the order is arbitrary: random variables associated with individuals
1 and 2 may be independent, while those associated with individuals
1 and $n$ could be strongly correlated. 

Let $\left(\Omega,\mathcal{F},\p\right)$ be the underlying probability
space, and $\c_{n}$ be a sub-$\sigma$-field of $\F$. Associated
with each $i\in[n]$ is a random vector $e_{i,n}\in\mathbb{R}^{p_{e}}$,
where $p_{e}\in\N$ is a constant. Suppose that $e_{i,n}$'s are conditionally
independent given $\c_{n}$, but they need not be identically distributed.
Denote $e_{n}=(e_{1,n},e_{2,n},\cdots,e_{n,n})$. Suppose that $Y_{1,n},\cdots,Y_{n,n}$
are $n$ random vectors generated by $e_{n}$:
\begin{equation}
Y_{j,n}=F_{j,n}(e_{n}),\label{eq: Y_i}
\end{equation}
where the functions $F_{1,n},\ldots,F_{n,n}$ can be random functions
(e.g. associated with a random network weights matrix, see Section~\ref{Sec: Example})
and we assume these functions are measurable with respect to $\c_{n}$.
The triangular array $\{Y_{j,n}:1\leq j\leq n,\ n\geq1\}$ exhibits
network dependence. Although $Y_{j,n}$ can be vector-valued, we restrict
attention to the real-valued case for simplicity and without loss
of generality. See Remark~\ref{remark 2.2} for details. 

Suppose that conditional on $\c_{n}$, $e^{*}_{i,n}$ is an independent
and identically distributed (i.i.d.) copy of $e_{i,n}$, and $e^{*}_{i,n}$
is independent of $e_{j,n}$ for all $j\neq i$. Denote $Y_{j,n,i}$
as the coupled version of $Y_{j,n}$ with $e_{i,n}$ replaced by $e^{*}_{i,n}$,
i.e., $Y_{j,n,i}\equiv F_{j,n}\left(e_{1,n},\cdots,e_{i-1,n},e^{*}_{i,n},e_{i+1,n},\cdots,e_{n,n}\right)$.
Now, we are ready to introduce the definition of functional dependence
measure.
\begin{defn}
[Functional dependence measure]\textbf{\label{def:(Functional-dependence-measure).}}
For $p\geq1$, define the (first-order) functional dependence measure
(FDM), also called the physical dependence measure, as 
\begin{equation}
\delta_{p,n}\left(j,i,\c_{n}\right)\equiv\left\Vert Y_{j,n}-Y_{j,n,i}\right\Vert _{L^{p},\c_{n}}.\label{eq: delta i j}
\end{equation}
When $\c_{n}$ is the trivial $\sigma$-field $\left\{ \Omega,\emptyset\right\} $,
we simplify the notation to $\delta_{p,n}\left(j,i\right)\equiv\delta_{p,n}\left(j,i,\c_{n}\right)$.
\end{defn}
\begin{rem}
The quantity $\delta_{p,n}\left(j,i,\c_{n}\right)$ in Eq.(\ref{eq: delta i j})
measures the influence of $e_{i,n}$ on $Y_{j,n}$: conditional on
$\c_{n}$, if $e_{i,n}$ is replaced by its i.i.d. version $e^{*}_{i,n}$,
the magnitude of the resulting change of $Y_{j,n}$ is $\delta_{p,n}\left(j,i,\c_{n}\right)$
under the norm $\left\Vert \cdot\right\Vert _{L^{p},\c_{n}}$. Although
this definition might superficially resemble a metric, it is not one
because it might not satisfy the triangle inequality. To illustrate,
consider two nodes $i$ and $j$ with weak dependence. Under a metric-based
concept of dependence, no node $k$ could be strongly correlated with
both. However, this configuration is entirely possible under the FDM
framework.
\end{rem}
\begin{rem}
\label{remark 2.2}When $Y_{j,n}=\left(Y^{(1)}_{j,n},\cdots,Y^{(p_{Y})}_{j,n}\right)^{\prime}$
is a $p_{Y}$-dimensional random vector, since $\|y\|\leq\|y\|_{1}$,
for all $i=1,\cdots,n$, 
\begin{align*}
 & \left\Vert Y_{j,n}-Y_{j,n,i}\right\Vert _{L^{p},\c_{n}}\leq\left\Vert \sum^{p_{Y}}_{k=1}\left|Y^{(k)}_{j,n}-Y^{(k)}_{j,n,i}\right|\right\Vert _{L^{p},\c_{n}}\leq\sum^{p_{Y}}_{k=1}\left\Vert Y^{(k)}_{j,n}-Y^{(k)}_{j,n,i}\right\Vert _{L^{p},\c_{n}},
\end{align*}
where $Y_{j,n,i}=\left(Y^{(1)}_{j,n,i},\cdots,Y^{(p_{Y})}_{j,n,i}\right)^{\prime}$
is the coupled version of $Y_{j,n}$ with $e_{i,n}$ replaced by its
i.i.d. copy $e^{*}_{i,n}$. Thus, without loss of generality, it suffices
to study the FDM for real-valued random variables.
\end{rem}
\textbf{Comparison with existing weak dependence concepts.} We now
compare our FDM with several well-known notions of weak dependence
in the literature.

(1) The index sets considered in \citet{wu2005nonlinear,wu2023application,ELMACHKOURI20131,jenish2009central,jenish2012spatial,kojevnikov_limit_2021}
are subsets of $\mathbb{Z}^{d}$, $\R^{d}$, or a general metric space.
In contrast, our definition does not require the index $i$ to belong
to any metric space. This makes our approach applicable to network
data, where individuals are not necessarily embedded in any metric
space.

(2) Compared with strong mixing, the FDM is easier to compute because
the coupled version $Y_{j,n,i}$ is constructed explicitly, and the
relevant $L^{p}$-norm is easier to evaluate. In contrast, the strong
mixing coefficient requires the calculation of a supremum over two
$\sigma$-fields, and thus quite challenging (\citet{wu2005nonlinear}
and \citet{xu2021Conditions}).

(3) Compared to the spatial NED, FDM is more conveniently to calculate
under any $L^{p}$-norm, while the $L^{p}$-NED property is often
tractable only for $p=2$, because it typically relies on the fact
that the conditional expectation is the best predictor under $L^{2}$-distance.
\begin{defn}
\label{def:FD}For any constants $p\geq1$ and $q\geq1$, $\left\{ Y_{j,n}\right\} $
is said to be $\left(L^{p},q\right)$-functionally dependent on $\left\{ e_{i,n}\right\} $
given $\c_{n}$ if 
\[
\Delta_{p,q}\left(\c_{n}\right)\equiv\frac{1}{n^{q}}\sum^{n}_{i=1}\left[\sum^{n}_{j=1}\delta_{p,n}\left(j,i,\c_{n}\right)\right]^{q}=o_{\p}(1)
\]
as $n\to\infty$. When $\c_{n}$ is the trivial $\sigma$-field $\left\{ \Omega,\emptyset\right\} $,
we simplify the notation as $\Delta_{p,q}\equiv\Delta_{p,q}\left(\c_{n}\right)$.
When $\Delta_{p,q}=o(1)$, we simply say that $\left\{ Y_{j,n}\right\} $
is $\left(L^{p},q\right)$-functionally dependent on $\left\{ e_{i,n}\right\} $.
\end{defn}
\begin{rem}
\label{Delta-implication}The term $\Delta_{p,q}\left(\c_{n}\right)$
plays a pivotal role throughout this paper. Since $\delta_{p,n}\left(j,i,\c_{n}\right)$
describes the impact of $e_{i,n}$ on $Y_{j,n}$, $\sum^{n}_{j=1}\delta_{p,n}\left(j,i,\c_{n}\right)$
is the total impact of $e_{i,n}$ on all $Y_{j,n}$'s, which can be
regarded as the ``influence power'' of $e_{i,n}$ on $Y_{j,n}$'s.
And $\frac{1}{n}\sum^{n}_{i=1}\left[\sum^{n}_{j=1}\delta_{p,n}\left(j,i,\c_{n}\right)\right]^{q}$
can be interpreted as the ``average'' $q$th power of the influence
powers of all $e_{i,n}$'s. Thus, if the average $q$th power of the
influence powers of all $e_{i,n}$'s on $Y_{j,n}$'s is almost surely
bounded given $\c_{n}$ for some $q>1$, then $\Delta_{p,q}\left(\c_{n}\right)=o_{\p}(1)$.
Critically, we allow some (but not all) individuals to have large
influence powers, i.e., we allow $\sup_{i\in[n]}\sum^{n}_{j=1}\delta_{p,n}\left(j,i,\c_{n}\right)$
to increase to $\infty$ as $n\to\infty$. As illustrated in Section~\ref{Sec: Example},
for SAR models, $\sum^{n}_{j=1}\delta_{p,n}\left(j,i,\c_{n}\right)$
is proportional to the summation of the absolute values of the elements
in the $i$th column of the matrix $L(I_{n}-L\left|\lambda W_{n}\right|)^{-1}$,
where $W_{n}$ is the network weights matrix, i.e., $\sum^{n}_{j=1}\delta_{p,n}\left(j,i,\c_{n}\right)\propto\sum^{n}_{j=1}\left|L\left[(I_{n}-L\left|\lambda W_{n}\right|)^{-1}\right]_{ji}\right|$.
Therefore, $\Delta_{p,q}\left(\c_{n}\right)=o_{\p}(1)$ generalizes
a standard assumption in many spatial econometric papers (\citealp{h._kelejian_asymptotic_2001,lee2004asymptotic,lee_gmm_2007,yu2008quasi},
among others): $\sup_{n}\left\Vert L(I_{n}-L\left|\lambda W_{n}\right|)^{-1}\right\Vert _{1}<\infty$,
i.e., the column sum norm of $L(I_{n}-L\left|\lambda W_{n}\right|)^{-1}$
is uniformly bounded in $n$. The condition $\Delta_{p,q}\left(\c_{n}\right)=o_{\p}(1)$
excludes the case that all $Y_{j,n}$'s are mainly affected by the
same very few $e_{i,n}$'s. Consider an extreme case that $Y_{j,n}=e_{1,n}$
for all $j=1,\dots,n$. Then $\sum^{n}_{j=1}\delta_{p,n}\left(j,1\right)\propto n$,
and thus $\Delta_{p,q}=\frac{1}{n^{q}}\sum^{n}_{i=1}\left[\sum^{n}_{j=1}\delta_{p,n}\left(j,i\right)\right]^{q}\propto1$.
So $\left\{ Y_{j,n}\right\} $ is not $\left(L^{p},q\right)$-functionally
dependent on $\left\{ e_{i,n}\right\} $ for any $p\geq1$ and $q\geq1$. 
\end{rem}

\section{\label{sec:Property}Properties of Functional Dependence}

In this section, we establish a moment inequality, a concentration
inequality, laws of large numbers (LLN), and central limit theorems
(CLTs) using the FDM. These tools are essential for deriving asymptotic
theory in statistical and econometric models. 

The basic idea of most proofs is to express $\sum^{n}_{j=1}\left(Y_{j,n}-\E_{\c_{n}}Y_{j,n}\right)$
as the summation of a martingale difference array (MDA) and then apply
the theories of MDA. Thus, we first define an increasing sequence
of $\sigma$-fields. For $i\in[n]$, let $\mathcal{F}_{i,n}\equiv\sigma\left(e_{j,n}:j\leq i,j\in[n]\right)$
be the $\sigma$-field generated by $e_{1,n},\cdots,e_{i-1,n},e_{i,n}$,
and let $\mathcal{F}_{0,n}\equiv\left\{ \Omega,\emptyset\right\} $. 

\subsection{Moment inequality and weak law of large numbers}

In this subsection, we establish a moment inequality in Theorem \ref{thm: Rosenthal}
for $\sum^{n}_{j=1}(Y_{j,n}-\E_{\c_{n}}Y_{j,n})$, which directly
yield a LLN and are useful for theoretical analysis in statistics
and econometrics. We begin with a crucial lemma.
\begin{lem}
\label{lem:predi < FMD}Let $P_{i}Y_{j,n}\equiv\E_{\c_{n}}(Y_{j,n}|\mathcal{F}_{i,n})-\E_{\c_{n}}(Y_{j,n}|\mathcal{F}_{i-1,n})$.
Then $\left\Vert P_{i}Y_{j,n}\right\Vert _{L^{p},\c_{n}}\leq\delta_{p,n}(j,i,\c_{n})$
almost surely (a.s.).
\end{lem}
\begin{rem}
The quantity $P_{i}Y_{j,n}$ can be regarded as the change in the
prediction of $Y_{j,n}$ when we have the new information $e_{i,n}$,
conditional on $\mathcal{F}_{i-1,n}$ and $\c_{n}$. Notice that $\left\{ P_{i}Y_{j,n},\mathcal{F}_{i,n}\right\} $
is a MDA under both the conditional expectation $\E_{\c_{n}}$ and
the unconditional expectation $\E$,\footnote{$\E_{\c_{n}}(P_{i}Y_{j,n}|\mathcal{F}_{i-1,n})=\E_{\c_{n}}\{[\E_{\c_{n}}(Y_{j,n}|\mathcal{F}_{i,n})-\E_{\c_{n}}(Y_{j,n}|\mathcal{F}_{i-1,n})]|\mathcal{F}_{i-1,n}\}=[\E_{\c_{n}}(Y_{j,n}|\mathcal{F}_{i-1,n})-\E_{\c_{n}}(Y_{j,n}|\mathcal{F}_{i-1,n})]=0$.
Hence, $\E(P_{i}Y_{j,n}|\mathcal{F}_{i-1,n})=\E\E_{\c_{n}}(P_{i}Y_{j,n}|\mathcal{F}_{i-1,n})=0$.} and $(Y_{j,n}-\E_{\c_{n}}Y_{j,n})=\sum^{n}_{i=1}P_{i}Y_{j,n}$. 
\end{rem}
With Lemma~\ref{lem:predi < FMD} and the Burkholder\textquoteright s
inequality for martingales (Lemma~\ref{lem: Burkholder}), we obtain
the following moment inequality. 
\begin{thm}
\label{thm: Rosenthal} Let $C_{p}\equiv\sqrt{p-1}$ when $p\geq2$
and $C_{p}\equiv\frac{1}{p-1}$ when $p\in(1,2)$. Then for any constant
$p>1$, $\frac{1}{n}\left\Vert \sum^{n}_{j=1}(Y_{j,n}-\E_{\c_{n}}Y_{j,n})\right\Vert _{L^{p},\c_{n}}\leq C_{p}\left\{ \Delta_{p,\min\{p,2\}}\left(\c_{n}\right)\right\} ^{1/\min\{p,2\}}$
a.s.
\end{thm}
A direct result of Theorem \ref{thm: Rosenthal} is a law of large
numbers for functionally dependent network random variables\@.
\begin{thm}
\label{thm:WLLN}If $\Delta_{p,\min\{p,2\}}\left(\c_{n}\right)=o_{\p}(1)$
for some constant $p>1$, then $\frac{1}{n}\sum^{n}_{j=1}(Y_{j,n}-\E_{\c_{n}}Y_{j,n})\xrightarrow{\p}0$.
\end{thm}
\begin{rem}
\label{rem:lln comparison}We compare our LLN with that of \citet[Theorem 3.1]{kojevnikov_limit_2021}.
A key assumption in their LLN (Assumption 3.2) implicitly requires
that the average dependence between two individuals decays to zero
as their geodesic distance grows, and that the geodesic distances
of most nodes are large. As they note in the discussion following
Assumption 3.2, this assumption rules out networks with small diameters,
which are common in social and financial networks. In contrast, our
LLN accommodates networks with small diameters. As discussed in Remark~\ref{Delta-implication},
our assumption $\Delta_{p,\min\{p,2\}}\left(\c_{n}\right)=o_{\p}(1)$
only requires that the average influence power of $e_{i,n}$'s on
$Y_{j,n}$'s be finite, a condition that is feasible even for networks
with small diameters (see Example~\ref{exa:ER model}).
\end{rem}

\subsection{Concentration inequality}

Concentration inequalities, also called exponential inequalities in
the literature, play an indispensable role in empirical process theory,
as well as in semi-parametric, non-parametric, and high-dimensional
statistics (\citealp{wainwright2019HighDimensional}). \citet{wu2005nonlinear}
and \citet{wu2016wu} establish two concentration inequalities for
functionally dependent stationary time series. In the literature,
although there are some concentration inequalities for spatial data
on irregular lattices (e.g., \citet{XU201896}, \citet{yuan2025Bernsteintype},
and \citet{wu2023application}), the concentration inequalities for
network data are rare. Hence, in this subsection, we follow the strategies
in \citet{wu2016wu} to establish a concentration inequality for network
data. 
\begin{thm}
\label{thm:exp inequality}Let $Z_{n}\equiv\frac{1}{\sqrt{n}}\sum^{n}_{j=1}Y_{j,n}$.
Assume $\E_{\c_{n}}Y_{j,n}=0$ for all $j=1,2,\ldots,n$ and $n\geq1$.
Assume that $\sup_{n\geq1}n\Delta_{p,2}\left(\c_{n}\right)<\infty$
a.s. for all $p\geq2$ and $\sup_{n\geq1}\sqrt{n}\Delta^{1/2}_{p,2}\left(\c_{n}\right)$
increases slower than $O(p^{\nu})$ for some constant $\nu\geq0$
in the sense that 
\begin{equation}
\sup_{p\geq2}\sup_{n\geq1}p^{-\nu}\sqrt{n}\Delta^{1/2}_{p,2}\left(\c_{n}\right)\leq\gamma_{0}\text{ a.s.}\label{eq: exp ineq gamma_0}
\end{equation}
for some finite constant $\gamma_{0}>0$. Let $\alpha=\frac{2}{1+2\nu}$.
Then for $t\in[0,t_{0})$, 
\[
m\left(t\right)\equiv\mathbb{E}\left[\exp\left(t\left|Z_{n}\right|^{\alpha}\right)\mid\c_{n}\right]\leq1+c_{\alpha}\left(1-\frac{t}{t_{0}}\right)^{-1/2}\frac{t}{t_{0}}\text{ a.s.},
\]
where $t_{0}=\left(e\alpha\gamma^{\alpha}_{0}\right)^{-1}$, $c_{\alpha}$
is a constant only depending on $\alpha$. Consequently, by letting
$t=\frac{t_{0}}{2}$, we have for all $x>0$, 
\begin{equation}
\mathbb{P}\left(\left|Z_{n}\right|\geq x\mid\c_{n}\right)\leq\exp\left(-tx^{\alpha}\right)m\left(t\right)\leq\left(1+\frac{\sqrt{2}c_{\alpha}}{2}\right)\exp\left(-\frac{x^{\alpha}}{2e\alpha\gamma^{\alpha}_{0}}\right),\quad\text{a.s.}\label{eq:exponential bound}
\end{equation}
\end{thm}
We now illustrate when the condition (\ref{eq: exp ineq gamma_0})
holds. Consider the SAR model from Section~\ref{Sec: Example}. From
Proposition~\ref{prop: SAR FDM}, we have $\delta_{p,n}\left(j,i\right)\leq2\left\Vert \epsilon\right\Vert _{L^{p}}S^{+}_{ji,n}$.\footnote{For simplicity, here we assume that $\c_{n}$ is the trivial $\sigma$-field
$\{\emptyset,\Omega\}$.} As a result, we obtain 
\begin{align*}
\Delta^{1/2}_{p,2} & =\sqrt{\frac{1}{n^{2}}\sum^{n}_{i=1}\left[\sum^{n}_{j=1}\delta_{p,n}\left(j,i\right)\right]^{2}}\leq\frac{2\left\Vert \epsilon\right\Vert _{L^{p}}}{\sqrt{n}}\sqrt{\frac{1}{n}\sum^{n}_{i=1}\left[\sum^{n}_{j=1}S^{+}_{ji,n}\right]^{2}}.
\end{align*}
If $\sup_{n\geq1}\frac{1}{n}\sum^{n}_{i=1}\left[\sum^{n}_{j=1}S^{+}_{ji,n}\right]^{2}<C^{2}<\infty$
for some constant $C>0$, then 
\[
\sup_{p\geq2}\sup_{n\geq1}p^{-\nu}\sqrt{n}\Delta^{1/2}_{p,2}\leq2C\sup_{p\geq2}p^{-\nu}\left\Vert \epsilon\right\Vert _{L^{p}}.
\]
Thus, the condition (\ref{eq: exp ineq gamma_0}) holds with $\nu=1$
if $\epsilon_{i,n}$'s are uniformly sub-exponential, with $\nu=\frac{1}{2}$
if $\epsilon_{i,n}$'s are uniformly sub-Gaussian, and with $\nu=0$
if $\epsilon_{i,n}$'s are uniformly bounded. Compared to the Hoeffding's
inequality (\citealp[Proposition 2.5]{wainwright2019HighDimensional}),
we see that when $\nu=0$, the decay rate on the right-hand-side of
(\ref{eq:exponential bound}) with respect to $x$ is the same as
in the independent case. When $\nu>0$, the decay rate is slower.

\subsection{Central limit theorems\label{subsec:Central-limit-theorems}}

In this section, we establish two central limit theorems (CLTs) based
on the FDM. Denote $Z_{i,n}\equiv\sum^{n}_{k=1}P_{i}Y_{k,n}$ for
$i\in[n]$, where $P_{i}Y_{k,n}=\E_{\c_{n}}(Y_{k,n}|\mathcal{F}_{i,n})-\E_{\c_{n}}(Y_{k,n}|\mathcal{F}_{i-1,n})$.
By construction, $\{Z_{i,n},\mathcal{F}_{i,n}\}^{n}_{i=1}$ is an
MDA under both $\p_{\mathcal{C}_{n}}$ and $\p$. Therefore, we will
prove our CLTs by applying a martingale CLT (Lemma~\ref{CLT MDA})
to the normalized sum $\sigma^{-1}_{n}\sum^{n}_{i=1}Z_{i,n}$. Before
stating the CLTs, we introduce a second-order functional dependence
measure, which is tailored to the analysis of the MDA increments $\{Z_{i,n}\}$
and, to the best of our knowledge, is new in the literature.

Fix $i\neq j$. Conditional on $\c_{n}$, let $e^{*}_{i,n}$ and $e^{*}_{j,n}$
be independent copies of $e_{i,n}$ and $e_{j,n}$, respectively,
such that $(e_{1,n},\dots,e_{n,n},e^{*}_{i,n},e^{*}_{j,n})$ are mutually
independent given $\c_{n}$. Define $Y_{k,n,\{i,j\}}$ as the coupled
version of $Y_{k,n}$ obtained by replacing $e_{i,n}$ and $e_{j,n}$
with $e^{*}_{i,n}$ and $e^{*}_{j,n}$:
\[
Y_{k,n,\{i,j\}}\equiv F_{k,n}(e_{1,n},\dots,e_{i-1,n},e^{*}_{i,n},e_{i+1,n},\dots,e_{j-1,n},e^{*}_{j,n},e_{j+1,n},\dots,e_{n,n})
\]
for $i<j$. When $j<i$, the same definition applies with the roles
of $i$ and $j$ swapped. Equivalently, the above display is understood
as replacing the $i$th and $j$th coordinates simultaneously.
\begin{defn}
[Second-order functional dependence measure]\textbf{\label{def:Second-order functional dependence measure}}
For $p\geq1$, define the second-order functional dependence measure
(second-order FDM) of the random field $\left\{ Y_{i,n}\right\} $
as 
\begin{equation}
\mathfrak{d}_{p,n}\left(i,j,\c_{n}\right)\equiv\left\Vert \sum^{n}_{k=1}\left\{ Y_{k,n}-Y_{k,n,i}-Y_{k,n,j}+Y_{k,n,\{i,j\}}\right\} \right\Vert _{L^{p},\c_{n}}\text{ for }i\neq j\label{eq: Delta ij}
\end{equation}
and 
\[
\mathfrak{d}_{p,n}\left(i,i,\c_{n}\right)\equiv\left\Vert \sum^{n}_{k=1}\left\{ Y_{k,n}-Y_{k,n,i}\right\} \right\Vert _{L^{p},\c_{n}}.
\]
When $\c_{n}$ is the trivial $\sigma$-field $\left\{ \Omega,\emptyset\right\} $,
we simplify the notation as $\mathfrak{d}_{p,n}\left(i,j\right)\equiv\mathfrak{d}_{p,n}\left(i,j,\c_{n}\right)$.
\end{defn}
Why is this second-order FDM needed? To apply the martingale CLT,
we need to establish the LLN for the squared increments $\{Z^{2}_{i,n}\}$.
A direct way to show this is to invoke Theorem~\ref{thm:WLLN}, which
requires controlling the (first-order) FDM of $\{Z^{2}_{i,n}\}$ (viewed
as functions of the innovations $e_{i,n}$). By applying the difference
of squares formula, controlling the FDM of $\{Z^{2}_{i,n}\}$ reduces
to controlling the (first-order) FDM of $\{Z_{i,n}\}$. It turns out
that this FDM can then be bounded in terms of the second-order FDM
of the original field $\{Y_{i,n}\}$, as summarized below.
\begin{lem}
\label{lem:fdm of martingale < 2nd FMD}Let $Z_{i,n}=\sum^{n}_{j=1}P_{i}Y_{j,n}$
for $i\in[n]$ and denote the FDM of $\{Z_{i,n}\}$ on $\{e_{i,n}\}$
as $\delta^{\sharp}_{p,n}(i,j,\c_{n})$ for $i,j\in[n]$. Then, $\delta^{\sharp}_{p,n}(i,j,\c_{n})\leq\mathfrak{d}_{p,n}\left(i,j,\c_{n}\right)=\mathfrak{d}_{p,n}\left(j,i,\c_{n}\right)$
a.s. for all $i,j\in[n]$. 
\end{lem}
Lemma~\ref{lem:fdm of martingale < 2nd FMD} can be viewed as a second-order
analogue of Lemma~\ref{lem:predi < FMD}. Indeed, Lemma~\ref{lem:predi < FMD}
implies $\left\Vert Z_{i,n}\right\Vert _{L^{p},\c_{n}}\leq\sum^{n}_{k=1}\delta_{p,n}(k,i,\c_{n})$;
that is, the $L^{p}$ norm of $Z_{i,n}$, which can be interpreted
as a ``zeroth-order'' FDM for $\{Z_{i,n}\}$, is controlled by the
(first-order) FDM of $\{Y_{i,n}\}$. In contrast, Lemma~\ref{lem:fdm of martingale < 2nd FMD}
shows that the first-order FDM of $\{Z_{i,n}\}$ is controlled by
the second-order FDM of $\{Y_{i,n}\}$. With this lemma in hand, we
are ready to state our CLTs.

\subsubsection{A univariate CLT}
\begin{thm}
\label{thm:CLT finite}Denote $\sigma^{2}_{n}\equiv\var_{\c_{n}}\left(\sum^{n}_{j=1}Y_{j,n}\right)$.
Let $p>2$ be a constant. Suppose that (1) $\sigma^{-2}_{n}=O_{\p}(n^{-1})$,
(2)
\begin{equation}
\frac{1}{n^{p/2}}\sum^{n}_{i=1}\left\{ \sum^{n}_{k=1}\delta_{p,n}(k,i,\c_{n})\right\} ^{p}=o_{\p}(1),\label{eq:l1 bounded for finite clt}
\end{equation}
and (3)
\begin{equation}
\frac{1}{n^{\min\{2,p/2\}}}\sum^{n}_{i=1}\left[\left\{ \sum^{n}_{k=1}\delta_{p,n}(k,i,\c_{n})\right\} \sum^{n}_{j=1}\mathfrak{d}_{p,n}\left(i,j,\c_{n}\right)\right]^{\min\{2,p/2\}}=o_{\p}(1)\label{eq:key condition for finite clt}
\end{equation}
as $n\to\infty$. Then 
\[
\frac{\sum^{n}_{j=1}\left\{ Y_{j,n}-\E_{\c_{n}}Y_{j,n}\right\} }{\sigma_{n}}\xrightarrow{d}N\left(0,1\right).
\]
\end{thm}
\textbf{}
\begin{rem}
While Theorem~\ref{thm:CLT finite} establishes asymptotic normality,
feasible inference additionally requires a consistent estimator of
$\sigma^{2}_{n}$. At this level of generality, we do not develop
a universal HAC-type estimator for the asymptotic variance under network
dependence. Rather, variance estimation is expected to be model-specific.
For example, for likelihood-based estimators, one may exploit the
information equality and estimate the variance using the Hessian matrix.
More generally, in parametric settings where the error distribution
belongs to a known family, simulation-based methods may be used to
approximate the asymptotic variance. A systematic treatment of feasible
variance estimation under the present abstract framework is left for
future work. 
\end{rem}
We next discuss how to verify the high-level conditions in Theorem~\ref{thm:CLT finite}.
First note that conditions (\ref{eq:l1 bounded for finite clt}) and
(\ref{eq:key condition for finite clt}) are invariant under relabeling
of the nodes $1,\ldots,n$. This is a natural requirement because
the validity of the CLT should depend on the dependence structure
of the network, not on an arbitrary ordering.

Condition (\ref{eq:l1 bounded for finite clt}) is based on the (first-order)
FDM and relatively easy to verify. Condition (\ref{eq:key condition for finite clt})
involves the second\nobreakdash-order FDM and is more demanding.
The next two lemmas provide two different ways to control the second-order
FDM. Lemma~\ref{lem:suffi condi for clt-smooth} exploits smoothness
of the functions $F_{j,n}$ (recall that $Y_{j,n}=F_{j,n}(e_{n})$),
whereas Lemma~\ref{lem:suffi condi for clt} does not require smoothness
and thus applies more generally.
\begin{lem}
\label{lem:suffi condi for clt-smooth}Consider the system (\ref{eq: Y_i}).
Suppose that $e_{1,n},\ldots,e_{n,n}$ are univariate, i.e., $p_{e}=1$.
Let $\left\Vert e\right\Vert _{L^{p},\c_{n}}\equiv\sup_{i\in[n]}\left\Vert e^{*}_{i,n}-e_{i,n}\right\Vert _{L^{p},\c_{n}}$,
where $e^{*}_{i,n}$ is an i.i.d. copy of $e_{i,n}$ conditional on
$\c_{n}$. Assume the (random) functions $F_{1,n}(\cdot),\ldots,F_{n,n}(\cdot)$
are measurable with respect to $\c_{n}$ and twice continuously differentiable
a.s. Then it holds that $\mathfrak{d}_{p,n}\left(i,i,\c_{n}\right)\leq\sum^{n}_{k=1}\delta_{p,n}(k,i,\c_{n}),\ \forall i\in[n],$
and 
\[
\mathfrak{d}_{p,n}\left(i,j,\c_{n}\right)\leq\left\Vert e\right\Vert ^{2}_{L^{p},\c_{n}}\sup_{e_{n}}\left|\sum^{n}_{k=1}\frac{\partial^{2}F_{k,n}(e_{n})}{\partial e_{i,n}\partial e_{j,n}}\right|,\ \forall i\neq j\text{ a.s.}
\]
 
\end{lem}
Lemma~\ref{lem:suffi condi for clt-smooth} is particularly convenient
for smooth models. In Section~\ref{Sec: Example}, we apply it to
the SAR model (see Lemma~\ref{lem:2nd FDM for SAR} and Proposition~\ref{prop:2nd FDM for SAR}).
The next lemma provides an alternative route that avoids smoothness
assumptions.
\begin{lem}
\label{lem:suffi condi for clt} Consider the system (\ref{eq: Y_i}). 
\begin{enumerate}[label=(\roman*)]
\item For any $i,j\in[n]$, it holds that 
\[
\mathfrak{d}_{p,n}\left(i,j,\c_{n}\right)\leq2\sum^{n}_{k=1}\min\left\{ \delta_{p,n}(k,i,\c_{n}),\delta_{p,n}(k,j,\c_{n})\right\} .
\]
\item For each $k\in[n]$, let $\pi_{k}(1),\ldots,\pi_{k}(n)$ be a permutation
of $1,\ldots,n$ such that 
\[
\delta_{p,n}\left(k,\pi_{k}(1),\c_{n}\right)\ge\delta_{p,n}\left(k,\pi_{k}(2),\c_{n}\right)\ge\cdots\ge\delta_{p,n}\left(k,\pi_{k}(n),\c_{n}\right).
\]
If $\sup_{i\in[n]}\sum^{n}_{k=1}\delta_{p,n}(k,i,\c_{n})=O_{\p}(1),$and
there exist two constants $\alpha>\frac{\min\{2,p/2\}}{\min\{2,p/2\}-1}>1$
and $C>0$ such that 
\begin{equation}
\delta_{p,n}\left(k,\pi_{k}(r),\c_{n}\right)\le Cr^{-\alpha},\qquad\forall\,k,r\in[n],\label{eq:suff condition for clt}
\end{equation}
with probability approaching one, then condition (\ref{eq:key condition for finite clt})
holds. 
\end{enumerate}
\end{lem}
Condition (\ref{eq:suff condition for clt}) requires that, for each
node $k$, the first-order dependence measures $\delta_{p,n}\left(k,\pi_{k}(r),\c_{n}\right)$
decay at a polynomial rate when ordered from largest to smallest.
In other words, although a node may be affected by many other nodes,
the strength of these influences must decline sufficiently fast across
the ordered neighbors.

\subsubsection{Comparison with the CLTs in the literature\label{subsec:Comparison-with-the}}

We now compare our CLT with existing results in the literature.

(1) The CLTs in \citet{wu2005nonlinear,wu2023application,ELMACHKOURI20131,jenish2009central,jenish2012spatial,kojevnikov_limit_2021}
all require individuals to be located in some metric space. Our CLT
imposes no such metric structure, making it applicable to network
data, where a natural embedding may be absent.

(2) The key condition for CLTs in the literature (e.g., \citealp{jenish2009central,jenish2012spatial,kojevnikov_limit_2021,wu2023application})
typically takes the form $\sum^{\infty}_{s=1}\nu_{s}\theta_{s}<\infty$,
where $s$ denotes distance, $\nu_{s}$ is the number of individuals
at distance approximately equal to $s$ from a given individual, and
$\theta_{s}$ is a weak dependence coefficient at distance $s$ (e.g.,
a strong mixing, NED, or $\psi$-dependence coefficient). For this
condition to hold, one typically needs $\theta_{s}$ to decay sufficiently
fast as $s$ increases, and the growth of $\nu_{s}$ must also be
controlled. As a result, these conditions are most natural in settings
where dependence weakens gradually over many distance shells. This
requirement can be restrictive for network data, since many relevant
networks exhibit small-world behavior and hence have small graph diameter.\footnote{For example, in the Erd\H{o}s-R\'{e}nyi model in Example~\ref{exa:ER model},
if the mean degree satisfies $D_{n}\propto n^{\alpha}$ ($0<\alpha<1)$,
then results in \citet{klee1981diameters} imply that the graph diameter
(based on the geodesic distance) is bounded by a constant with probability
approaching one. Consequently, the graph radius is also of constant
order. In such settings, existing weak dependence concepts based on
decay with network distance become less compelling, because only finitely
many distance shells are asymptotically relevant.} In contrast, our condition~(\ref{eq:suff condition for clt}) only
requires that, for any node $i$, the magnitude of the impact it receives
from other nodes (i.e., the FDM $\delta_{p,n}\left(k,\pi_{k}(r),\c_{n}\right)$)
decreases as a power function of $r$ when these impacts are ordered
from largest to smallest. This allows our CLT to accommodate networks
with small or moderate diameters, a common feature of social and financial
networks.\footnote{Financial networks, ranging from interbank markets to asset holding
networks, typically exhibit small diameters. For example, consider
the holding-based network of Chinese mutual funds. Funds $i$ and
$j$ are regarded as connected if they allocate at least 5\% of their
portfolios to the same stocks. Using data from the second half of
2015, \citet{jiang2026NAVaR} find that this network has a diameter
of only 4.}

(3) \citet{kuersteiner2019limit} develops a CLT for spatial mixingale
processes using a model\nobreakdash-dependent random metric, relaxing
the assumption of a fixed metric. \citet{kuersteiner2019limit} assumes
the network to be sparse that ``rules out a buildup of a mass of
nodes with very similar features is captured by a summability condition
of the probabilities that two nodes are close in an appropriate sense''.
In contrast, our approach does not require any form of underlying
metric and allows for networks with small diameters.

(4) Going beyond the conventional notion of weak dependence (e.g.,
mixing and NED), \citet{leung_normal_2019} propose the \textquotedblleft stabilization\textquotedblright{}
conditions that impose weak dependence on node degrees. While their
motivation is similar, they focus on strategic network formation,
whereas our concept applies to various types of networks.

(5) \citet{lee2022qml} study a CLT for a linear-quadratic form of
independent random variables in the presence of dominant units (also
called popular units in their paper). Their CLT is mainly used for
linear SAR type models, but ours is applicable to nonlinear models
and some nonlinear estimators (e.g., Huber or quantile regression
estimators) of linear SAR models.

\subsubsection{A multivariate CLT }

Using the Cram\'er-Wold device, we can generalize Theorem \ref{thm:CLT finite}
to multivariate case. Now, $Y_{j,n}$'s are random vectors taking
values in $\R^{p_{Y}}$ ($p_{Y}\geq1$). Denote $\Sigma_{n}\equiv\var_{\c_{n}}\left(\sum^{n}_{j=1}\text{\ensuremath{Y_{j,n}}}\right)$,
and let $I_{p_{Y}}$ be the $p_{Y}\times p_{Y}$ identity matrix. 
\begin{thm}
\label{thm: multi CLT finite}Suppose that (1) $\{\min\eig(\Sigma_{n})\}^{-1}=O_{\p}(n^{-1})$,
(2) $\frac{1}{n^{p/2}}\sum^{n}_{i=1}\left\{ \sum^{n}_{k=1}\delta_{p,n}(k,i,\c_{n})\right\} ^{p}=o_{\p}(1),$
and (3) $\frac{1}{n^{\min\{2,p/2\}}}\sum^{n}_{i=1}\left[\left\{ \sum^{n}_{k=1}\delta_{p,n}(k,i,\c_{n})\right\} \sum^{n}_{j=1}\mathfrak{d}_{p,n}\left(i,j,\c_{n}\right)\right]^{\min\{2,p/2\}}=o_{\p}(1)$
as $n\to\infty$. Then, $\Sigma^{-1/2}_{n}\sum^{n}_{j=1}\left\{ Y_{j,n}-\E_{\c_{n}}Y_{j,n}\right\} \xrightarrow{d}N\left(0,I_{p_{Y}}\right).$
\end{thm}

\section{\label{Sec: Example}Application of functional dependence to SAR
models}

To illustrate the application of the FDM, in this section, we calculate
the FDM for SAR models and establish the consistency of the maximum
likelihood estimator for the SAR Tobit model using FDM. 

\subsection{\label{subsec:SAR-model}SAR models}

Let $F:\mathbb{R}\to\mathbb{R}$ be a Lipschitz function, i.e., $\left|F\left(x\right)-F\left(y\right)\right|\leq L\left|x-y\right|$
for some constant $L>0$ and all $x,y\in\mathbb{R}$. A (possibly
nonlinear) SAR model can be written as 
\begin{equation}
Y_{j,n}=F\left(\lambda w_{j\cdot,n}Y_{n}+X'_{j,n}\beta+\epsilon_{j,n}\right)\label{eq:SAR model}
\end{equation}
for $j=1,\ldots,n$, where $w_{j\cdot,n}$ is the $j$th row of a
non-zero spatial/network weights matrix $W_{n}=\left(w_{ji,n}\right)_{n\times n}$,
$Y_{n}=\left(Y_{1,n},Y_{2,n},\cdots,Y_{n,n}\right)'$, $X_{j,n}\in\mathbb{R}^{p}$
is the exogenous regressor, $\epsilon_{j,n}$ is the disturbance term,
and $\lambda$ and $\beta$ are model parameters. When $F\left(x\right)=x$,
Eq.\eqref{eq:SAR model} is the standard (linear) SAR model; when
$F\left(x\right)=\max\left(0,x\right)$, Eq.\eqref{eq:SAR model}
becomes a SAR Tobit model. Now, we investigate the FDM of SAR model.
The following two assumptions on $F(\cdot)$ and $\{\epsilon_{j,n}\}$
are needed.
\begin{assumption}
\label{ass: zeta}$F$ is a Lipschitz function with a Lipschitz constant
$L>0$, and $\zeta\equiv L\left|\lambda\right|\sup_{n}\left\Vert W_{n}\right\Vert _{\infty}<1$.
\end{assumption}
\begin{rem}
Assumption \ref{ass: zeta} is also considered by \citet{wu2023application},
and it ensures the existence and uniqueness of the solution of Eq.\eqref{eq:SAR model}.
It is analogous to the stationarity condition for autoregressive models
in time series. 
\end{rem}
\begin{assumption}
\label{ass: x epsi moment}Denote $\c_{n}\equiv\bigvee^{n}_{j=1}\sigma\left(X_{j,n}\right)\bigvee\sigma(W_{n})$.
$\epsilon_{j,n}$'s are conditionally independent given $\c_{n}$
and $\left\Vert \epsilon\right\Vert _{L^{p},\c_{n}}\equiv\sup_{j,n}\left\Vert \epsilon_{j,n}\right\Vert _{L^{p},\c_{n}}<\infty$
a.s. for some $p>1$.
\end{assumption}
Note that Assumption~\ref{ass: x epsi moment} allows the weights
matrix $W_{n}$ to be random. Let $e_{j,n}=X'_{j,n}\beta+\epsilon_{j,n}$
in the SAR model \eqref{eq:SAR model}. Then $Y_{i,n}$'s are generated
by the underlying random variables $e_{1,n},\ldots,e_{n,n}$. From
Assumption \ref{ass: x epsi moment}, $e_{j,n}$'s are conditionally
independent on $\c_{n}$. Denote $\left|W_{n}\right|\equiv(|w_{ij,n}|)_{n\times n}$
and 
\begin{equation}
S^{+}_{n}\equiv L\left(I_{n}-L\left|\lambda W_{n}\right|\right)^{-1}=(S^{+}_{ji,n})_{n\times n},\label{eq:S_n}
\end{equation}
where $I_{n}$ denotes the $n\times n$ identity matrix.
\begin{prop}
\label{prop: SAR FDM}Under Assumptions \ref{ass: zeta}-\ref{ass: x epsi moment},
$\delta_{p,n}\left(j,i,\c_{n}\right)\leq2\left\Vert \epsilon\right\Vert _{L^{p},\c_{n}}S^{+}_{ji,n}$
a.s. 
\end{prop}
As a direct consequence of Proposition~\ref{prop: SAR FDM} and Theorem~\ref{thm:WLLN},
we have the following LLN for the dependent variable $Y_{i,n}$ generated
by the SAR model (\ref{eq:SAR model}). 
\begin{prop}
\label{prop:lln for sar}Suppose Assumptions \ref{ass: zeta}-\ref{ass: x epsi moment}
hold. Let $\mu_{\min\{p,2\}}\equiv\left\{ \frac{1}{n}\sum^{n}_{i=1}\left\Vert S^{+}_{\cdot i,n}\right\Vert ^{\min\{p,2\}}_{1}\right\} ^{\frac{1}{\min\{p,2\}}}$,
where $S^{+}_{\cdot i,n}$ denotes the $i$th column of $S^{+}_{n}$.
If $\mu_{\min\{p,2\}}=o_{\p}(n^{\frac{\min\{p,2\}-1}{\min\{p,2\}}})$,
then for the SAR model (\ref{eq:SAR model}), it holds that $\frac{1}{n}\sum^{n}_{i=1}(Y_{i,n}-\E_{\c_{n}}Y_{i,n})\xrightarrow{\p}0$.
\end{prop}
Proposition \ref{prop:lln for sar} directly follows from Proposition~\ref{prop: SAR FDM}
and Theorem~\ref{thm:WLLN}, and we omit the proof. Since $\left\Vert S^{+}_{n}\right\Vert _{1}\geq\mu_{\min\{p,2\}}$,
a sufficient condition for LLN is $\left\Vert S^{+}_{n}\right\Vert _{1}=o_{\p}(n^{\frac{\min\{p,2\}-1}{\min\{p,2\}}})$.

Next, we bound the second-order FDM for SAR models in order to establish
CLT for the dependent variable $Y_{i,n}$ generated by the SAR model
(\ref{eq:SAR model}). To do so, we make the following assumption
on the smoothness of the function $F$.
\begin{assumption}
\label{assu:F second order derivative}The function $F:\mathbb{R}\to\mathbb{R}$
is twice continuously differentiable. Moreover, its second derivative
is uniformly bounded, i.e., $\|F''\|_{\infty}\equiv\sup_{x\in\mathbb{R}}|F''(x)|<\infty.$
\end{assumption}
Relying on Lemma~\ref{lem:suffi condi for clt-smooth}, we have the
following lemma on bounding the second-order FDM for SAR models. 
\begin{lem}
\label{lem:2nd FDM for SAR}Under Assumptions~\ref{ass: zeta}--\ref{assu:F second order derivative},
the second-order FDM for $\{Y_{i,n}\}$ under the SAR model (\ref{eq:SAR model})
satisfies $\mathfrak{d}_{p,n}\left(i,j,\c_{n}\right)\leq\frac{4\|F''\|_{\infty}}{L^{3}}\left\Vert \epsilon\right\Vert ^{2}_{L^{p},\c_{n}}\sum^{n}_{k=1}\left\Vert S^{+}_{\cdot k,n}\right\Vert _{1}S^{+}_{ki,n}S^{+}_{kj,n}\ \text{ for }i\neq j$
and $\mathfrak{d}_{p,n}\left(i,i,\c_{n}\right)\leq2\left\Vert \epsilon\right\Vert _{L^{p},\c_{n}}\sum^{n}_{k=1}S^{+}_{ki,n}\ \text{ for }i\in[n]\text{ a.s.}$
\end{lem}
Based on Lemma~\ref{lem:2nd FDM for SAR}, we impose the following
assumption to verify condition (\ref{eq:key condition for finite clt}).
\begin{assumption}
\label{assu:S column condition}For any $q>1$, define $\mu_{q}\equiv\left\{ \frac{1}{n}\sum^{n}_{i=1}\left\Vert S^{+}_{\cdot i,n}\right\Vert ^{q}_{1}\right\} ^{1/q}$
and $\mu_{\infty}\equiv\left\Vert S^{+}_{n}\right\Vert _{1}$, where
$S^{+}_{\cdot i,n}$ denotes the $i$th column of $S^{+}_{n}$. Let
$\tilde{p}\equiv\min\{2,p/2\}$. The matrix $S^{+}_{n}$ satisfies
(i) $\mu^{2\tilde{p}}_{2\tilde{p}}=o_{\p}(n^{\tilde{p}-1})$ and (ii)
there exists a pair of conjugate indices $u,v\in[1,\infty]$ such
that $1/u+1/v=1$ (with the convention $1/\infty=0$) and $\mu^{(2\tilde{p}-1)+1/u}_{u(2\tilde{p}-1)+1}\cdot\mu^{\tilde{p}}_{v\tilde{p}}=o_{\p}(n^{\tilde{p}-1}).$
\end{assumption}
Notice that $\mu_{q}$ increases to $\mu_{\infty}$ as $q\to\infty$.
We provide a sufficient condition for Assumption~\ref{assu:S column condition}.
If $\mu_{\infty}=o_{\p}(n^{(\tilde{p}-1)/(3\tilde{p}-1)})$, we have
$\mu^{2\tilde{p}}_{2\tilde{p}}\leq\mu^{2\tilde{p}}_{\infty}=o(n^{2\tilde{p}(\tilde{p}-1)/(3\tilde{p}-1)})=o_{\p}(n^{\tilde{p}-1}).$
Besides, by letting $u=\infty$ and $v=1$, we have $\mu^{(2\tilde{p}-1)+1/u}_{u(2\tilde{p}-1)+1}\cdot\mu^{\tilde{p}}_{v\tilde{p}}\leq\mu^{2\tilde{p}-1}_{\infty}\mu^{\tilde{p}}_{1}\leq\mu^{2\tilde{p}-1}_{\infty}\mu^{\tilde{p}}_{\infty}=o_{\p}(n^{\tilde{p}-1}).$
Thus, a sufficient condition for Assumption~\ref{assu:S column condition}
is $\|S^{+}_{n}\|_{1}=o_{\p}\bigl(n^{(\tilde{p}-1)/(3\tilde{p}-1)}\bigr)$.
An important special case is $\sup_{n}\|S^{+}_{n}\|_{1}=O_{\p}(1)$,
which is widely imposed in the literature; see, for example, \citet{lee2004asymptotic,lee_gmm_2007,yu2008quasi,xu2015maximum}.
Moreover, since $\frac{\tilde{p}-1}{3\tilde{p}-1}\le\frac{\min\{p,2\}-1}{\min\{p,2\}},$
the condition $\|S^{+}_{n}\|_{1}=o_{\p}(n^{(\tilde{p}-1)/(3\tilde{p}-1)})$
is stronger than the corresponding condition used for the LLN. Hence,
by Proposition~\ref{prop:lln for sar}, it also implies the LLN.

Next, we state the CLT result for the SAR model. 
\begin{prop}
\label{prop:2nd FDM for SAR}Suppose Assumptions~\ref{ass: zeta}--\ref{assu:S column condition}
hold with $p>2$. Then conditions (\ref{eq:l1 bounded for finite clt})--(\ref{eq:key condition for finite clt})
hold for the dependent variable $Y_{i,n}$ generated by the SAR model
(\ref{eq:SAR model}). In addition, if $\sigma^{-2}_{n}\equiv\left\{ \var_{\c_{n}}\left(\sum^{n}_{j=1}Y_{j,n}\right)\right\} ^{-1}=O_{\p}(n^{-1})$,
then $\sigma^{-1}_{n}\sum^{n}_{i=1}\left\{ Y_{i,n}-\E_{\c_{n}}Y_{i,n}\right\} \xrightarrow{d}N\left(0,1\right)$.
\end{prop}
We next verify Assumption~\ref{assu:S column condition} in several
commonly used network settings. We start with a deterministic design
that allows for a small number of highly influential units. We then
consider three random graph models: the Erd\H{o}s-R\'{e}nyi model,
the triangle model, and the stochastic block model. Together, these
examples show that Assumption~\ref{assu:S column condition} holds
under a range of network structures of practical interest.
\begin{example}
[Dominant popular units]\label{exa:popular units}We first consider
the ``dominant popular units'' structure for the network weights
matrix $W_{n}$ proposed by \citet{lee2022qml}. In this example,
the network matrix is nonrandom and may contain a small group of highly
influential units. This setting serves as a useful benchmark because
it isolates, in a deterministic way, the effect of column concentration
generated by dominant units. Following \citet{lee2022qml}, partition
the network weights matrix as 
\begin{equation}
W_{n}=\begin{pmatrix}W_{n,11} & W_{n,1B}\\
W_{n,B1} & W_{n,BB}
\end{pmatrix},\label{eq:popular units matrix-1}
\end{equation}
where the first block corresponds to the $m_{n}$ dominant units,
with $m_{n}=O(n^{\eta_{m}})$ for some $0\le\eta_{m}<1$. We assume
that the column sums associated with the dominant units in $W_{n}$
are of order $O(n^{\delta})$ for some $0\le\delta<1$; that is, the
column sums of $\left(W'_{n,11},W'_{n,B1}\right)'$ have magnitude
$O(n^{\delta})$.
\end{example}
Within this framework, it is useful to distinguish two topological
scenarios according to whether non-dominant units can feed back into
the dominant ones.
\begin{casenv}
\item \textbf{General dominant network ($W_{n,1B}\neq0$).} In this case,
non-dominant units may influence the dominant units. As emphasized
in Proposition 3(a) of \citet{lee2022qml}, this feedback channel
can amplify propagation through the hubs and therefore makes Assumption~\ref{assu:S column condition}
harder to satisfy. 
\item \textbf{Strict Dominant Structure ($W_{n,1B}=0$).} In this case,
non-dominant units do not influence the dominant units. This removes
the feedback-amplification channel and allows a sharper bound on the
column sum of $S^{+}_{n}$; see Proposition 3(b) of \citet{lee2022qml}.
\end{casenv}
The next proposition makes this comparison precise by characterizing
when Assumption~\ref{assu:S column condition} holds under each of
the two dominant-unit configurations.
\begin{prop}[Dominant popular units]
\label{prop:verify clt for popular units}Under the dominant popular
units structure in Example~\ref{exa:popular units}, suppose that
$\eta_{m}+\delta<1$ and that $\left\Vert \left(I_{n-m_{n}}-L\left|\lambda W_{n,BB}\right|\right)^{-1}\right\Vert _{1}$
and $\left\Vert \left(I_{n-m_{n}}-L\left|\lambda W_{n,BB}\right|\right)^{-1}\right\Vert _{\infty}$
are uniformly bounded. Then:
\begin{enumerate}[label=(\roman*)]
\item for the general dominant network ($W_{n,1B}\neq0$), Assumption~\ref{assu:S column condition}
holds with $p=4$ if $0\leq\eta_{m}+\delta<1/5$;
\item for the strict dominant structure ($W_{n,1B}=0$), Assumption~\ref{assu:S column condition}
holds with $p=4$ if $(\eta_{m},\delta)\in\{(\eta_{m},\delta):0\leq\eta_{m}<\frac{1}{3},0\leq\delta<\frac{3-5\eta_{m}-\sqrt{(1-\eta_{m})(5-7\eta_{m})}}{2}\}$.
\end{enumerate}
\end{prop}
Proposition~\ref{prop:verify clt for popular units} shows that the
strict dominant structure permits a wider range of $(\eta_{m},\delta)$
than the general case. This reflects the fact that ruling out feedback
from non-dominant units to dominant units makes the concentration
of column influence easier to control. In particular, under the strict
dominant structure, when $\eta_{m}=0$ we require $\delta<(3-\sqrt{5})/2\approx0.382$,
whereas when $\eta_{m}=1/5$ we require $\delta<(5-3\sqrt{2})/5\approx0.152$.

We next consider network weights matrices generated by random graph
models.
\begin{example}
[Erd\H{o}s-R\'{e}nyi model]\label{exa:ER model}Suppose that the
network is generated by an undirected Erd\H{o}s-R\'{e}nyi graph without
self-links. For each unordered pair of distinct individuals $\{i,j\}$,
a link is formed independently with probability $\frac{D_{n}}{n-1}$,
where $D_{n}\in[0,n-1]$ is a deterministic sequence that parameterizes
the expected degree. This normalization is natural because each node
has $n-1$ potential neighbors, so if $d_{i,n}$ denotes the degree
of node $i$, then $d_{i,n}\sim\mathrm{Binomial}(n-1,\frac{D_{n}}{n-1})$
and hence $\E[d_{i,n}]=D_{n}.$ Let $A_{n}=(A_{ij,n})_{n\times n}$
denote the resulting adjacency matrix, with $A_{ii,n}=0$ and $A_{ij,n}=A_{ji,n}$.
The network weights matrix $W_{n}=(w_{ij,n})_{n\times n}$ is defined
by row-normalizing $A_{n}$: 
\[
w_{ij,n}=\begin{cases}
A_{ij,n}/d_{i,n}, & d_{i,n}>0,\\
0, & d_{i,n}=0.
\end{cases}
\]

\end{example}
The ER model provides a convenient baseline for the random-graph setting.
The next proposition shows that Assumption~\ref{assu:S column condition}
generally holds for the ER model.
\begin{prop}[Erd\H{o}s-R\'{e}nyi model]
\label{prop:ER_S_column_condition}Recall that $S^{+}_{n}=L\left(I_{n}-L\left|\lambda W_{n}\right|\right)^{-1}$.
In Example~\ref{exa:ER model}, if $L|\lambda|<1$, then Assumption~\ref{assu:S column condition}
holds. 
\end{prop}
Proposition~\ref{prop:ER_S_column_condition} shows that Assumption~\ref{assu:S column condition}
holds for the Erd\H{o}s-R\'{e}nyi model without any further restriction
on the growth rate of $D_{n}$. This highlights that the proposed
dependence concept can accommodate highly connected networks with
small or moderate diameter.
\begin{example}
[Triangle model]\label{exa:triangle model}We next consider a random
graph model with local clustering. Related subgraph-based random graph
models with links and triangles are studied by \citet{chandrasekhar2025Network}.
Suppose that each unordered triple of distinct individuals forms a
triangle with probability $T_{n}/\binom{n}{3}$, where $T_{n}$ denotes
the expected number of triangles in the network. Let $A^{(\mathrm{T})}_{n}=(A^{(\mathrm{T})}_{ij,n})_{n\times n}$
denote the adjacency matrix generated by these triangles, where 
\[
A^{(\mathrm{T})}_{ij,n}=\begin{cases}
1, & \text{if nodes }i\text{ and }j\text{ belong to at least one selected triangle},\\
0, & \text{otherwise}.
\end{cases}
\]
In addition, let $A^{(\mathrm{ER})}_{n}$ be an independent Erd\H{o}s-R\'{e}nyi
adjacency matrix with link probability $\frac{D_{n}}{n-1}$, so that
$D_{n}$ denotes the expected degree contributed by the background
links. We define the overall adjacency matrix by $A_{n}=(A_{ij,n})=(\max\{A^{(\mathrm{T})}_{ij,n},A^{(\mathrm{ER})}_{ij,n}\})_{n\times n},$
and let $W_{n}$ be the associated row-normalized weights matrix.
The following proposition verifies Assumption~\ref{assu:S column condition}
for this triangle network design under a simple restriction on the
overall level of expected degree. For $1\le i<j<k\le n$, let $B_{\{i,j,k\},n}\equiv1$
when the triangle generating process generates a triangle $\{i,j,k\}$,
and $B_{\{i,j,k\},n}\equiv0$ otherwise. 
\end{example}
\begin{prop}[Triangle model]
\label{prop:triangle_S_column_condition} In Example~\ref{exa:triangle model},
suppose that the triangle indicators $B_{\{i,j,k\},n}$ are i.i.d.
$\mathrm{Bernoulli}\!\left(\frac{T_{n}}{\binom{n}{3}}\right)$, and
the Erd\H{o}s-R\'{e}nyi edges $\{A^{(\mathrm{ER})}_{ij,n}:1\le i<j\le n\}$
are i.i.d. $\mathrm{Bernoulli}\!\left(\frac{D_{n}}{n-1}\right)$,
independent of the triangle indicators. If $L|\lambda|<1$ and 
\begin{equation}
D_{n}+\frac{6T_{n}}{n}\le C_{0}\log n\label{eq:triangle_sparse_condition}
\end{equation}
for some constant $C_{0}>0$ and all sufficiently large $n$, then
Assumption~\ref{assu:S column condition} holds. 
\end{prop}
This result shows that Assumption~\ref{assu:S column condition}
continues to hold when the network exhibits local clustering, provided
that the overall degree level remains sufficiently controlled. We
next turn to a different source of network heterogeneity, namely latent
community structure.
\begin{example}
[Stochastic block model]\label{exa:Stochastic block models}We finally
consider the stochastic block model (SBM), which allows link probabilities
to differ within and across latent communities. The SBM is widely
used in the analysis of network community structure; see, for example,
\citet{abbe2018community} and \citet{wu2023distributed}. Suppose
there are $M_{n}$ blocks in total, and each individual $i$ is independently
assigned to a latent block $g_{i,n}\in\{1,\ldots,M_{n}\}$ with probability
$\p(g_{i,n}=m)=1/M_{n}$, $m=1,\ldots,M_{n}$. Let $D_{\mathrm{wb},n}$
and $D_{\mathrm{bb},n}$ denote the within-block and between-block
expected degrees, respectively. Conditional on the block assignments,
links are formed independently, with 
\[
\p(A_{ij,n}=1\mid g_{i,n}=g_{j,n})=\frac{D_{\mathrm{wb},n}}{n/M_{n}-1},\qquad\p(A_{ij,n}=1\mid g_{i,n}\neq g_{j,n})=\frac{D_{\mathrm{bb},n}}{n-n/M_{n}},
\]
for $1\le i<j\le n$, and $A_{ii,n}=0$, $A_{ij,n}=A_{ji,n}$. The
normalization is chosen so that $D_{\mathrm{wb},n}$ and $D_{\mathrm{bb},n}$
correspond to the expected numbers of within-block and between-block
neighbors, respectively. Let $A_{n}=(A_{ij,n})_{n\times n}$ denote
the resulting adjacency matrix, and let $W_{n}$ be the associated
row-normalized weights matrix.

The next proposition verifies Assumption~\ref{assu:S column condition}
under this SBM specification.
\end{example}
\begin{prop}[Stochastic block model]
\label{prop:SBM_S_column_condition} In Example~\ref{exa:Stochastic block models},
suppose that $M_{n}\ge2$, $\frac{D_{\mathrm{wb},n}}{n/M_{n}-1}\in[0,1]$,
and $\frac{D_{\mathrm{bb},n}}{n-n/M_{n}}\in[0,1]$. If $M_{n}\log M_{n}=o(n)$
and $L|\lambda|<1$, then Assumption~\ref{assu:S column condition}
holds.
\end{prop}
This result shows that Assumption~\ref{assu:S column condition}
continues to hold in the presence of community structure. It is also
worth noting that, beyond the requirement that the edge probabilities
lie in $[0,1]$, the proposition does not impose a separate growth
condition on $D_{\mathrm{wb},n}$ or $D_{\mathrm{bb},n}$.

Taken together, the above examples show that Assumption~\ref{assu:S column condition}
holds in a broad class of network settings, including deterministic
designs with influential units as well as standard random graph models.
As a result, the CLT developed in Proposition~\ref{prop:2nd FDM for SAR}
applies under a range of dependence patterns commonly used in empirical
and theoretical work on networks.

\subsection{Application to the MLE of the SAR Tobit model}

In this section, we investigate the consistency of the maximum likelihood
estimator (MLE) for the SAR Tobit model in \citet{xu2015maximum}.
Specifically, for each node $i\in[n]$, the model is defined as: 
\[
Y_{i,n}=\max\{0,Y^{*}_{i,n}\},\qquad Y^{*}_{i,n}=\lambda_{0}w_{i\cdot,n}Y_{n}+X^{\prime}_{i,n}\beta_{0}+\epsilon_{i,n},
\]
where the exogenous regressors $X_{i,n}$ and the network weights
matrix $W_{n}$ generate the sub-$\sigma$-field $\c_{n}\equiv\bigvee^{n}_{j=1}\sigma\left(X_{j,n}\right)\bigvee\sigma(W_{n})$.
Conditional on $\mathcal{C}_{n}$, the innovations $\epsilon_{1,n},\dots,\epsilon_{n,n}$
are assumed to be i.i.d. following a normal distribution $N(0,\sigma^{2}_{0})$.
Fix a parameter value $\theta=(\lambda,\beta,\sigma)$. Let $z_{i,n}(\theta)=(Y_{i,n}-\lambda w_{i\cdot,n}Y_{n}-X^{\prime}_{i,n}\beta)/\sigma$
and $G_{n}(Y_{n})=\text{diag}(1(Y_{1,n}>0),\dots,1(Y_{n,n}>0))$.
The corresponding log-likelihood function can be written as
\begin{align*}
\ell_{n}(\theta)= & \sum^{n}_{i=1}1(Y_{i,n}=0)\log\Phi(z_{i,n}(\theta))-\frac{1}{2}\log(2\pi\sigma^{2})\sum^{n}_{i=1}1(Y_{i,n}>0)\\
 & +\log\det(I_{n}-\lambda G_{n}(Y_{n})W_{n}G_{n}(Y_{n}))-\frac{1}{2}\sum^{n}_{i=1}1(Y_{i,n}>0)z_{i,n}(\theta)^{2},
\end{align*}
where $\Phi(\cdot)$ is the cumulative distribution function of the
standard normal distribution. Our goal here is not to re-establish
the full consistency argument, but rather to show that the key pointwise
convergence step in that argument can be obtained without relying
on an underlying metric space. In the framework of \citet{xu2015maximum},
the metric-space assumption is used to establish the pointwise convergence
of $\frac{1}{n}\ell_{n}(\theta)$. Accordingly, we focus on proving
that $\frac{1}{n}\ell_{n}(\theta)=\frac{1}{n}\mathbb{E}_{\mathcal{C}_{n}}\ell_{n}(\theta)+o_{\mathbb{P}}(1)$,
which is one ingredient in the consistency proof. The following assumptions
are required for the subsequent analysis.
\begin{assumption}
\label{assu:consis}
\begin{enumerate}[label=(\roman*)]
\item Conditional on $\c_{n}$, $\epsilon_{1,n},\dots,\epsilon_{n,n}$
are i.i.d. normal.
\item $0<|\lambda_{0}|<1$ and $\|W_{n}\|_{\infty}\le1$\textup{ a.s. for
all $n$.}
\item $\mu_{2}=\left(\frac{1}{n}\sum^{n}_{i=1}\left\Vert S^{+}_{\cdot i,n}\right\Vert ^{2}_{1}\right)^{\frac{1}{2}}=o_{\p}(\sqrt{n})$,
where $S^{+}_{n}\equiv(I_{n}-|\lambda_{0}W_{n}|)^{-1}$.
\item Conditional on $\c_{n}$, elements in $X_{i,n}$ are uniformly bounded
for all $i$, $n$.
\item $\sup_{i,n}\sup_{y}f_{Y^{*}_{i,n}\mid\mathcal{C}_{n}}(y)\le C_{f}<\infty$
a.s. for some constant $C_{f}>0$, where $f_{Y^{*}_{i,n}\mid\mathcal{C}_{n}}(y)$
is the conditional density of $Y^{*}_{i,n}$.
\end{enumerate}
\end{assumption}
See \citet{xu2015maximum} for primitive conditions for Assumption
\ref{assu:consis}(v). We then obtain the following result.
\begin{prop}
\label{prop:tobit consistency}Under Assumption~\ref{assu:consis},
for any fixed $\theta$ such that $\left|\lambda\right|<1$ and $\sigma>0$,
we have that $\frac{1}{n}\ell_{n}(\theta)=\frac{1}{n}\E_{\c_{n}}\ell_{n}(\theta)+o_{\p}(1)$.
\end{prop}
Proposition~\ref{prop:tobit consistency} shows that this pointwise
convergence of the log-likelihood can be established within the FDM
framework, without imposing any underlying metric-space structure.
This provides one of the probabilistic ingredients used in the consistency
analysis of the SAR Tobit MLE.

\section{\label{sec: Transformation}Functional Dependence Measure Under Transformations}

To study the asymptotic properties of an estimator, we usually need
to deal with various functions of random variables, e.g., $Y^{2}_{i,n}$
or $\Phi(Y_{i,n})$, where $\Phi(\cdot)$ is a distribution function
of some random variable. Therefore, to broaden the applicability of
our theory, we study the FDM under different transformations, and
investigate whether the important inequalities and the CLTs remain
valid. As in Section \ref{sec:Func depend}, denote $Y_{j,n}=F_{j,n}(e_{n})\in\mathbb{R}$,
$Y_{j,n,i}\equiv F_{j,n}\left(e_{1,n},\cdots,e_{i-1,n},e^{*}_{i,n},e_{i+1,n},\cdots,e_{n,n}\right)$,
and $\delta_{p,n}\left(j,i,\c_{n}\right)\equiv\left\Vert Y_{j,n}-Y_{j,n,i}\right\Vert _{L^{p},\c_{n}}$
as the FDM of $\left\{ Y_{j,n}\right\} $ conditional on $\c_{n}$.
We add the superscript $``(Y)"$ in $\delta^{(Y)}_{p,n}\left(j,i,\c_{n}\right)$
to differentiate it from the FDM of other random variables.

We first consider Lipschitz-type functions $H_{j,n}:\mathbb{R}\rightarrow\mathbb{R}$.
Suppose for all $\left(y,y^{\bullet}\right)\in\mathbb{R}\times\mathbb{R}$
and all $j$ and $n\geq1$: 
\begin{equation}
\left|H_{j,n}\left(y\right)-H_{j,n}\left(y^{\bullet}\right)\right|\leq B_{j,n}\left(y,y^{\bullet}\right)\left|y-y^{\bullet}\right|.\label{eq:lip condi}
\end{equation}
In Propositions \ref{prop: Lips}-\ref{prop:unbdd lip 2}, we denote
$Z_{j,n}\equiv H_{j,n}(Y_{j,n})$.
\begin{prop}
\label{prop: Lips} Suppose $H_{j,n}\left(\cdot\right)$ satisfies
Eq.(\ref{eq:lip condi}) with $\sup_{n,j}\sup_{y,y^{\bullet}}B_{j,n}\left(y,y^{\bullet}\right)\leq C<\infty$
for some constant $C$. Then 
\[
\delta^{(Z)}_{p,n}\left(j,i,\c_{n}\right)\leq C\delta^{(Y)}_{p,n}\left(j,i,\c_{n}\right)\ a.s.
\]
\end{prop}
Obviously, if the FDM of $\{Y_{j,n}\}$ satisfies the conditions in
Theorems~\ref{thm: Rosenthal}, \ref{thm:exp inequality}, or \ref{thm:CLT finite},
then the same holds for the FDM of $\{Z_{j,n}\}$.

Next, we consider unbounded $B_{j,n}\left(y,y^{\bullet}\right)$,
e.g., $H_{j,n}\left(y\right)=y^{2}$. 
\begin{prop}
\label{prop:unbdd lip 1} Suppose $H_{j,n}\left(\cdot\right)$ satisfies
Eq.(\ref{eq:lip condi}) with $B_{j,n}\left(y,y^{\bullet}\right)\leq C_{1}(\left|y\right|^{a}+\left|y^{\bullet}\right|^{a}+1)$
for some finite constants $C_{1}>0$ and $a\geq1$, constants $p,q,r\geq1$
satisfying $p^{-1}=q^{-1}+r^{-1}$, and $\left\Vert Y\right\Vert _{L^{ar},\c_{n}}\equiv\sup_{n,j}\left\Vert Y_{j,n}\right\Vert _{L^{ar},\c_{n}}<\infty$
a.s. Then
\[
\delta^{(Z)}_{p,n}\left(j,i,\c_{n}\right)\leq C_{1}(2\left\Vert Y\right\Vert ^{a}_{L^{ar},\c_{n}}+1)\delta^{(Y)}_{q,n}\left(j,i,\c_{n}\right)\ a.s.
\]
\end{prop}
Hence, if $\frac{1}{n^{\widetilde{q}}}\sum^{n}_{i=1}\left[\sum^{n}_{j=1}\delta^{(Y)}_{q,n}\left(j,i,\c_{n}\right)\right]^{\widetilde{q}}=o_{\p}(1)$
and $\left\Vert Y\right\Vert _{L^{ar},\c_{n}}<\infty$ a.s., then
\[
\frac{1}{n^{\widetilde{q}}}\sum^{n}_{i=1}\left[\sum^{n}_{j=1}\delta^{(Z)}_{p,n}\left(j,i,\c_{n}\right)\right]^{\widetilde{q}}=o_{\p}(1).
\]
So Theorem~\ref{thm: Rosenthal} is applicable. Similarly, condition
(\ref{eq:l1 bounded for finite clt}) in Theorem~\ref{thm:CLT finite}
can be verified for $\{Z_{j,n}\}$. In Proposition~\ref{prop:unbdd lip 1},
there is a trade-off between $p$, $q$ and $r$. If a larger $p$
is desired, then larger $q$ or $r$ is required. By imposing further
conditions on $\delta^{(Y)}_{p,n}\left(j,i,\c_{n}\right)$, we can
avoid the trade-off between $p$, $q$ and $r$, which is presented
in the next proposition. 

\begin{prop}
\label{prop:unbdd lip 2} Suppose $H_{j,n}\left(\cdot\right)$ satisfies
condition (\ref{eq:lip condi}) with $B_{j,n}\left(y,y^{\bullet}\right)\leq C_{1}(\left|y\right|^{a}+\left|y^{\bullet}\right|^{a}+1)$
for some finite constants $C_{1}>0$ and $a\geq1$. If $\left\Vert Y\right\Vert _{L^{q},\c_{n}}\equiv\sup_{n,j}\left\Vert Y_{j,n}\right\Vert _{L^{q},\c_{n}}<\infty$
a.s. for some $q>\max\left(\frac{ap}{p-1},ap+p\right)$. Then
\[
\delta^{(Z)}_{p,n}\left(j,i,\c_{n}\right)\leq C_{2}\left(\c_{n}\right)\left[\delta^{(Y)}_{p,n}\left(j,i,\c_{n}\right)\right]^{\left(q-ap-p\right)/\left(pq-ap-p\right)}\ a.s.,
\]
where $C_{2}\left(\c_{n}\right)<\infty$ a.s.
\end{prop}
If $\frac{1}{n^{q}}\sum^{n}_{i=1}\left\{ \sum^{n}_{j=1}\left[\delta^{(Y)}_{p,n}\left(j,i,\c_{n}\right)\right]^{\left(q-ap-p\right)/\left(pq-ap-p\right)}\right\} ^{q}=o_{\p}(1)$
, then, by Proposition~\ref{prop:unbdd lip 2},
\[
\frac{1}{n^{q}}\sum^{n}_{i=1}\left\{ \sum^{n}_{j=1}\delta^{(Z)}_{p,n}\left(j,i,\c_{n}\right)\right\} ^{q}=o_{\p}(1)
\]
So Theorem~\ref{thm: Rosenthal} is applicable. Moreover, condition
(\ref{eq:l1 bounded for finite clt}) in Theorem~\ref{thm:CLT finite}
can be verified using similar arguments.

In addition to Lipschitz transformation, another important nonlinear
transformation is $1\left(y>0\right)$, which is useful in discrete
choice and censor data model. See, e.g., \citet{xu2015maximum,XU201896}. 
\begin{prop}
\label{prop:1 >0} Denote $Z_{j,n}\equiv1\left(Y_{j,n}>0\right)$.
Denote the density function of $Y_{j,n}$ conditional on $\c_{n}$
by $f_{j,n}(y\mid\c_{n})$. If $\sup_{n\geq1,1\leq j\leq n}\sup_{y}f_{j,n}(y\mid\c_{n})<C_{1}<\infty$
a.s. for some constant $C_{1}>0$ , then 
\[
\delta^{(Z)}_{p,n}\left(j,i,\c_{n}\right)\leq C\left[\delta^{(Y)}_{p,n}\left(j,i,\c_{n}\right)\right]^{1/\left(p+1\right)},
\]
 for some constant $C$ not depending on $i$, $j$, nor $n$. 
\end{prop}
Moreover, we are interested in the sum or the product of two random
variables. Let $\left\{ Y_{j,n}\right\} $ and $\left\{ Z_{j,n}\right\} $
be two sets of random variables, and they are both functions of some
conditionally independent random vectors $e_{i,n}$ given $\c_{n}$,
where $i\in[n]$. Denote the FDMs of $\left\{ Y_{j,n}+Z_{j,n}\right\} $
and $\left\{ Y_{j,n}Z_{j,n}\right\} $ by $\delta^{(Y+Z)}_{p,n}\left(j,i,\c_{n}\right)$
and $\delta^{(YZ)}_{p,n}\left(j,i,\c_{n}\right)$, respectively. For
summation, the conclusion is a direct result of Minkowski's inequality,
and we summarize it below.
\begin{prop}
\label{prop:+ FDM} $\delta^{(Y+Z)}_{p,n}\left(j,i,\c_{n}\right)\leq\delta^{(Y)}_{p,n}\left(j,i,\c_{n}\right)+\delta^{(Z)}_{p,n}\left(j,i,\c_{n}\right)$
a.s.
\end{prop}
The FDM of product is more complicated. For simplicity, we assume
the original random fields are real-valued, since if $Y$ is functionally
dependent, then all the elements of $Y$ are still functionally dependent
and vice versa. Analogous to Propositions~\ref{prop:unbdd lip 1}
and \ref{prop:unbdd lip 2}, we have two versions of conclusions,
presented below.
\begin{prop}
\label{prop:FDM prod 1} Suppose constants $p,q_{1},q_{2},r_{1},r_{2}>1$
satisfy $p^{-1}=q^{-1}_{1}+r^{-1}_{1}=q^{-1}_{2}+r^{-1}_{2}$, $\left\Vert Y\right\Vert _{L^{r_{2}},\c_{n}}\equiv\sup_{n,j}\left\Vert Y_{j,n}\right\Vert _{L^{r_{2}},\c_{n}}<\infty$
a.s. and $\left\Vert Z\right\Vert _{L^{r_{1}},\c_{n}}\equiv\sup_{n,j}\left\Vert Z_{j,n}\right\Vert _{L^{r_{1}},\c_{n}}<\infty$
a.s. Then 
\[
\delta^{(YZ)}_{p,n}\left(j,i,\c_{n}\right)\leq\left\Vert Z\right\Vert _{L^{r_{1}},\c_{n}}\delta^{(Y)}_{q_{1},n}\left(j,i,\c_{n}\right)+\left\Vert Y\right\Vert _{L^{r_{2}},\c_{n}}\delta^{(Z)}_{q_{2},n}\left(j,i,\c_{n}\right)\ a.s.
\]
\end{prop}
Let us examine whether $\{Y_{j,n}Z_{j,n}\}$ satisfies the conditions
in Theorem \ref{thm: Rosenthal}. Denote $\Delta^{(YZ)}_{p,2}\left(\c_{n}\right)\equiv\frac{1}{n^{2}}\sum^{n}_{i=1}[\sum^{n}_{j=1}\delta^{(YZ)}_{p,n}\left(j,i,\c_{n}\right)]^{2}$,
and $\Delta^{(Y)}_{q_{1},2}\left(\c_{n}\right)$ and $\Delta^{(Z)}_{q_{2},2}\left(\c_{n}\right)$
are defined similarly. Then by Proposition~\ref{prop:FDM prod 1},
\begin{align*}
 & \Delta^{(YZ)}_{p,2}\left(\c_{n}\right)\leq\frac{1}{n^{2}}\sum^{n}_{i=1}\left[\left\Vert Z\right\Vert _{L^{r_{1}},\c_{n}}\sum^{n}_{j=1}\delta^{(Y)}_{q_{1},n}\left(j,i,\c_{n}\right)+\left\Vert Y\right\Vert _{L^{r_{2}},\c_{n}}\sum^{n}_{j=1}\delta^{(Z)}_{q_{2},n}\left(j,i,\c_{n}\right)\right]^{2}\\
\leq & \frac{1}{n^{2}}\sum^{n}_{i=1}\left\{ 2\left\Vert Z\right\Vert ^{2}_{L^{r_{1}},\c_{n}}\left[\sum^{n}_{j=1}\delta^{(Y)}_{q_{1},n}\left(j,i,\c_{n}\right)\right]^{2}+2\left\Vert Y\right\Vert ^{2}_{L^{r_{2}},\c_{n}}\left[\sum^{n}_{j=1}\delta^{(Z)}_{q_{2},n}\left(j,i,\c_{n}\right)\right]^{2}\right\} \\
= & 2\left\Vert Z\right\Vert ^{2}_{L^{r_{1}},\c_{n}}\Delta^{(Y)}_{q_{1},2}\left(\c_{n}\right)+2\left\Vert Y\right\Vert ^{2}_{L^{r_{2}},\c_{n}}\Delta^{(Z)}_{q_{2},2}\left(\c_{n}\right)=o_{\p}(1),
\end{align*}
where the second inequality follows from the fact that $\left(a+b\right)^{2}\leq2a^{2}+2b^{2}$
for arbitrary $a$, $b\in\mathbb{R}$. Consequently, when $\max\left\{ \Delta^{(Y)}_{q_{1},2}\left(\c_{n}\right),\Delta^{(Z)}_{q_{2},2}\left(\c_{n}\right)\right\} =o_{\p}(1)$,
$\{Y_{j,n}Z_{j,n}\}$ also satisfies the conditions in Theorem~\ref{thm: Rosenthal}.

Similar to Proposition \ref{prop:unbdd lip 1}, there is a trade-off
between $p$, $q_{1},r_{1}$and $p,q_{2},r_{2}$. Since $q_{1}>p$
and $q_{2}>p$, when three or more random fields are multiplied, larger
values for $q_{1}$ and $q_{2}$ are required, which may not always
be feasible. However, by introducing additional conditions, it may
be possible to avoid such trade-offs, as demonstrated below.
\begin{prop}
\label{prop:FDM prod 2} Suppose $\left\Vert Y\right\Vert _{L^{q},\c_{n}}\equiv\sup_{n,j}\left\Vert Y_{j,n}\right\Vert _{L^{q},\c_{n}}<\infty$
a.s., $\left\Vert Z\right\Vert _{L^{q},\c_{n}}\equiv\sup_{n,j}\left\Vert Z_{j,n}\right\Vert _{L^{q},\c_{n}}<\infty$
a.s. for some $q>\max(p/(p-1),2p)$ and $p>1$. Then
\[
\delta^{(YZ)}_{p,n}\left(j,i,\c_{n}\right)\le C_{1}\left(\c_{n}\right)\left[\delta^{(Y)}_{p,n}\left(j,i,\c_{n}\right)\right]^{\frac{q-2p}{pq-2p}}+C_{2}\left(\c_{n}\right)\left[\delta^{(Z)}_{p,n}\left(j,i,\c_{n}\right)\right]^{\frac{q-2p}{pq-2p}}
\]
a.s., where $C_{1}\left(\c_{n}\right)<\infty$ a.s. and $C_{2}\left(\c_{n}\right)<\infty$
a.s.
\end{prop}
Using Proposition \ref{prop:FDM prod 2} together with the inequality
$(a+b)^{r}\leq\max\{1,2^{r-1}\}(a^{r}+b^{r})$ for any $a,b,r\geq0$,
we can show that if 
\[
\frac{1}{n^{2}}\sum^{n}_{i=1}\left[\sum^{n}_{j=1}\left(\delta^{(Y)}_{p,n}\left(i,j,\c_{n}\right)\right)^{\left(q-2p\right)/\left(pq-2p\right)}\right]^{2}=o_{\p}(1)
\]
and 
\[
\frac{1}{n^{2}}\sum^{n}_{i=1}\left[\sum^{n}_{j=1}\left(\delta^{(Z)}_{p,n}\left(i,j,\c_{n}\right)\right)^{\left(q-2p\right)/\left(pq-2p\right)}\right]^{2}=o_{\p}(1),
\]
then
\[
\Delta^{(YZ)}_{p,2}\left(\c_{n}\right)=\frac{1}{n^{2}}\sum^{n}_{i=1}\left[\sum^{n}_{j=1}\delta^{(YZ)}_{p,n}\left(i,j\right)\right]^{2}=o_{\p}(1).
\]
Then, Theorem \ref{thm: Rosenthal} is applicable. We can verify condition
(\ref{eq:l1 bounded for finite clt}) in Theorem~\ref{thm:CLT finite}
under similar conditions.

In practice, one may use either Proposition \ref{prop:FDM prod 1}
or \ref{prop:FDM prod 2} to study the FDM of $\{Y_{j,n}Z_{j,n}\}$.
Proposition \ref{prop:FDM prod 1} requires a stronger moment condition,
while Proposition \ref{prop:FDM prod 2} requires that most of $\delta^{(Y)}_{p,n}\left(i,j,\c_{n}\right)$
are smaller, as $\frac{\left(q-2p\right)}{pq-2p}<1$ implies that
$\left[\delta^{(Y)}_{p,n}\left(i,j,\c_{n}\right)\right]^{\left(q-2p\right)/\left(pq-2p\right)}>\delta^{(Y)}_{p,n}\left(i,j,\c_{n}\right)$.\footnote{Note that most of $\delta^{(Y)}_{p,n}\left(i,j,\c_{n}\right)$ are
smaller than 1 asymptotically. }

\section{Conclusion\label{sec:Conclusion}}

This paper develops limit theorems for random variables with network
structure, without requiring individuals to be located in a Euclidean
or other metric space. This sets our approach apart from most existing
limit theories in network statistics and econometrics, which typically
rely on weak dependence concepts such as strong mixing, near-epoch
dependence, or $\psi$-dependence. To establish these results, we
generalize the functional dependence measure proposed by \citet{wu2005nonlinear}.
Using this framework, we derive several inequalities, a law of large
numbers, and central limit theorems. We also demonstrate the applicability
of these results by verifying their conditions for spatial autoregressive
models, which are widely used in network data analysis. Finally, we
study the behavior of the functional dependence measure under various
transformations commonly encountered in applications. Extending this
framework to panel data with network dependence is a natural direction
for future research.

\newpage{}

\appendix
{\LARGE\textbf{\centerline{Appendices}}}{\LARGE\par}

\section{Some Useful Lemmas\label{sec:Some-Useful-Lemmas}}
\begin{lem}
\label{lem: Burkholder}Let $\left(\Omega,\mathcal{F},\p\right)$
be a probability space and $\c$ is a sub-$\sigma$-field of $\F$.
Let $X_{1},X_{2},\ldots,X_{n}$ be a zero-mean martingale difference
array under the conditional expectation $\E_{\c}$. Let $p>1$ be
a constant. Let $C_{p}\equiv\sqrt{p-1}$ when $p\geq2$ and $C_{p}\equiv(p-1)^{-1}$
when $p\in(1,2)$. Then for any $p>1$, it holds that 
\begin{equation}
\left\Vert \sum^{n}_{i=1}X_{i}\right\Vert _{L^{p},\c}\leq C_{p}\left(\sum^{n}_{i=1}\left\Vert X_{i}\right\Vert ^{\min\{p,2\}}_{L^{p},\c}\right)^{1/\min\{p,2\}}\ \text{a.s.}\label{eq:burkholder}
\end{equation}
\end{lem}
\begin{lem}
\citep[Theorem 3.2]{hallMartingaleLimitTheory1980}. \label{CLT MDA}
Given a triangular MDA $\{X_{nk},\mathcal{F}_{nk}:1\leq k\leq k_{n},n\geq1\}$,
if (a) $\sum^{k_{n}}_{k=1}X^{2}_{nk}\xrightarrow{\p}1$, (b) $\max_{1\leq k\leq k_{n}}\left|X_{nk}\right|\xrightarrow{\p}0$,
and (c) $\E\max_{1\leq k\leq k_{n}}$ $X^{2}_{nk}<K$ for all $n\geq1$,
where $K>0$ is a constant, then $S_{n}=\sum^{k_{n}}_{k=1}X_{nk}\xrightarrow{d}N(0,1)$. 
\end{lem}
\begin{lem}
\label{lem:general pred<FDM}Let $\left(\Omega,\mathcal{F},\p\right)$
be a probability space and $\c_{n}$ is a sub-$\sigma$-field of $\F$.
Let $Q_{n}=g_{n}(e_{1,n},\ldots,e_{n,n})$, where the function $g_{n}$
can be a random function and we assume it is measurable with respect
to $\c_{n}$. Let $Q_{n,i}=g_{n}(e_{1,n},\ldots,e^{*}_{i,n},\ldots,e_{n,n})$
be a coupled version of $Q_{n}$, where $e^{*}_{i,n}$ is a conditional
i.i.d. copy of $e_{i,n}$ given $\c_{n}$. Let $\mathcal{I}\subset[n]$
be nonempty. Then for any $i\in\mathcal{I}$, 
\[
\left\Vert \E_{\c_{n}}\left(Q_{n}|\sigma(e_{k,n}:k\in\mathcal{I})\right)-\E_{\c_{n}}\left(Q_{n}|\sigma(e_{k,n}:k\in\mathcal{I}\backslash\{i\})\right)\right\Vert _{L^{p},\c_{n}}\leq\|Q_{n}-Q_{n,i}\|_{L^{p},\c_{n}}.
\]
\end{lem}

\section{Proofs for Section \ref{sec:Property} \label{sec:Proofs-for-Section}}

\textbf{Proof of Lemma \ref{lem:predi < FMD}.} This directly follows
from Lemma \ref{lem:general pred<FDM}. $\hfill\qedsymbol$

\textbf{Proof of Theorem \ref{thm: Rosenthal}.} Notice that $(Y_{j,n}-\E_{\c_{n}}Y_{j,n})=\sum^{n}_{i=1}P_{i}Y_{j,n}$.
As a result, 
\begin{equation}
\sum^{n}_{j=1}(Y_{j,n}-\E_{\c_{n}}Y_{j,n})=\sum^{n}_{j=1}\sum^{n}_{i=1}P_{i}Y_{j,n}=\sum^{n}_{i=1}\left(\sum^{n}_{j=1}P_{i}Y_{j,n}\right).\label{eq: MDA decomposition}
\end{equation}
Because $\left\{ \sum^{n}_{j=1}P_{i}Y_{j,n}:i\in[n]\right\} $ is
a MDA and measurable with respect to $\c_{n}$, 
\begin{equation}
\begin{aligned} & \left\Vert \sum^{n}_{j=1}(Y_{j,n}-\E_{\c_{n}}Y_{j,n})\right\Vert _{L^{p},\c_{n}}=\left\Vert \sum^{n}_{i=1}\left(\sum^{n}_{j=1}P_{i}Y_{j,n}\right)\right\Vert _{L^{p},\c_{n}}\\
\leq & C_{p}\left(\sum^{n}_{i=1}\left\Vert \sum^{n}_{j=1}P_{i}Y_{j,n}\right\Vert ^{\min\{p,2\}}_{L^{p},\c_{n}}\right)^{\frac{1}{\min\{p,2\}}}\leq C_{p}\left[\sum^{n}_{i=1}\left(\sum^{n}_{j=1}\delta_{p,n}(j,i,\c_{n})\right)^{\min\{p,2\}}\right]^{\frac{1}{\min\{p,2\}}}\\
\leq & C_{p}\left\{ \Delta_{p,\min\{p,2\}}\left(\c_{n}\right)\right\} ^{\frac{1}{\min\{p,2\}}}n
\end{aligned}
\label{eq: Rosenthal 1 pf}
\end{equation}
a.s., where the first equality follows from Eq.(\ref{eq: MDA decomposition}),
the first inequality follows from Lemma~\ref{lem: Burkholder}, the
second one follows from conditional Minkowski's inequality and Lemma
\ref{lem:predi < FMD}, and the last one follows from the definition
of $\Delta_{p,\min\{p,2\}}\left(\c_{n}\right)$. $\hfill\qedsymbol$

\textbf{Proof of Theorem \ref{thm:WLLN}.} By Theorem~\ref{thm: Rosenthal},
we have 
\[
\frac{1}{n}\left\Vert \sum^{n}_{j=1}(Y_{j,n}-\E_{\c_{n}}Y_{j,n})\right\Vert _{L^{p},\c_{n}}=o_{\p}(1).
\]
It directly follows from \citet[Lemma 6.1]{chernozhukov2018Double}
that 
\[
\frac{1}{n}\sum^{n}_{j=1}(Y_{j,n}-\E_{\c_{n}}Y_{j,n})=o_{\p}(1).
\]
The proof is completed.$\hfill\qedsymbol$

\textbf{Proof of Theorem \ref{thm:exp inequality}.} The idea of the
proof is borrowed from that in \citet{wu2016wu}. From Theorem \ref{thm: Rosenthal},
for any $p\geq2$, 
\[
\left\Vert \sum^{n}_{j=1}Y_{j,n}\right\Vert _{L^{p},\c_{n}}\leq\sqrt{p-1}\sqrt{\Delta_{p,2}\left(\c_{n}\right)}n\text{ a.s.}
\]
Consequently, $\left\Vert Z_{n}\right\Vert _{L^{p},\c_{n}}\leq\sqrt{p-1}\sqrt{n}\Delta^{1/2}_{p,2}\left(\c_{n}\right)$
a.s. for $p\geq2$. Recall a Taylor's formula from \citet{wu2016wu}:
$\left(1-s\right)^{-1/2}=1+\sum^{\infty}_{k=1}a_{k}s^{k}$, where
$\left|s\right|<1$ and $a_{k}=\left(2k\right)!/\left(2^{2k}\left(k!\right)^{2}\right),a_{0}=1$.
By Stirling's formula, $a_{k}\sim(\pi k)^{-1/2}$ as $k\to\infty$.
Hence, $k!\sim\sqrt{2}\left(k/e\right)^{k}a^{-1}_{k}$ and $\frac{a_{k}}{a_{k-1}}\to1$,
and there exist constants $c_{1},c_{2}>0$ such that $k!\geq c_{1}\left(k/e\right)^{k}a^{-1}_{k}$
and $a_{k}\leq c_{2}a_{k-1}$ hold for all $k\geq1$. By Eq.\eqref{eq: exp ineq gamma_0},
when $\alpha k\geq2$, we have $\sup_{n\geq1}\sqrt{n}\Delta^{1/2}_{\alpha k,2}\left(\c_{n}\right)\leq\gamma_{0}\left(\alpha k\right)^{\nu}$.
As a result, when $\alpha k\geq2$, 
\begin{align*}
 & \frac{t^{k}\left\Vert Z_{n}\right\Vert ^{\alpha k}_{L^{\alpha k},\c_{n}}}{k!}\leq\frac{t^{k}\left(\alpha k-1\right)^{\alpha k/2}\left[\sqrt{n}\Delta^{1/2}_{\alpha k,2}\left(\c_{n}\right)\right]^{\alpha k}}{c_{1}\left(k/e\right)^{k}a^{-1}_{k}}\leq\frac{t^{k}\left(\alpha k-1\right)^{\alpha k/2}\gamma^{\alpha k}_{0}\left(\alpha k\right)^{\alpha k\nu}}{c_{1}\left(k/e\right)^{k}a^{-1}_{k}},\\
= & \frac{a_{k}t^{k}}{c_{1}t^{k}_{0}}\frac{\left(\alpha k-1\right)^{\alpha k/2}}{\left(\alpha k\right)^{\alpha k/2}}\leq\frac{a_{k}}{c_{1}\sqrt{e}}\frac{t^{k}}{t^{k}_{0}}\ \ \text{a.s.},
\end{align*}
where the equality follows from $t_{0}=\left(e\alpha\gamma^{\alpha}_{0}\right)^{-1}$
and $\nu=\frac{1}{\alpha}-\frac{1}{2}$, and the last inequality is
due to the fact that $\left(x-1\right)^{x/2}/x^{x/2}\leq e^{-1/2}$
for all $x\geq2$. When $0<\alpha k<2$ and $k\geq1$, we have $\left\Vert Z_{n}\right\Vert _{L^{\alpha k},\c_{n}}\leq\left\Vert Z_{n}\right\Vert _{L^{2},\c_{n}}\leq\sqrt{n}\Delta^{1/2}_{2,2}\left(\c_{n}\right)\leq2^{\nu}\gamma_{0}$
a.s. and 
\begin{align*}
\frac{t^{k}\left\Vert Z_{n}\right\Vert ^{\alpha k}_{L^{\alpha k},\c_{n}}}{k!} & \leq\frac{t^{k}2^{\nu\alpha k}\gamma^{\alpha k}_{0}}{c_{1}\left(k/e\right)^{k}a^{-1}_{k}}=\frac{a_{k}t^{k}}{c_{1}t^{k}_{0}}\frac{2^{\nu\alpha k}}{\left(\alpha k\right)^{k}}\leq\frac{2^{2/\alpha-1}}{\min\left\{ \alpha,\alpha^{2/\alpha}\right\} }\frac{a_{k}t^{k}}{c_{1}t^{k}_{0}}\text{\, a.s.,}
\end{align*}
where the equality is because $t_{0}=\left(e\alpha\gamma^{\alpha}_{0}\right)^{-1}$
and the last inequality follows from $2^{\nu\alpha k}\leq2^{2\nu}$
(as $\nu=\frac{1}{\alpha}-\frac{1}{2}\geq0$), and $\left(\alpha k\right)^{k}\geq\alpha1(\alpha\geq1)+\alpha^{2/\alpha}1(\alpha<1)\geq\min\left\{ \alpha,\alpha^{2/\alpha}\right\} $.
Using $e^{x}=1+\sum^{\infty}_{k=1}\frac{x^{k}}{k!}$ and the above
two displayed inequalities, we obtain 
\begin{align*}
 & m\left(t\right)=1+\sum^{\infty}_{k=1}\frac{t^{k}\E\left[\left|Z_{n}\right|^{\alpha k}\mid\c_{n}\right]}{k!}=1+\sum_{1\leq k<2/\alpha}\frac{t^{k}\left\Vert Z_{n}\right\Vert ^{\alpha k}_{L^{\alpha k},\c_{n}}}{k!}+\sum_{k\geq2/\alpha}\frac{t^{k}\left\Vert Z_{n}\right\Vert ^{\alpha k}_{L^{\alpha k},\c_{n}}}{k!}\\
\leq & 1+\sum_{1\leq k<2/\alpha}\frac{2^{2/\alpha-1}}{\min\left\{ \alpha,\alpha^{2/\alpha}\right\} }\frac{a_{k}t^{k}}{c_{1}t^{k}_{0}}+\sum_{k\geq2/\alpha}\frac{a_{k}}{c_{1}\sqrt{e}}\frac{t^{k}}{t^{k}_{0}}\leq1+c'_{\alpha}\sum^{\infty}_{k=1}a_{k}\frac{t^{k}}{t^{k}_{0}}\\
\leq & 1+c'_{\alpha}\sum^{\infty}_{k=1}c_{2}a_{k-1}\frac{t^{k}}{t^{k}_{0}}=1+c_{\alpha}\frac{t}{t_{0}}\sum^{\infty}_{k=0}a_{k}\frac{t^{k}}{t^{k}_{0}}=1+c_{\alpha}\frac{t/t_{0}}{\left(1-t/t_{0}\right)^{1/2}}\text{\, a.s.,}
\end{align*}
where $c'_{\alpha},c_{\alpha}\geq0$ are constants depending only
on $\alpha$. By conditional Markov's inequality, letting $t=t_{0}/2$,
\begin{align*}
 & \p\left(\left|Z_{n}\right|\geq x\mid\c_{n}\right)=\p\left(\exp\left(t\left|Z_{n}\right|^{\alpha}\right)\geq\exp\left(tx^{\alpha}\right)\mid\c_{n}\right)\\
\leq & \exp\left(-tx^{\alpha}\right)m\left(t\right)\leq\left(1+\frac{\sqrt{2}c_{\alpha}}{2}\right)\exp\left(-\frac{x^{\alpha}}{2e\alpha\gamma^{\alpha}_{0}}\right)
\end{align*}
a.s.$\hfill\qedsymbol$

\textbf{Proof of Lemma \ref{lem:fdm of martingale < 2nd FMD}.} Let
\begin{align*}
\mathcal{G}_{i,j} & =\begin{cases}
\sigma(e_{1,n},\dots,e^{*}_{j,n},\dots,e_{i,n}) & \text{if}\ j\leq i\\
\F_{i,n} & \text{if}\ j>i
\end{cases}
\end{align*}
and $Z_{i,n,j}=\sum^{n}_{k=1}\left[\E_{\c_{n}}\left(Y_{k,n,j}\mid\mathcal{G}_{i,j}\right)-\E_{\c_{n}}\left(Y_{k,n,j}\mid\mathcal{G}_{i-1,j}\right)\right].$
Here $Z_{i,n,j}$ is the coupled version of $Z_{i,n}$ with $e_{j,n}$
being replaced by its i.i.d. copy $e^{*}_{j,n}$ conditional on $\c_{n}$.
Note that for any $j\leq i-1$, we have 
\begin{align*}
 & \left\Vert Z_{i,n}-Z_{i,n,j}\right\Vert _{L^{p},\c_{n}}=\biggl\Vert\sum^{n}_{k=1}\Bigl[\E_{\c_{n}}\left(Y_{k,n}\mid\F_{i,n}\right)-\E_{\c_{n}}\left(Y_{k,n}\mid\F_{i-1,n}\right)-\E_{\c_{n}}\left(Y_{k,n,j}\mid\mathcal{G}_{i,j}\right)\\
 & \ +\E_{\c_{n}}\left(Y_{k,n,j}\mid\mathcal{G}_{i-1,j}\right)\Bigr]\biggr\Vert_{L^{p},\c_{n}}=\biggl\Vert\sum^{n}_{k=1}\Bigl[\E_{\c_{n}}\left(Y_{k,n}\mid\F_{i,n}\vee\sigma(e^{*}_{j,n})\right)-\E_{\c_{n}}\left(Y_{k,n,i}\mid\F_{i,n}\vee\sigma(e^{*}_{j,n})\right)\\
 & \ -\E_{\c_{n}}\left(Y_{k,n,j}\mid\mathcal{G}_{i,j}\vee\sigma(e_{j,n})\right)+\E_{\c_{n}}\left(Y_{k,n,\{i,j\}}\mid\mathcal{G}_{i-1,j}\vee\sigma(e_{i,n},e_{j,n})\right)\Bigr]\biggl\Vert_{L^{p},\c_{n}},\\
 & =\biggl\Vert\sum^{n}_{k=1}\Bigl[\E_{\c_{n}}\left(Y_{k,n}-Y_{k,n,j}-Y_{k,n,i}+Y_{k,n,\{i,j\}}\mid\F_{i,n}\vee\sigma(e^{*}_{j,n})\right)\Bigr]\biggl\Vert_{L^{p},\c_{n}}\\
 & \leq\biggl\Vert\sum^{n}_{k=1}\left[Y_{k,n}-Y_{k,n,j}-Y_{k,n,i}+Y_{k,n,\{i,j\}}\right]\biggl\Vert_{L^{p},\c_{n}}=\mathfrak{d}_{p,n}(i,j,\c_{n}),
\end{align*}
where the second equality follows from the following facts: 
\[
\E_{\c_{n}}\left(Y_{k,n}\mid\F_{i,n}\right)=\E_{\c_{n}}\left(Y_{k,n}\mid\F_{i,n}\vee\sigma(e^{*}_{j,n})\right),
\]
\[
\E_{\c_{n}}\left(Y_{k,n,j}\mid\mathcal{G}_{i,j}\right)=\E_{\c_{n}}\left(Y_{k,n,j}\mid\mathcal{G}_{i,j}\vee\sigma(e_{j,n})\right),
\]
\[
\E_{\c_{n}}\left(Y_{k,n}\mid\F_{i-1,n}\right)=\E_{\c_{n}}\left(Y_{k,n,i}\mid\F_{i-1,n}\right)=\E_{\c_{n}}\left(Y_{k,n,i}\mid\F_{i,n}\vee\sigma(e^{*}_{j,n})\right),
\]
\[
\E_{\c_{n}}\left(Y_{k,n,j}\mid\mathcal{G}_{i-1,j}\right)=\E_{\c_{n}}\left(Y_{k,n,\{i,j\}}\mid\mathcal{G}_{i-1,j}\right)=\E_{\c_{n}}\left(Y_{k,n,\{i,j\}}\mid\mathcal{G}_{i-1,j}\vee\sigma(e_{i,n},e_{j,n})\right),
\]
the third equality follows $\F_{i,n}\vee\sigma(e^{*}_{j,n})=\mathcal{G}_{i,j}\vee\sigma(e_{j,n})=\mathcal{G}_{i-1,j}\vee\sigma(e_{i,n},e_{j,n})$,
and the inequality follows from the conditional Jensen's inequality. 

For $i=j$, we have 
\begin{align*}
 & \left\Vert Z_{i,n}-Z_{i,n,j}\right\Vert _{L^{p},\c_{n}}=\biggl\Vert\sum^{n}_{k=1}\Bigl[\E_{\c_{n}}\left(Y_{k,n}\mid\F_{i,n}\right)-\E_{\c_{n}}\left(Y_{k,n}\mid\F_{i-1,n}\right)-\E_{\c_{n}}\left(Y_{k,n,j}\mid\mathcal{G}_{i,j}\right)\\
 & \ +\E_{\c_{n}}\left(Y_{k,n,j}\mid\F_{i-1,n}\right)\Bigr]\biggr\Vert_{L^{p},\c_{n}}=\biggl\Vert\sum^{n}_{k=1}\Bigl[\E_{\c_{n}}\left(Y_{k,n}\mid\F_{i,n}\right)-\E_{\c_{n}}\left(Y_{k,n,i}\mid\mathcal{G}_{i,i}\right)\Bigr]\biggr\Vert_{L^{p},\c_{n}}\\
 & =\biggl\Vert\sum^{n}_{k=1}\Bigl[\E_{\c_{n}}\left(Y_{k,n}\mid\F_{i,n}\vee\sigma(e^{*}_{i,n})\right)-\E_{\c_{n}}\left(Y_{k,n,i}\mid\mathcal{G}_{i,i}\vee\sigma(e_{i,n})\right)\Bigr]\biggr\Vert_{L^{p},\c_{n}}\\
 & =\biggl\Vert\sum^{n}_{k=1}\Bigl[\E_{\c_{n}}\left(Y_{k,n}-Y_{k,n,i}\mid\F_{i,n}\vee\sigma(e^{*}_{i,n})\right)\biggr\Vert_{L^{p},\c_{n}}\leq\mathfrak{d}_{p,n}(i,i,\c_{n}),
\end{align*}
where the first equality follows from $\mathcal{G}_{i-1,i}=\F_{i-1,n}$,
the third equality follows from $\E_{\c_{n}}\left(Y_{k,n}\mid\F_{i,n}\right)=\E_{\c_{n}}\left(Y_{k,n}\mid\F_{i,n}\vee\sigma(e^{*}_{i,n})\right)$
and $\E_{\c_{n}}\left(Y_{k,n,i}\mid\mathcal{G}_{i,i}\right)=\E_{\c_{n}}\left(Y_{k,n,i}\mid\mathcal{G}_{i,i}\vee\sigma(e_{i,n})\right)$,
and the last inequality follows from the conditional Jensen's inequality. 

For $j>i$, it is obvious that $\left\Vert Z_{i,n}-Z_{i,n,j}\right\Vert _{L^{p},\c_{n}}=0\leq\mathfrak{d}_{p,n}(i,j,\c_{n})$. 

Thus, we have shown that $\delta^{\sharp}_{p,n}(i,j,\c_{n})=\left\Vert Z_{i,n}-Z_{i,n,j}\right\Vert _{L^{p},\c_{n}}\leq\mathfrak{d}_{p,n}\left(i,j,\c_{n}\right)$.
Since $\mathfrak{d}_{p,n}\left(i,j,\c_{n}\right)=\mathfrak{d}_{p,n}\left(j,i,\c_{n}\right)$,
the proof is completed.$\hfill\qedsymbol$

\textbf{Proof of Theorem \ref{thm:CLT finite}.} Recall: $P_{i}Y_{j,n}\equiv\E_{\c_{n}}(Y_{j,n}|\mathcal{F}_{i,n})-\E_{\c_{n}}(Y_{j,n}|\mathcal{F}_{i-1,n})$
and $Z_{i,n}\equiv\sum^{n}_{j=1}P_{i}Y_{j,n}$. The proof of this
theorem is based on a CLT for MDA (Lemma~\ref{CLT MDA}). Notice
that 
\[
\sigma^{-1}_{n}\sum^{n}_{j=1}\left(Y_{j,n}-\E_{\c_{n}}Y_{j,n}\right)=\sigma^{-1}_{n}\sum^{n}_{j=1}\sum^{n}_{i=1}P_{i}Y_{j,n}=\sum^{n}_{i=1}\sigma^{-1}_{n}Z_{i,n}=\sum^{n}_{i=1}X_{i,n},
\]
where $X_{i,n}\equiv\sigma^{-1}_{n}Z_{i,n}$. And $\left\{ X_{i,n},\F_{i,n}:i=1,\ldots,n,n\geq1\right\} $
is a MDA under the unconditional probability measure $\p$. To apply
Lemma \ref{CLT MDA}, we need to show that (a) $\sum^{n}_{i=1}X^{2}_{i,n}\xrightarrow{\p}1$,
(b) $\max_{i=1,\ldots,n}\left|X_{i,n}\right|\xrightarrow{\p}0$, and
(c) $\E\left(\max_{i=1,\ldots,n}X^{2}_{i,n}\right)<K$ for some constant
$K>0$ for all $n\geq1$.

\textbf{Condition (a):} Since $\{X_{i,n},i=1,\ldots,n\}$ is a MDA
under the conditional probability measure $\p_{\c_{n}}$ and $\sigma_{n}$
is measurable with respect to $\c_{n}$, we have 
\[
\sigma^{2}_{n}=\var_{\c_{n}}\left(\sum^{n}_{j=1}Y_{j,n}\right)=\var_{\c_{n}}\left[\sum^{n}_{i=1}\left(\sum^{n}_{j=1}P_{i}Y_{j,n}\right)\right]=\sigma^{2}_{n}\sum^{n}_{i=1}\var_{\c_{n}}(X_{i,n}),
\]
and thus 
\begin{equation}
\sum^{n}_{i=1}\E_{\c_{n}}X^{2}_{i,n}=\sum^{n}_{i=1}\var_{\c_{n}}(X_{i,n})=1\label{eq:sum of varian is 1}
\end{equation}
a.s. Then, condition (a) is equivalent to $\sigma^{-2}_{n}\sum^{n}_{i=1}\left(Z^{2}_{i,n}-\E_{\c_{n}}Z^{2}_{i,n}\right)\xrightarrow{\p}0$.
By \citet[Lemma 6.1]{chernozhukov2018Double}, it suffices to show
that $\sigma^{-2}_{n}\left\Vert \sum^{n}_{i=1}\left(Z^{2}_{i,n}-\E_{\c_{n}}Z^{2}_{i,n}\right)\right\Vert _{L^{p/2},\c_{n}}=o_{\p}(1)$.
We intend to apply Theorem~\ref{thm: Rosenthal}. Let 
\begin{align*}
\mathcal{G}_{i,j} & =\begin{cases}
\sigma(e_{1,n},\dots,e^{*}_{j,n},\dots,e_{i,n}) & \text{if}\ j\leq i\\
\F_{i,n} & \text{if}\ j>i
\end{cases}
\end{align*}
and $Z_{i,n,j}=\sum^{n}_{k=1}\left[\E_{\c_{n}}\left(Y_{k,n,j}\mid\mathcal{G}_{i,j}\right)-\E_{\c_{n}}\left(Y_{k,n,j}\mid\mathcal{G}_{i-1,j}\right)\right].$
Here $Z_{i,n,j}$ is the coupled version of $Z_{i,n}$ with $e_{j,n}$
being replaced by its i.i.d. copy $e^{*}_{j,n}$ conditional on $\c_{n}$.
The FDM $\delta^{*}_{p/2}(i,j,\c_{n})\equiv\left\Vert Z^{2}_{i,n}-Z^{2}_{i,n,j}\right\Vert _{L^{p/2},\c_{n}}$
measures the impact of $e_{j,n}$ on $Z^{2}_{i,n}$ under the $L^{p/2}$-norm.
Then, by Theorem~\ref{thm: Rosenthal}, it suffices to show 
\begin{equation}
\frac{1}{\sigma^{\min\{p,4\}}_{n}}\sum^{n}_{i=1}\left[\sum^{n}_{j=1}\delta^{*}_{p/2}\left(j,i,\c_{n}\right)\right]^{\min\{p/2,2\}}=o_{\p}(1)\label{eq:condition for LLN for Z^2}
\end{equation}
as $n\to\infty$. To show Eq.(\ref{eq:condition for LLN for Z^2}),
we control the FDM $\delta^{*}_{p/2}(i,j,\c_{n})$. Note that we have

\begin{align*}
 & \delta^{*}_{p/2}(i,j,\c_{n})=\left\Vert Z^{2}_{i,n}-Z^{2}_{i,n,j}\right\Vert _{L^{p/2},\c_{n}}=\left\Vert \left(Z_{i,n}-Z_{i,n,j}\right)\left(Z_{i,n}+Z_{i,n,j}\right)\right\Vert _{L^{p/2},\c_{n}}\\
\leq & \left\Vert Z_{i,n}+Z_{i,n,j}\right\Vert _{L^{p},\c_{n}}\left\Vert Z_{i,n}-Z_{i,n,j}\right\Vert _{L^{p},\c_{n}}\\
\leq & \left(\left\Vert Z_{i,n,j}\right\Vert _{L^{p},\c_{n}}+\left\Vert Z_{i,n}\right\Vert _{L^{p},\c_{n}}\right)\left\Vert Z_{i,n}-Z_{i,n,j}\right\Vert _{L^{p},\c_{n}}\\
= & 2\left\Vert Z_{i,n}\right\Vert _{L^{p},\c_{n}}\left\Vert Z_{i,n}-Z_{i,n,j}\right\Vert _{L^{p},\c_{n}}=2\left\Vert \sum^{n}_{j=1}P_{i}Y_{j,n}\right\Vert _{L^{p},\c_{n}}\left\Vert Z_{i,n}-Z_{i,n,j}\right\Vert _{L^{p},\c_{n}}\\
\leq & 2\left(\sum^{n}_{j=1}\left\Vert P_{i}Y_{j,n}\right\Vert _{L^{p},\c_{n}}\right)\left\Vert Z_{i,n}-Z_{i,n,j}\right\Vert _{L^{p},\c_{n}}\leq2\left(\sum^{n}_{j=1}\delta_{p,n}(j,i,\c_{n})\right)\mathfrak{d}_{p,n}\left(i,j,\c_{n}\right),
\end{align*}
where the first inequality follows from Hölder's inequality, the final
inequality follows from Lemma~\ref{lem:predi < FMD}, Lemma~\ref{lem:fdm of martingale < 2nd FMD}
and the definition of $\mathfrak{d}_{p,n}\left(i,j,\c_{n}\right)$,
and the third equality is from the fact that $Z_{i,n,j}$ and $Z_{i,n}$
are identically distributed conditional on $\c_{n}$. Thus, it follows
from Conditions (1) and (3) in this theorem that Eq.(\ref{eq:condition for LLN for Z^2})
holds. We conclude that Condition (a) is satisfied.

\textbf{Condition (b):} Note that 
\begin{align*}
 & \left\{ \E_{\c_{n}}\left(\max_{i=1,\cdots,n}|X_{i,n}|\right)\right\} ^{p}\leq\E_{\c_{n}}\left[\left(\max_{i=1,\cdots,n}|X_{i,n}|\right)^{p}\right]\leq\E_{\c_{n}}\sum^{n}_{i=1}|X_{i,n}|^{p}\\
= & \sigma^{-p}_{n}\sum^{n}_{i=1}\left\Vert \sum^{n}_{j=1}P_{i}Y_{j,n}\right\Vert ^{p}_{L^{p},\c_{n}}\leq\sigma^{-p}_{n}\sum^{n}_{i=1}\left(\sum^{n}_{j=1}\delta_{p,n}(j,i,\c_{n})\right)^{p}=o_{\p}(1),
\end{align*}
where the first inequality follows from conditional Jensen's inequality,
the third one is by Lemma~\ref{lem:predi < FMD}, and the last step
is by Conditions (1) and (2) in this theorem. Then Condition (b) follows
from conditional Markov's inequality and \citet[Lemma 6.1 (a)]{chernozhukov2018Double}. 

\textbf{Condition (c):} Condition (c) is satisfied because
\[
\E\left(\max_{i=1,\ldots,n}X^{2}_{i,n}\right)\leq\E\left(\sum^{n}_{i=1}X^{2}_{i,n}\right)=\E\left[\E_{\c_{n}}\left(\sum^{n}_{i=1}X^{2}_{i,n}\right)\right]=\E\left[\sum^{n}_{i=1}\E_{\c_{n}}\left(X^{2}_{i,n}\right)\right]=1,
\]
where the last step follows from Eq.(\ref{eq:sum of varian is 1}).
Thus the conclusion follows.$\hfill\qedsymbol$

\textbf{Proof of Lemma \ref{lem:suffi condi for clt-smooth}.} It
follows from the definition of $\mathfrak{d}_{p,n}\left(i,i,\c_{n}\right)$
and Minkowski's inequality that
\[
\mathfrak{d}_{p,n}\left(i,i,\c_{n}\right)\leq\sum^{n}_{k=1}\left\Vert Y_{k,n}-Y_{k,n,i}\right\Vert _{L^{p},\c_{n}}=\sum^{n}_{k=1}\delta_{p,n}(k,i,\c_{n}).
\]
Now, fix any $i\neq j$. Let $a\equiv(e^{*}_{i,n}-e_{i,n})u_{i}$
and $b\equiv(e^{*}_{j,n}-e_{j,n})u_{j}$, where $u_{i}$ denotes the
$i$th standard basis vector in $\R^{n}$. Consequently, we have the
following integral representation of the second-order increment: 
\begin{align*}
 & F_{n}(e_{n}+a+b)-F_{n}(e_{n}+a)-F_{n}(e_{n}+b)+F_{n}(e_{n})\\
= & \int^{1}_{0}\int^{1}_{0}\frac{\partial^{2}F_{n}(e_{n}+sa+tb)}{\partial e_{i,n}\partial e_{j,n}}\,ds\,dt\,(e^{*}_{i,n}-e_{i,n})(e^{*}_{j,n}-e_{j,n}),
\end{align*}
where $F_{n}(e_{n})\equiv(F_{1,n}(e_{n}),\ldots,F_{n,n}(e_{n}))^{\prime}$.
Therefore, 
\begin{align}
 & \mathfrak{d}_{p,n}\left(i,j,\c_{n}\right)=\left\Vert \sum^{n}_{k=1}\left\{ Y_{k,n}-Y_{k,n,i}-Y_{k,n,j}+Y_{k,n,\{i,j\}}\right\} \right\Vert _{L^{p},\c_{n}}\nonumber \\
= & \left\Vert \sum^{n}_{k=1}\left\{ F_{k,n}(e_{n}+a+b)-F_{k,n}(e_{n}+a)-F_{k,n}(e_{n}+b)+F_{k,n}(e_{n})\right\} \right\Vert _{L^{p},\c_{n}}\nonumber \\
\leq & \left\Vert \left|\int^{1}_{0}\int^{1}_{0}\sum^{n}_{k=1}\frac{\partial^{2}F_{k,n}(e_{n}+sa+tb)}{\partial e_{i,n}\partial e_{j,n}}\,ds\,dt\right|\times\left|e^{*}_{i,n}-e_{i,n}\right|\left|e^{*}_{j,n}-e_{j,n}\right|\right\Vert _{L^{p},\c_{n}}\nonumber \\
\leq & \left\Vert \sup_{e_{n}}\left|\sum^{n}_{k=1}\frac{\partial^{2}F_{k,n}(e_{n})}{\partial e_{i,n}\partial e_{j,n}}\right|\left|e^{*}_{i,n}-e_{i,n}\right|\left|e^{*}_{j,n}-e_{j,n}\right|\right\Vert _{L^{p},\c_{n}}\nonumber \\
\leq & \sup_{e_{n}}\left|\sum^{n}_{k=1}\frac{\partial^{2}F_{k,n}(e_{n})}{\partial e_{i,n}\partial e_{j,n}}\right|\left\Vert e^{*}_{i,n}-e_{i,n}\right\Vert _{L^{p},\c_{n}}\left\Vert e^{*}_{j,n}-e_{j,n}\right\Vert _{L^{p},\c_{n}}\nonumber \\
\leq & \left\Vert e\right\Vert ^{2}_{L^{p},\c_{n}}\sup_{e_{n}}\left|\sum^{n}_{k=1}\frac{\partial^{2}F_{k,n}(e_{n})}{\partial e_{i,n}\partial e_{j,n}}\right|.\label{eq: 2nd FDM and 2nd derivative}
\end{align}
The first equality utilizes the coupling construction of $Y_{k,n}$,
while the first inequality stems from the integral representation
of the second-order increment. The third inequality follows from the
conditional independence of $(e^{*}_{i,n}-e_{i,n})$ and $(e^{*}_{j,n}-e_{j,n})$
given $\mathcal{C}_{n}$ and the fact that the functions $F_{1,n},\ldots,F_{n,n}$
are measurable with respect to $\c_{n}$. The proof is completed.$\hfill\qedsymbol$

\textbf{Proof of Lemma \ref{lem:suffi condi for clt}.} For any $i\neq j$,
we have 
\begin{align*}
 & \mathfrak{d}_{p,n}\left(i,j,\c_{n}\right)=\left\Vert \sum^{n}_{k=1}\left\{ Y_{k,n}-Y_{k,n,i}-Y_{k,n,j}+Y_{k,n,\{i,j\}}\right\} \right\Vert _{L^{p},\c_{n}}\leq\sum^{n}_{k=1}\Bigl\Vert Y_{k,n}-Y_{k,n,i}\\
 & \quad-Y_{k,n,j}+Y_{k,n,\{i,j\}}\Bigr\Vert_{L^{p},\c_{n}}\leq\sum^{n}_{k=1}\min\Bigl\{\left\Vert Y_{k,n}-Y_{k,n,i}\right\Vert _{L^{p},\c_{n}}+\left\Vert Y_{k,n,j}-Y_{k,n,\{i,j\}}\right\Vert _{L^{p},\c_{n}},\\
 & \quad\left\Vert Y_{k,n}-Y_{k,n,j}\right\Vert _{L^{p},\c_{n}}+\left\Vert Y_{k,n,i}-Y_{k,n,\{i,j\}}\right\Vert _{L^{p},\c_{n}}\Bigr\}.
\end{align*}
Conditional on $\c_{n}$, $Y_{k,n}-Y_{k,n,i}$ has the same distribution
as $Y_{k,n,j}-Y_{k,n,\{i,j\}}$, and $Y_{k,n}-Y_{k,n,j}$ has the
same distribution as $Y_{k,n,i}-Y_{k,n,\{i,j\}}$. Hence $\left\Vert Y_{k,n,j}-Y_{k,n,\{i,j\}}\right\Vert _{L^{p},\c_{n}}=\left\Vert Y_{k,n}-Y_{k,n,i}\right\Vert _{L^{p},\c_{n}}=\delta_{p,n}(k,i,\c_{n})$
and similarly $\left\Vert Y_{k,n,i}-Y_{k,n,\{i,j\}}\right\Vert _{L^{p},\c_{n}}=\left\Vert Y_{k,n}-Y_{k,n,j}\right\Vert _{L^{p},\c_{n}}=\delta_{p,n}(k,j,\c_{n})$.
Therefore, $\mathfrak{d}_{p,n}\left(i,j,\c_{n}\right)\le2\sum^{n}_{k=1}\min\{\delta_{p,n}(k,i,\c_{n}),\delta_{p,n}(k,j,\c_{n})\}$.
For $i=j$, we have 
\begin{align*}
\mathfrak{d}_{p,n}\left(i,i,\c_{n}\right) & =\left\Vert \sum^{n}_{k=1}\left\{ Y_{k,n}-Y_{k,n,i}\right\} \right\Vert _{L^{p},\c_{n}}\le\sum^{n}_{k=1}\left\Vert Y_{k,n}-Y_{k,n,i}\right\Vert _{L^{p},\c_{n}}\\
 & =\sum^{n}_{k=1}\delta_{p,n}(k,i,\c_{n})\le2\sum^{n}_{k=1}\min\{\delta_{p,n}(k,i,\c_{n}),\delta_{p,n}(k,i,\c_{n})\}.
\end{align*}
This proves part (i).

We next prove part (ii). Let $\tilde{p}\equiv\min\{2,p/2\}>1$. By
part (i), condition (\ref{eq:key condition for finite clt}) will
follow if 
\[
\frac{1}{n^{\tilde{p}}}\sum^{n}_{i=1}\left[\left\{ \sum^{n}_{k=1}\delta_{p,n}(k,i,\c_{n})\right\} \sum^{n}_{j=1}\sum^{n}_{k=1}\min\{\delta_{p,n}(k,i,\c_{n}),\delta_{p,n}(k,j,\c_{n})\}\right]^{\tilde{p}}=o_{\p}(1).
\]
Since $\sup_{i\in[n]}\sum^{n}_{k=1}\delta_{p,n}(k,i,\c_{n})=O_{\p}(1)$,
it suffices to show that 
\begin{equation}
\sup_{i\in[n]}\sum^{n}_{j=1}\sum^{n}_{k=1}\min\{\delta_{p,n}(k,i,\c_{n}),\delta_{p,n}(k,j,\c_{n})\}=o_{\p}\left(n^{(\tilde{p}-1)/\tilde{p}}\right).\label{eq: proof suff condi clt}
\end{equation}

Let $\kappa_{n}\equiv\left(\frac{n}{\log n}\right)^{1/\alpha}$. By
Eq.(\ref{eq:suff condition for clt}), 
\begin{align}
 & \sup_{k\in[n]}\sum_{r\ge\kappa_{n}}\delta_{p,n}\left(k,\pi_{k}(r),\c_{n}\right)\le C\sup_{k\in[n]}\sum_{r\ge\kappa_{n}}r^{-\alpha}\nonumber \\
\le & C\int^{\infty}_{\kappa_{n}-1}x^{-\alpha}\,dx=\frac{C}{\alpha-1}(\kappa_{n}-1)^{-(\alpha-1)}=o\left(n^{-1/\tilde{p}}\right)\label{eq:partial sum of delta diminishes}
\end{align}
with probability approaching one, because $\alpha>\tilde{p}/(\tilde{p}-1)$.

Now, for any fixed $i\in[n]$, 
\begin{align*}
 & \sum^{n}_{j=1}\sum^{n}_{k=1}\min\{\delta_{p,n}(k,i,\c_{n}),\delta_{p,n}(k,j,\c_{n})\}=\sum^{n}_{k=1}\sum^{n}_{r=1}\min\{\delta_{p,n}(k,i,\c_{n}),\delta_{p,n}(k,\pi_{k}(r),\c_{n})\}\\
\le & \sum^{n}_{k=1}\sum_{r<\kappa_{n}}\delta_{p,n}(k,i,\c_{n})+\sum^{n}_{k=1}\sum_{r\ge\kappa_{n}}\delta_{p,n}(k,\pi_{k}(r),\c_{n}).
\end{align*}
Taking the supremum over $i\in[n]$ gives 
\begin{align*}
 & \sup_{i\in[n]}\sum^{n}_{j=1}\sum^{n}_{k=1}\min\{\delta_{p,n}(k,i,\c_{n}),\delta_{p,n}(k,j,\c_{n})\}\\
\le & \sup_{i\in[n]}\sum^{n}_{k=1}\sum_{r<\kappa_{n}}\delta_{p,n}(k,i,\c_{n})+\sum^{n}_{k=1}\sum_{r\ge\kappa_{n}}\delta_{p,n}(k,\pi_{k}(r),\c_{n})\\
\le & \kappa_{n}\sup_{i\in[n]}\sum^{n}_{k=1}\delta_{p,n}(k,i,\c_{n})+n\sup_{k\in[n]}\sum_{r\ge\kappa_{n}}\delta_{p,n}(k,\pi_{k}(r),\c_{n}).
\end{align*}
By assumption and Eq.(\ref{eq:partial sum of delta diminishes}),
the right-hand side is 
\[
O_{\p}(1)\cdot\kappa_{n}+n\cdot o_{\p}\left(n^{-1/\tilde{p}}\right)=o_{\p}\left(n^{(\tilde{p}-1)/\tilde{p}}\right),
\]
since $\kappa_{n}=\left(\frac{n}{\log n}\right)^{1/\alpha}=o\left(n^{(\tilde{p}-1)/\tilde{p}}\right)$.
Thus Eq.(\ref{eq: proof suff condi clt}) holds, and therefore condition
(\ref{eq:key condition for finite clt}) follows. The proof is completed.
$\hfill\qedsymbol$

\bibliographystyle{dcu}
\bibliography{r}

\newpage{}
\begin{center}
{\LARGE\textbf{Online Supplementary Appendices to ``Limit Theorems
for Network Data without Metric Structure''}}{\LARGE\par}
\par\end{center}

\begin{center}
Wen Jiang\ \ \ Yachen Wang\ \ \ Zeqi Wu\ \ \ Xingbai Xu 
\par\end{center}

This Supplement Appendix contains the proofs of Lemmas \ref{lem: Burkholder}-\ref{lem:general pred<FDM},
as well as the proofs for Sections~\ref{Sec: Example} and \ref{sec: Transformation}.

\section{Proofs for Lemmas \ref{lem: Burkholder}-\ref{lem:general pred<FDM}}

\textbf{Proof of Lemma \ref{lem: Burkholder}.} Fix a regular conditional
probability measure $\{\mathbb{P}_{\omega}(\cdot):\omega\in\Omega\}$
for $\mathbb{P}(\cdot\mid\c)$. For almost every $\omega$, the array
$\{X_{i}\}$ is a martingale difference sequence under $\mathbb{P}_{\omega}$,
so the cited unconditional inequality applies under $\mathbb{P}_{\omega}$.
Translating back to conditional norms yields the stated result. 

When $p>2$, the conclusion follows from Theorem 2.1 in \citet{rio2009moment}.
When $p=2$, since $\E_{\c}\left(X_{i}X_{j}\right)=0$ for all $i\neq j$,
$\left\Vert \sum^{n}_{i=1}X_{i}\right\Vert _{L^{2},\c}=\left(\sum^{n}_{i=1}\left\Vert X_{i}\right\Vert ^{2}_{L^{2},\c}\right)^{1/2}$.

When $1<p<2$, by Theorem 3.1 in \citet{burkholder1988sharp}, we
have 
\[
\left\Vert \sum^{n}_{i=1}X_{i}\right\Vert _{L^{p},\c}\leq C_{p}\left\Vert \left(\sum^{n}_{i=1}X^{2}_{i}\right)^{1/2}\right\Vert _{L^{p},\c}
\]
a.s. Notice that for any non-negative real numbers $a_{1},\ldots,a_{n}$
and $q>r>1$, we have $\left(\sum^{n}_{i=1}a^{r}_{i}\right)^{1/r}\geq\left(\sum^{n}_{i=1}a^{q}_{i}\right)^{1/q}$
(Proposition 9.1.5 in \citet{dennis2009matrix}, p.599). Thus, from
the above two inequalities, we have 
\begin{align*}
 & \left\Vert \sum^{n}_{i=1}X_{i}\right\Vert _{L^{p},\c}\leq C_{p}\left\Vert \left(\sum^{n}_{i=1}X^{2}_{i}\right)^{1/2}\right\Vert _{L^{p},\c}\leq C_{p}\left\Vert \left(\sum^{n}_{i=1}|X_{i}|^{p}\right)^{1/p}\right\Vert _{L^{p},\c}\\
= & C_{p}\left(\E_{\c}\sum^{n}_{i=1}|X_{i}|^{p}\right)^{1/p}=C_{p}\left(\sum^{n}_{i=1}\left\Vert X_{i}\right\Vert ^{p}_{L^{p},\c}\right)^{1/p}
\end{align*}
a.s.$\hfill\qedsymbol$

\textbf{Proof of Lemma \ref{CLT MDA}.} See Theorem 3.2 in \citet{hallMartingaleLimitTheory1980}
and the Remarks in \citet[p.59]{hallMartingaleLimitTheory1980}. $\hfill\qedsymbol$

\textbf{Proof of Lemma \ref{lem:general pred<FDM}.} It follows that
\begin{align*}
 & \left\Vert \E_{\c_{n}}\left(Q_{n}|\sigma(e_{k,n}:k\in\mathcal{I})\right)-\E_{\c_{n}}\left(Q_{n}|\sigma(e_{k,n}:k\in\mathcal{I}\backslash\{i\})\right)\right\Vert _{L^{p},\c_{n}}\\
= & \left\Vert \E_{\c_{n}}\left(Q_{n}|\sigma(e_{k,n}:k\in\mathcal{I})\right)-\E_{\c_{n}}\left(Q_{n,i}|\sigma(e_{k,n}:k\in\mathcal{I}\backslash\{i\})\right)\right\Vert _{L^{p},\c_{n}}\\
= & \left\Vert \E_{\c_{n}}\left(Q_{n}|\sigma(e_{k,n}:k\in\mathcal{I})\right)-\E_{\c_{n}}\left(Q_{n,i}|\sigma(e_{k,n}:k\in\mathcal{I})\right)\right\Vert _{L^{p},\c_{n}}\\
= & \left\Vert \E_{\c_{n}}\left(Q_{n}-Q_{n,i}|\sigma(e_{k,n}:k\in\mathcal{I})\right)\right\Vert _{L^{p},\c_{n}}\leq\left\Vert Q_{n}-Q_{n,i}\right\Vert _{L^{p},\c_{n}},
\end{align*}
where the first equality follows from the fact that $e^{*}_{i,n}$
is an i.i.d. (conditional on $\c_{n}$) copy of $e_{i,n}$, the second
equality follows from Corollary 2 in \citet[p.231]{chowProbabilityTheoryIndependence1988},
and the inequality follows from conditional Jensen's inequality. $\hfill\qedsymbol$

\section{Proofs for Section \ref{Sec: Example}}

\textbf{Proof of Proposition \ref{prop: SAR FDM}. }In this proof,
for any vector or matrix $A=\left(a_{ij}\right)_{n\times m}$, we
denote $\left|A\right|\equiv\left(\left|a_{ij}\right|\right)_{n\times m}$.
Direct calculations show that $|A+B|\leq^{*}|A|+|B|$ and $|AB|\leq^{*}|A||B|$,
where $A=\left(a_{ij}\right)_{m\times n}\leq^{*}B=\left(b_{ij}\right)_{m\times n}$
means $\forall i,j:a_{ij}\leq b_{ij}$. To shorten formulas, denote
$e_{i,n}\equiv X_{i,n}'\beta+\epsilon_{i,n}$ and $e_{n}\equiv X_{n}\beta+\epsilon_{n}$.
Denote the solution of \eqref{eq:SAR model} as $Y_{n}\left(e_{n}\right)$.
Then $Y_{n}\left(e_{n}\right)=F\left(\lambda W_{n}Y_{n}\left(e_{n}\right)+e_{n}\right)$.
For any two realizations of $e_{n}$, denoted by $e^{\left(1\right)}_{n}$
and $e^{\left(2\right)}_{n}$, consider $Y^{\left(1\right)}_{n}=Y_{n}\left(e^{\left(1\right)}_{n}\right)$
and $Y^{\left(2\right)}_{n}=Y_{n}\left(e^{\left(2\right)}_{n}\right)$.
So, for any $1\leq i\leq n$, 
\begin{align*}
 & \left|Y^{\left(1\right)}_{j,n}-Y^{\left(2\right)}_{j,n}\right|=\left|F\left(\lambda w_{j.,n}Y^{\left(1\right)}_{n}+e^{\left(1\right)}_{j,n}\right)-F\left(\lambda w_{j.,n}Y^{\left(2\right)}_{n}+e^{\left(2\right)}_{j,n}\right)\right|\\
\leq & L\left|\lambda\right|\sum^{n}_{i=1}\left|w_{ji,n}\right|\left|Y^{\left(1\right)}_{i,n}-Y^{\left(2\right)}_{i,n}\right|+L\left|e^{\left(1\right)}_{j,n}-e^{\left(2\right)}_{j,n}\right|.
\end{align*}
The above inequality can be written in a matrix form:
\begin{equation}
\text{\ensuremath{\left(I_{n}-L\left|\lambda\right|\left|W_{n}\right|\right)}}\left|Y^{\left(1\right)}_{n}-Y^{\left(2\right)}_{n}\right|\leq^{*}L\left|e^{(1)}_{n}-e^{(2)}_{n}\right|.\label{eq:SAR vector inequality}
\end{equation}
Since $S^{+}_{n}=L\left(I_{n}-L\left|\lambda W_{n}\right|\right)^{-1}=L\sum^{\infty}_{l=0}\left(L\left|\lambda W_{n}\right|\right)^{l}$,
all entries of $S^{+}_{n}$ are nonnegative. As a result, we can multiply
$\frac{1}{L}S^{+}_{n}$ on both sides of \eqref{eq:SAR vector inequality}:
$\left|Y^{\left(1\right)}_{n}-Y^{\left(2\right)}_{n}\right|\leq^{*}S^{+}_{n}\left|e^{(1)}_{n}-e^{(2)}_{n}\right|$.
So, for any $1\leq i\leq n$, 
\begin{equation}
\left|Y^{\left(1\right)}_{j,n}-Y^{\left(2\right)}_{j,n}\right|\leq\sum^{n}_{i=1}S^{+}_{ji,n}\left|e^{\left(1\right)}_{i,n}-e^{\left(2\right)}_{i,n}\right|.\label{eq:SAR lip property}
\end{equation}

Denote $Y_{j,n,i}$ as the coupled version of $Y_{j,n}$ with $e_{i,n}$
replaced by its i.i.d. copy $e^{*}_{i,n}\equiv X_{i,n}'\beta+\epsilon^{*}_{i,n}$
(conditional on $\c_{n}$). By Eq.(\ref{eq:SAR lip property}), 
\begin{align*}
 & \delta_{p,n}\left(j,i,\c_{n}\right)=\left\Vert Y_{j,n}-Y_{j,n,i}\right\Vert _{L^{p},\c_{n}}\leq S^{+}_{ji,n}\left\Vert e_{i,n}-e^{*}_{i,n}\right\Vert _{L^{p},\c_{n}}\\
= & S^{+}_{ji,n}\left\Vert X_{i,n}'\beta+\epsilon_{i,n}-X_{i,n}'\beta-\epsilon^{*}_{i,n}\right\Vert _{L^{p},\c_{n}}=S^{+}_{ji,n}\left\Vert \epsilon_{i,n}-\epsilon^{*}_{i,n}\right\Vert _{L^{p},\c_{n}}\leq2\left\Vert \epsilon\right\Vert _{L^{p},\c_{n}}S^{+}_{ji,n}.
\end{align*}
$\hfill\qedsymbol$

\textbf{Proof of Lemma \ref{lem:2nd FDM for SAR}.} In this proof,
for any vector or matrix $A=\left(a_{ij}\right)_{n\times m}$, we
denote $\left|A\right|\equiv\left(\left|a_{ij}\right|\right)_{n\times m}$.
Direct calculations show that $|A+B|\leq^{*}|A|+|B|$ and $|AB|\leq^{*}|A||B|$,
where $A=\left(a_{ij}\right)_{m\times n}\leq^{*}B=\left(b_{ij}\right)_{m\times n}$
means $\forall i,j:a_{ij}\leq b_{ij}$. To shorten formulas, we introduce
some notation. Let $e_{i,n}\equiv X_{i,n}'\beta+\epsilon_{i,n}$ and
$e_{n}\equiv X_{n}\beta+\epsilon_{n}$. Denote the solution of \eqref{eq:SAR model}
as $Y_{n}\left(e_{n}\right)$. 

First of all, it follows from the definition of $\mathfrak{d}_{p,n}\left(i,i,\c_{n}\right)$
and Proposition~\ref{prop: SAR FDM} that
\[
\mathfrak{d}_{p,n}\left(i,i,\c_{n}\right)\leq\sum^{n}_{k=1}\left\Vert Y_{k,n}-Y_{k,n,i}\right\Vert _{L^{p},\c_{n}}=\sum^{n}_{k=1}\delta_{p,n}(k,i,\c_{n})\leq2\left\Vert \epsilon\right\Vert _{L^{p},\c_{n}}\sum^{n}_{k=1}S^{+}_{ki,n}.
\]

Now, we consider the case that $i\neq j$. Since $F$ is twice continuously
differentiable, the implicit function theorem implies that $Y_{n}(e_{n})$
is also twice continuously differentiable, provided that the Jacobian
$I_{n}-\lambda\Lambda(e_{n})W_{n}$ is invertible (a condition to
be established later). By Lemma~\ref{lem:suffi condi for clt-smooth},
we have 
\begin{align}
\mathfrak{d}_{p,n}\left(i,j,\c_{n}\right) & \leq\sup_{i\in[n],n\geq1}\left\Vert e_{i,n}-e^{*}_{i,n}\right\Vert ^{2}_{L^{p},\c_{n}}\sup_{e_{n}}\left|\sum^{n}_{k=1}\frac{\partial^{2}Y_{k,n}(e_{n})}{\partial e_{i,n}\partial e_{j,n}}\right|\nonumber \\
 & \leq4\left\Vert \epsilon\right\Vert ^{2}_{L^{p},\c_{n}}\sup_{e_{n}}\left|\sum^{n}_{k=1}\frac{\partial^{2}Y_{k,n}(e_{n})}{\partial e_{i,n}\partial e_{j,n}}\right|,\label{eq:2nd FDM sar bound}
\end{align}
where the last step follows from $e^{*}_{i,n}=X_{i,n}'\beta+\epsilon^{*}_{i,n}$
and $e_{i,n}=X_{i,n}'\beta+\epsilon_{i,n}$. Now, it remains to bound
$\sup_{e_{n}}\left|\sum^{n}_{k=1}\frac{\partial^{2}Y_{k,n}(e_{n})}{\partial e_{i,n}\partial e_{j,n}}\right|$.

Let $\eta_{j,n}=\lambda w_{j\cdot,n}Y_{n}+X'_{j,n}\beta+\epsilon_{j,n}$
and $\eta_{n}=(\eta_{1,n},\ldots,\eta_{n,n})^{\prime}=\lambda W_{n}Y_{n}+X_{n}\beta+\epsilon_{n}$.
Notice that $\eta_{n}$ is also a function of $e_{n}$: $\eta_{n}=\eta_{n}(e_{n})=\lambda W_{n}Y_{n}(e_{n})+e_{n}$,
and $Y_{j,n}=F(\eta_{j,n})$. Denote
\[
\Lambda(e_{n})\equiv\mathrm{diag}(F'(\eta_{n}(e_{n}))),\text{ and}\qquad\Gamma(e_{n})\equiv\mathrm{diag}(F''(\eta_{n}(e_{n}))).
\]
\footnote{Throughout this paper, when scalar functions (such as $F$, $F'$,
and $F''$) applied to vectors, it means to operate component-wise.
Accordingly, $\Lambda(e_{n})$ and $\Gamma(e_{n})$ are diagonal matrices
whose $i$th diagonal entries are $F'(\eta_{i,n}(e_{n}))$ and $F''(\eta_{i,n}(e_{n}))$,
respectively.}Differentiating both sides of $Y_{n}(e_{n})=F\left(\eta_{n}(e_{n})\right)$
with respect to $e_{i,n}$ yields
\begin{equation}
\frac{\partial Y_{n}(e_{n})}{\partial e_{i,n}}=\Lambda(e_{n})\Big(\lambda W_{n}\frac{\partial Y_{n}(e_{n})}{\partial e_{i,n}}+u_{i}\Big),\label{eq: Y e deri}
\end{equation}
where $u_{i}$ denotes the $i$th standard basis vector in $\R^{n}$.
By Assumption~\ref{ass: zeta} and the mean value theorem, we have
$\sup_{x}\left|F'(x)\right|\leq L$ and thus $|\Lambda(e_{n})|\le^{*}LI_{n}$.
By Assumption~\ref{ass: zeta}, $\left\Vert \lambda\Lambda(e_{n})W_{n}\right\Vert _{\infty}\leq\zeta<1$.
Thus, $I_{n}-\lambda\Lambda(e_{n})W_{n}$ is invertible and by Neumann
expansion we have 
\begin{align}
 & \left|(I_{n}-\lambda\Lambda(e_{n})W_{n})^{-1}\right|=\left|\sum^{\infty}_{\ell=0}\left\{ \lambda\Lambda(e_{n})W_{n}\right\} ^{\ell}\right|\le^{*}\sum^{\infty}_{\ell=0}\left|\lambda\Lambda(e_{n})\right|^{\ell}\left|W_{n}\right|^{\ell}\nonumber \\
\le^{*} & \sum^{\infty}_{\ell=0}\left|\lambda L\right|^{\ell}\left|W_{n}\right|^{\ell}=\frac{1}{L}L\left(I_{n}-L\left|\lambda W_{n}\right|\right)^{-1}=\frac{1}{L}S^{+}_{n}.\label{eq:derive < S}
\end{align}
As a result, by Eq.(\ref{eq: Y e deri}),
\[
\frac{\partial Y_{n}(e_{n})}{\partial e_{i,n}}=(I_{n}-\lambda\Lambda(e_{n})W_{n})^{-1}\Lambda(e_{n})u_{i}
\]
and 
\begin{align}
 & \left|\frac{\partial Y_{n}(e_{n})}{\partial e_{i,n}}\right|=\left|(I_{n}-\lambda\Lambda(e_{n})W_{n})^{-1}\Lambda(e_{n})u_{i}\right|\le^{*}\frac{S^{+}_{n}}{L}\left|\Lambda(e_{n})\right|u_{i}\le^{*}S^{+}_{n}u_{i}=S^{+}_{\cdot i,n}.\label{eq:DY_bound_col}
\end{align}
Next, for $i\neq j$, differentiating both sides of $\frac{\partial Y_{n}(e_{n})}{\partial e_{i,n}}=\Lambda(e_{n})\frac{\partial\eta_{n}(e_{n})}{\partial e_{i,n}}$
with respect to $e_{j,n}$ yields  
\begin{align*}
 & \frac{\partial^{2}Y_{n}(e_{n})}{\partial e_{i,n}\partial e_{j,n}}=\Gamma(e_{n})\mathrm{diag}\left\{ \frac{\partial\eta_{n}(e_{n})}{\partial e_{j,n}}\right\} \frac{\partial\eta_{n}(e_{n})}{\partial e_{i,n}}+\Lambda(e_{n})\frac{\partial^{2}\eta_{n}(e_{n})}{\partial e_{i,n}\partial e_{j,n}}\\
= & \Gamma(e_{n})\frac{\partial\eta_{n}(e_{n})}{\partial e_{j,n}}\odot\frac{\partial\eta_{n}(e_{n})}{\partial e_{i,n}}+\lambda\Lambda(e_{n})W_{n}\frac{\partial^{2}Y_{n}(e_{n})}{\partial e_{i,n}\partial e_{j,n}},
\end{align*}
where $\odot$ represents the Hadamard product, defined as the entry-wise
multiplication of two vectors of the same dimension, and in the last
step we have used the fact that $\frac{\partial^{2}\eta_{n}(e_{n})}{\partial e_{i,n}\partial e_{j,n}}=\lambda W_{n}\frac{\partial^{2}Y_{n}(e_{n})}{\partial e_{i,n}\partial e_{j,n}}$.
Rearranging the above display gives that 
\[
\frac{\partial^{2}Y_{n}(e_{n})}{\partial e_{i,n}\partial e_{j,n}}=\left(I_{n}-\lambda\Lambda(e_{n})W_{n}\right)^{-1}\Gamma(e_{n})\frac{\partial\eta_{n}(e_{n})}{\partial e_{j,n}}\odot\frac{\partial\eta_{n}(e_{n})}{\partial e_{i,n}}.
\]
By Assumption~\eqref{assu:F second order derivative}, we have $\left|\Gamma(e_{n})\right|\le^{*}\|F''\|_{\infty}I_{n}$.
Then it follows from (\ref{eq:derive < S}) that 
\[
\left|\frac{\partial^{2}Y_{n}(e_{n})}{\partial e_{i,n}\partial e_{j,n}}\right|\le^{*}\frac{\|F''\|_{\infty}}{L}S^{+}\left\{ \left|\frac{\partial\eta_{n}(e_{n})}{\partial e_{j,n}}\right|\odot\left|\frac{\partial\eta_{n}(e_{n})}{\partial e_{i,n}}\right|\right\} .
\]
Now, we analyze $\left|\frac{\partial\eta_{n}(e_{n})}{\partial e_{i,n}}\right|$.
Note that $\frac{\partial\eta_{n}(e_{n})}{\partial e_{i,n}}=\lambda W_{n}\frac{\partial Y_{n}(e_{n})}{\partial e_{i,n}}+u_{i}$,
hence by \eqref{eq:DY_bound_col}, 
\[
\left|\frac{\partial\eta_{n}(e_{n})}{\partial e_{i,n}}\right|\le^{*}\left|\lambda W_{n}\right|S^{+}_{\cdot i,n}+u_{i}=\left(I_{n}-L\left|\lambda W_{n}\right|\right)^{-1}u_{i}=\frac{1}{L}S^{+}_{\cdot i,n},
\]
where we used the identity 
\begin{align*}
 & \left|\lambda W_{n}\right|S^{+}_{\cdot i,n}+u_{i}=\left|\lambda W_{n}\right|L\left(I_{n}-L\left|\lambda W_{n}\right|\right)^{-1}u_{i}+u_{i}\\
= & \left\{ \left|\lambda W_{n}\right|L\left(I_{n}-L\left|\lambda W_{n}\right|\right)^{-1}+I_{n}\right\} u_{i}=\left(I_{n}-L\left|\lambda W_{n}\right|\right)^{-1}u_{i}.
\end{align*}
The above two equations yield
\[
\left|\frac{\partial^{2}Y_{n}(e_{n})}{\partial e_{i,n}\partial e_{j,n}}\right|\le^{*}\frac{\|F''\|_{\infty}}{L^{3}}S^{+}_{n}\left\{ \left|S^{+}_{\cdot i,n}\right|\odot\left|S^{+}_{\cdot j,n}\right|\right\} .
\]
As a result, we have 
\[
\sum^{n}_{k=1}\left|\frac{\partial^{2}Y_{k,n}(e_{n})}{\partial e_{i,n}\partial e_{j,n}}\right|\leq\frac{\|F''\|_{\infty}}{L^{3}}\sum^{n}_{k=1}\left\Vert S^{+}_{\cdot k,n}\right\Vert _{1}S^{+}_{ki,n}S^{+}_{kj,n}.
\]
Taking the supremum over $e_{n}$ we obtain that 
\[
\sup_{e_{n}}\left|\sum^{n}_{k=1}\frac{\partial^{2}Y_{k,n}(e_{n})}{\partial e_{i,n}\partial e_{j,n}}\right|\leq\sup_{e_{n}}\sum^{n}_{k=1}\left|\frac{\partial^{2}Y_{k,n}(e_{n})}{\partial e_{i,n}\partial e_{j,n}}\right|\leq\frac{\|F''\|_{\infty}}{L^{3}}\sum^{n}_{k=1}\left\Vert S^{+}_{\cdot k,n}\right\Vert _{1}S^{+}_{ki,n}S^{+}_{kj,n}.
\]
Combining the above display with (\ref{eq:2nd FDM sar bound}), the
proof is completed.$\hfill\qedsymbol$

\textbf{Proof of Proposition \ref{prop:2nd FDM for SAR}.} Recall
that $\widetilde{p}=\min\{2,p/2\}$. By Assumption~\ref{assu:S column condition}
and Proposition~\ref{prop: SAR FDM}, it suffices to show that 
\[
\frac{1}{n^{p/2}}\sum^{n}_{i=1}\left\Vert S^{+}_{\cdot i,n}\right\Vert ^{p}_{1}=o_{\p}(1).
\]
Notice that for any non-negative real numbers $a_{1},\ldots,a_{n}$
and $q>r>1$, we have $\left(\sum^{n}_{i=1}a^{r}_{i}\right)^{1/r}\geq\left(\sum^{n}_{i=1}a^{q}_{i}\right)^{1/q}$
(Proposition 9.1.5 in \citet{dennis2009matrix}, p.599). By this inequality,
we have 
\[
\frac{1}{n^{p/2}}\sum^{n}_{i=1}\left\Vert S^{+}_{\cdot i,n}\right\Vert ^{p}_{1}\leq\frac{1}{n^{p/2}}\left\{ \sum^{n}_{i=1}\left\Vert S^{+}_{\cdot i,n}\right\Vert ^{2\widetilde{p}}_{1}\right\} ^{\frac{p}{2\widetilde{p}}}=\left\{ \frac{1}{n^{\widetilde{p}-1}}\frac{1}{n}\sum^{n}_{i=1}\left\Vert S^{+}_{\cdot i,n}\right\Vert ^{2\widetilde{p}}_{1}\right\} ^{\frac{p}{2\widetilde{p}}}=o_{\p}(1),
\]
where the last step follows from Assumption~\ref{assu:S column condition}.
Thus condition (\ref{eq:l1 bounded for finite clt}) holds.

Now, we focus on establishing condition (\ref{eq:key condition for finite clt}).
By Lemma~\ref{lem:2nd FDM for SAR} and Proposition~\ref{prop: SAR FDM},
it suffices to show that 

\[
\frac{1}{n^{\widetilde{p}}}\sum^{n}_{i=1}\left[\left(\sum^{n}_{k=1}S^{+}_{ki,n}\right)\left\{ \sum^{n}_{k=1}S^{+}_{ki,n}+\sum^{n}_{j=1}\sum^{n}_{k=1}\left\Vert S^{+}_{\cdot k,n}\right\Vert _{1}S^{+}_{ki,n}S^{+}_{kj,n}\right\} \right]^{\widetilde{p}}=o_{\p}(1).
\]
Since $\left\Vert S^{+}_{\cdot i,n}\right\Vert _{1}=\sum^{n}_{k=1}S^{+}_{ki,n}$,
it suffices to show that (i)
\begin{equation}
\frac{1}{n^{\widetilde{p}}}\sum^{n}_{i=1}\left\Vert S^{+}_{\cdot i,n}\right\Vert ^{2\widetilde{p}}_{1}=o_{\p}(1)\label{eq:SAR clt con1}
\end{equation}
and (ii)
\begin{equation}
\frac{1}{n^{\widetilde{p}}}\sum^{n}_{i=1}\left[\left\Vert S^{+}_{\cdot i,n}\right\Vert _{1}\sum^{n}_{j=1}\sum^{n}_{k=1}\left\Vert S^{+}_{\cdot k,n}\right\Vert _{1}S^{+}_{ki,n}S^{+}_{kj,n}\right]^{\widetilde{p}}=o_{\p}(1).\label{eq:SAR clt con2}
\end{equation}
Eq.(\ref{eq:SAR clt con1}) directly follows from Assumption~\ref{assu:S column condition}.
It remains to show Eq.(\ref{eq:SAR clt con2}). Since
\[
\sum^{n}_{j=1}\sum^{n}_{k=1}\left\Vert S^{+}_{\cdot k,n}\right\Vert _{1}S^{+}_{ki,n}S^{+}_{kj,n}=\sum^{n}_{k=1}\left\Vert S^{+}_{\cdot k,n}\right\Vert _{1}S^{+}_{ki,n}\left(\sum^{n}_{j=1}S^{+}_{kj,n}\right)\leq\left\Vert S^{+}_{n}\right\Vert _{\infty}\sum^{n}_{k=1}\left\Vert S^{+}_{\cdot k,n}\right\Vert _{1}S^{+}_{ki,n},
\]
we have 
\begin{align*}
 & \sum^{n}_{i=1}\left[\left\Vert S^{+}_{\cdot i,n}\right\Vert _{1}\sum^{n}_{j=1}\sum^{n}_{k=1}\left\Vert S^{+}_{\cdot k,n}\right\Vert _{1}S^{+}_{ki,n}S^{+}_{kj,n}\right]^{\widetilde{p}}\leq\left\Vert S^{+}_{n}\right\Vert ^{\widetilde{p}}_{\infty}\sum^{n}_{i=1}\left\Vert S^{+}_{\cdot i,n}\right\Vert ^{2\widetilde{p}}_{1}\left[\sum^{n}_{k=1}\frac{S^{+}_{ki,n}}{\left\Vert S^{+}_{\cdot i,n}\right\Vert _{1}}\left\Vert S^{+}_{\cdot k,n}\right\Vert _{1}\right]^{\widetilde{p}}\\
\leq & \left\Vert S^{+}_{n}\right\Vert ^{\widetilde{p}}_{\infty}\sum^{n}_{i=1}\left\Vert S^{+}_{\cdot i,n}\right\Vert ^{2\widetilde{p}}_{1}\sum^{n}_{k=1}\frac{S^{+}_{ki,n}}{\left\Vert S^{+}_{\cdot i,n}\right\Vert _{1}}\left\Vert S^{+}_{\cdot k,n}\right\Vert ^{\widetilde{p}}_{1}=\left\Vert S^{+}_{n}\right\Vert ^{\widetilde{p}}_{\infty}\sum^{n}_{i=1}\sum^{n}_{k=1}\left\Vert S^{+}_{\cdot i,n}\right\Vert ^{2\widetilde{p}-1}_{1}\left\Vert S^{+}_{\cdot k,n}\right\Vert ^{\widetilde{p}}_{1}S^{+}_{ki,n},
\end{align*}
where the second inequality follows from $x\mapsto x^{\widetilde{p}}$
is convex and Jensen's inequality (observe that $\sum^{n}_{k=1}\frac{S^{+}_{ki,n}}{\left\Vert S^{+}_{\cdot i,n}\right\Vert _{1}}=1$).
Note that $(u,v)$ is a pair of conjugate indices such that $1/u+1/v=1$.
Applying Hölder's inequality, we obtain:
\begin{align*}
 & \sum^{n}_{i=1}\sum^{n}_{k=1}\left\Vert S^{+}_{\cdot i,n}\right\Vert ^{2\widetilde{p}-1}_{1}\left\Vert S^{+}_{\cdot k,n}\right\Vert ^{\widetilde{p}}_{1}S^{+}_{ki,n}\le\left(\sum^{n}_{i=1}\sum^{n}_{k=1}\left\Vert S^{+}_{\cdot i,n}\right\Vert ^{u(2\widetilde{p}-1)}_{1}S^{+}_{ki,n}\right)^{\frac{1}{u}}\left(\sum^{n}_{i=1}\sum^{n}_{k=1}\left\Vert S^{+}_{\cdot k,n}\right\Vert ^{v\widetilde{p}}_{1}S^{+}_{ki,n}\right)^{\frac{1}{v}}\\
\leq & \left(\sum^{n}_{i=1}\left\Vert S^{+}_{\cdot i,n}\right\Vert ^{u(2\widetilde{p}-1)+1}_{1}\right)^{1/u}\left(\left\Vert S^{+}_{n}\right\Vert _{\infty}\sum^{n}_{k=1}\left\Vert S^{+}_{\cdot k,n}\right\Vert ^{v\widetilde{p}}_{1}\right)^{1/v}=\left\Vert S^{+}_{n}\right\Vert ^{1/v}_{\infty}n\mu^{(2\widetilde{p}-1)+1/u}_{u(2\widetilde{p}-1)+1}\cdot\mu^{\widetilde{p}}_{v\widetilde{p}},
\end{align*}
where the last step follows from the definition of $\mu_{p}$ and
$1/u+1/v=1$. Combining the above two displays, we have 
\begin{align*}
 & \sum^{n}_{i=1}\left[\left\Vert S^{+}_{\cdot i,n}\right\Vert _{1}\sum^{n}_{j=1}\sum^{n}_{k=1}\left\Vert S^{+}_{\cdot k,n}\right\Vert _{1}S^{+}_{ki,n}S^{+}_{kj,n}\right]^{\widetilde{p}}\leq\left\Vert S^{+}_{n}\right\Vert ^{\widetilde{p}+1/v}_{\infty}n\mu^{(2\widetilde{p}-1)+1/u}_{u(2\widetilde{p}-1)+1}\cdot\mu^{\widetilde{p}}_{v\widetilde{p}}.
\end{align*}
By $S^{+}_{n}=L\left(I_{n}-L\left|\lambda W_{n}\right|\right)^{-1}=L\sum^{\infty}_{l=0}\left(L\left|\lambda W_{n}\right|\right)^{l}$
and Assumption~\ref{ass: zeta}, we have 
\[
\left\Vert S^{+}_{n}\right\Vert _{\infty}\leq L\sum^{\infty}_{l=0}\left(L\left|\lambda\right|\right)^{l}\left\Vert W_{n}\right\Vert ^{l}_{\infty}\leq L\sum^{\infty}_{l=0}\zeta^{l}=\frac{L}{1-\zeta}.
\]
Thus, Eq.(\ref{eq:SAR clt con2}) follows from the above two displays
and Assumption~\ref{assu:S column condition}. The proof is completed.$\hfill\qedsymbol$

\textbf{Proof of Proposition \ref{prop:verify clt for popular units}.}
We first consider the General Dominant Network ($W_{n,1B}\neq0$).
By Proposition 3(a) of \citet{lee2022qml}, the column sums of $S^{+}_{n}$
are bounded globally by $O(n^{\delta+\eta_{m}})$. As discussed below
Assumption~\ref{assu:S column condition}, for $p=4$, a sufficient
condition for Assumption~\ref{assu:S column condition} is $\left\Vert S^{+}_{n}\right\Vert _{1}=o(n^{1/5})$,
which holds if $\delta+\eta_{m}<1/5$. 

Now, we consider the strict dominant structure ($W_{n,1B}=0$). By
Proposition 3(b) of \citet{lee2022qml}, we have the first $m_{n}$
columns of $S^{+}_{n}$ have sums of order $O(n^{\delta+\eta_{m}})$,
while the column sums of the remaining $n-m_{n}$ nonpopular units
remain uniformly bounded by $O(1)$. Note that 
\[
\mu^{q}_{q}\leq\frac{1}{n}\left[O(n^{\eta_{m}}\cdot n^{q(\delta+\eta_{m})})+O(n\cdot1)\right]=O(n^{q(\delta+\eta_{m})+\eta_{m}-1}+1).
\]
Since we are considering the case that $p=4$ ($\widetilde{p}=2$),
Assumption~\ref{assu:S column condition}(i) requires $4(\delta+\eta_{m})+\eta_{m}-2<0$,
i.e., $4\delta+5\eta_{m}<2$, and Assumption~\ref{assu:S column condition}(ii)
requires that there exists a pair $u,v\in[1,\infty]$ such that $1/u+1/v=1$
and 
\[
\frac{\max\left\{ (3u+1)(\delta+\eta_{m})+\eta_{m}-1,0\right\} }{u}+\frac{\max\left\{ 2v(\delta+\eta_{m})+\eta_{m}-1,0\right\} }{v}<1.
\]
First, we show that it must be the case that $2\delta+3\eta_{m}<1$.
Suppose $2\delta+3\eta_{m}\geq1$, then 
\begin{align*}
 & \frac{\max\left\{ (3u+1)(\delta+\eta_{m})+\eta_{m}-1,0\right\} }{u}+\frac{\max\left\{ 2v(\delta+\eta_{m})+\eta_{m}-1,0\right\} }{v}\\
= & \frac{(3u+1)(\delta+\eta_{m})+\eta_{m}-1}{u}+\frac{2v(\delta+\eta_{m})+\eta_{m}-1}{v}\\
= & 3(\delta+\eta_{m})+\frac{\delta+2\eta_{m}-1}{u}+2(\delta+\eta_{m})+\frac{\eta_{m}-1}{v}\\
= & 5(\delta+\eta_{m})+\frac{\delta+2\eta_{m}-1}{u}+(\eta_{m}-1)\left(1-\frac{1}{u}\right)\\
= & 5\delta+6\eta_{m}-1+\frac{\delta+\eta_{m}}{u},
\end{align*}
where the first step follows from $2v(\delta+\eta_{m})+\eta_{m}-1\geq2\delta+3\eta_{m}-1\geq0$
and $(3u+1)(\delta+\eta_{m})+\eta_{m}-1\geq2\delta+3\eta_{m}-1\geq0$.
Thus, Assumption~\ref{assu:S column condition}(ii) requires
\[
5\delta+6\eta_{m}-1+\frac{\delta+\eta_{m}}{u}<1,
\]
which implies that $5\delta+6\eta_{m}<2$. However,
\[
5\delta+6\eta_{m}=\delta+4\delta+6\eta_{m}=\delta+2(2\delta+3\eta_{m})\geq\delta+2\geq2,
\]
which yields a contradiction. Thus, we claim that it must be the case
that $2\delta+3\eta_{m}<1$. Now, choose $v=(1-\eta_{m})/(2\delta+2\eta_{m})\in[1,\infty]$
so that $2v(\delta+\eta_{m})+\eta_{m}-1=0$, and set $u=(1-\eta_{m})/(1-2\delta-3\eta_{m})\in[1,\infty]$
so that $1/u+1/v=1$. Under this choice, Assumption~\ref{assu:S column condition}(ii)
reduces to 
\[
\left\{ 3\frac{1-\eta_{m}}{1-2\delta-3\eta_{m}}+1\right\} (\delta+\eta_{m})+\eta_{m}-1<\frac{1-\eta_{m}}{1-2\delta-3\eta_{m}}.
\]
Simplifying yields
\[
\delta<\frac{3-5\eta_{m}-\sqrt{(1-\eta_{m})(5-7\eta_{m})}}{2}.
\]
Note that 
\[
\frac{3-5\eta_{m}-\sqrt{(1-\eta_{m})(5-7\eta_{m})}}{2}<\frac{1-3\eta_{m}}{2}
\]
for all $0\leq\eta_{m}<1/3$. Thus, we obtain 
\[
0\leq\delta<\frac{3-5\eta_{m}-\sqrt{(1-\eta_{m})(5-7\eta_{m})}}{2},\text{ for }0\leq\eta_{m}<\frac{1}{3}.
\]
The proof is completed.$\hfill\qedsymbol$

We record a lemma for a general undirected row-normalized network.
This lemma will be repeatedly used below.
\begin{lem}[A bound for $\mu_{\infty}$ under undirected row-normalized networks]
\label{lem:undirected_row_normalized_Splus} For each $n$, let $A_{n}=(A_{ji,n})_{n\times n}$
be a symmetric matrix with $A_{jj,n}=0$ and $A_{ji,n}\in\{0,1\}$.
Define 
\[
d_{j,n}\equiv\sum_{k\neq j}A_{jk,n},\qquad w_{ji,n}\equiv\begin{cases}
A_{ji,n}/d_{j,n}, & d_{j,n}>0,\\
0, & d_{j,n}=0,
\end{cases}
\]
and let $W_{n}=(w_{ji,n})_{n\times n}$. Assume $L|\lambda|<1$ and
define 
\[
S^{+}_{n}\equiv L\big(I_{n}-L|\lambda|W_{n}\big)^{-1}.
\]
Then, for every realization of $A_{n}$, the following statements
hold.
\begin{enumerate}[label=(\roman*)]
\item For every integer $m\ge1$, 
\begin{equation}
\mathcal{D}_{n}W^{m}_{n}=(W^{m}_{n})'\mathcal{D}_{n},\qquad\mathcal{D}_{n}\equiv\mathrm{diag}(d_{1,n},\dots,d_{n,n}).\label{eq:generic_DWm_balance}
\end{equation}
\item For every $i\in[n]$ and every integer $m\ge1$, 
\begin{equation}
\sum^{n}_{j=1}(W^{m}_{n})_{ji}\le d_{i,n}.\label{eq:generic_column_sum_Wm}
\end{equation}
\item For every $i\in[n]$, 
\begin{equation}
\|S^{+}_{\cdot i,n}\|_{1}\le\frac{L}{1-L|\lambda|}(1+d_{i,n}).\label{eq:generic_column_sum_Splus}
\end{equation}
\item If $d_{\infty,n}\equiv\max_{1\le i\le n}d_{i,n},$then 
\begin{equation}
\mu_{\infty}\equiv\|S^{+}_{n}\|_{1}\le\frac{L}{1-L|\lambda|}(1+d_{\infty,n}).\label{eq:generic_mu_infty}
\end{equation}
\end{enumerate}
\end{lem}
\begin{proof}
We prove the four parts one by one.

\textbf{Proof of Part 1.} Since the graph is undirected, $A_{n}=A^{\prime}_{n}$.
For every $j,i\in[n]$, by definition of $W_{n}$, 
\[
(\mathcal{D}_{n}W_{n})_{ji}=d_{j,n}w_{ji,n}=A_{ji,n}.
\]
On the other hand, 
\[
(W_{n}'\mathcal{D}_{n})_{ji}=w_{ij,n}d_{i,n}=A_{ij,n}=A_{ji,n},
\]
where the last equality uses symmetry of $A_{n}$. Therefore, 
\[
\mathcal{D}_{n}W_{n}=W_{n}'\mathcal{D}_{n}.
\]
We now show \eqref{eq:generic_DWm_balance} by induction on $m$.

When $m=1$, the statement is exactly the identity above. Suppose
\eqref{eq:generic_DWm_balance} holds for some $m\ge1$. Then 
\[
\mathcal{D}_{n}W^{m+1}_{n}=(\mathcal{D}_{n}W^{m}_{n})W_{n}=(W^{m}_{n})'\mathcal{D}_{n}W_{n}=(W^{m}_{n})'W_{n}'\mathcal{D}_{n}=(W^{m+1}_{n})'\mathcal{D}_{n}.
\]
Hence \eqref{eq:generic_DWm_balance} holds for all $m\ge1$.

\textbf{Proof of Part 2.} Fix $i\in[n]$ and $m\ge1$. We first consider
the case $d_{i,n}=0$. Since the graph is undirected, node $i$ is
isolated, so 
\[
A_{ji,n}=A_{ij,n}=0,\qquad\forall j\in[n].
\]
Hence the $i$th column of $W_{n}$ is identically zero, and therefore
the $i$th column of $W^{m}_{n}$ is also identically zero. Thus 
\[
\sum^{n}_{j=1}(W^{m}_{n})_{ji}=0=d_{i,n},
\]
and \eqref{eq:generic_column_sum_Wm} holds.

Now consider the case $d_{i,n}>0$. If $d_{j,n}=0$, then the $j$th
row of $W_{n}$ is zero by definition, and hence the $j$th row of
$W^{m}_{n}$ is also zero. Therefore, 
\[
(W^{m}_{n})_{ji}=0\qquad\text{whenever }d_{j,n}=0.
\]
Thus, 
\[
\sum^{n}_{j=1}(W^{m}_{n})_{ji}=\sum_{j:d_{j,n}>0}(W^{m}_{n})_{ji}.
\]
Taking the $(j,i)$th entry of \eqref{eq:generic_DWm_balance} gives
\[
d_{j,n}(W^{m}_{n})_{ji}=d_{i,n}(W^{m}_{n})_{ij}.
\]
Hence, for every $j$ with $d_{j,n}>0$, 
\[
(W^{m}_{n})_{ji}=\frac{d_{i,n}}{d_{j,n}}(W^{m}_{n})_{ij}.
\]
Therefore, 
\[
\sum^{n}_{j=1}(W^{m}_{n})_{ji}=\sum_{j:d_{j,n}>0}\frac{d_{i,n}}{d_{j,n}}(W^{m}_{n})_{ij}\le d_{i,n}\sum_{j:d_{j,n}>0}(W^{m}_{n})_{ij}\le d_{i,n}\sum^{n}_{j=1}(W^{m}_{n})_{ij}.
\]
It remains to bound the last row sum. Each row sum of $W_{n}$ is
either $1$ (if the degree is positive) or $0$ (if the degree is
zero), so $\left\Vert W_{n}\right\Vert _{\infty}\leq1$. Hence $\left\Vert W^{m}_{n}\right\Vert _{\infty}\leq\left\Vert W_{n}\right\Vert ^{m}_{\infty}\leq1$
and therefore $\sum^{n}_{j=1}(W^{m}_{n})_{ij}\le1.$ This proves Eq.\eqref{eq:generic_column_sum_Wm}.

\textbf{Proof of Part 3.} Since $\|W_{n}\|_{\infty}\le1$ and $L|\lambda|<1$,
the Neumann series is valid: 
\[
S^{+}_{n}=L\sum^{\infty}_{m=0}(L|\lambda|)^{m}W^{m}_{n}.
\]
Fix $i\in[n]$. Because all entries are nonnegative, 
\[
\|S^{+}_{\cdot i,n}\|_{1}=\sum^{n}_{j=1}S^{+}_{ji,n}=L\sum^{\infty}_{m=0}(L|\lambda|)^{m}\sum^{n}_{j=1}(W^{m}_{n})_{ji}.
\]
For $m=0$, $W^{0}_{n}=I_{n}$, so $\sum^{n}_{j=1}(W^{0}_{n})_{ji}=\sum^{n}_{j=1}(I_{n})_{ji}=1$.
For $m\ge1$, Part 2 gives $\sum^{n}_{j=1}(W^{m}_{n})_{ji}\le d_{i,n}$.
Therefore, 
\begin{align*}
\|S^{+}_{\cdot i,n}\|_{1} & \le L+L\sum^{\infty}_{m=1}(L|\lambda|)^{m}d_{i,n}=L+Ld_{i,n}\frac{L|\lambda|}{1-L|\lambda|}\le\frac{L}{1-L|\lambda|}(1+d_{i,n}),
\end{align*}
which proves Eq.\eqref{eq:generic_column_sum_Splus}.

\textbf{Proof of Part 4.} Taking the maximum over $i$ in Eq.\eqref{eq:generic_column_sum_Splus},
we obtain 
\[
\mu_{\infty}=\|S^{+}_{n}\|_{1}=\max_{1\le i\le n}\|S^{+}_{\cdot i,n}\|_{1}\le\frac{L}{1-L|\lambda|}(1+d_{\infty,n}).
\]
This is exactly Eq.\eqref{eq:generic_mu_infty}.
\end{proof}

\textbf{Proof of Proposition \ref{prop:ER_S_column_condition}.} We
split the proof into two regimes.

\textbf{Step 1: choose a tolerance level for the dense regime.} Since
$L|\lambda|<1$, we may choose an $\eta\in(0,1)$ such that 
\begin{equation}
L|\lambda|\frac{1+\eta}{1-\eta}<1.\label{eq:ER_eta_choice}
\end{equation}
Next choose a constant $C_{0}>\max\left\{ 1,\frac{1}{c\eta^{2}}\right\} $,
where $c>0$ is the absolute constant from the Exercise 2.3.5 in \citet{vershynin2018HighDimensional}.
For each $n$, exactly one of the following two deterministic alternatives
holds: 
\[
\text{(i) }D_{n}\le C_{0}\log n,\qquad\qquad\text{(ii) }D_{n}>C_{0}\log n.
\]

\textbf{Step 2: sparse regime $D_{n}\le C_{0}\log n$.} Fix $i\in[n]$.
Since the graph is undirected and links are independent, 
\[
d_{i,n}=\sum_{j\neq i}A_{ij,n}\sim\mathrm{Binomial}\left(n-1,\frac{D_{n}}{n-1}\right),\qquad\E d_{i,n}=D_{n}.
\]
Let $K\equiv e^{2}C_{0}.$ Then $K\log n\ge e^{2}D_{n}>D_{n}$. Applying
the Chernoff inequality (Theorem 2.3.1 in \citet{vershynin2018HighDimensional}),
\[
\p(X\ge t)\le e^{-\mu}\left(\frac{e\mu}{t}\right)^{t}\leq\left(\frac{e\mu}{t}\right)^{t},\qquad t>\mu,
\]
with $X=d_{i,n}$, $\mu=D_{n}$, and $t=K\log n$, we obtain 
\[
\p(d_{i,n}\ge K\log n)\le\left(\frac{eD_{n}}{K\log n}\right)^{K\log n}.
\]
Since $D_{n}\le C_{0}\log n$ and $K=e^{2}C_{0}$, 
\[
\frac{eD_{n}}{K\log n}\le\frac{eC_{0}\log n}{e^{2}C_{0}\log n}=\frac{1}{e}.
\]
Hence 
\[
\p(d_{i,n}\ge K\log n)\le\left(\frac{1}{e}\right)^{K\log n}=n^{-K}.
\]
Now let $d_{\infty,n}\equiv\max_{1\le i\le n}d_{i,n}$ and $\mathcal{E}_{1,n}\equiv\left\{ d_{\infty,n}<K\log n\right\} $.
By the union bound, 
\begin{equation}
\p\left(\mathcal{E}_{1,n}\right)=1-\p(d_{\infty,n}\ge K\log n)\geq1-\sum^{n}_{i=1}\p(d_{i,n}\ge K\log n)\geq1-n\cdot n^{-K}\to1,\label{eq:ER_max_degree_log}
\end{equation}
because $K=e^{2}C_{0}>1$. Besides, by Eq.\eqref{eq:generic_mu_infty},
on $\mathcal{E}_{1,n}$ we have $\mu_{\infty}=\|S^{+}_{n}\|_{1}\leq\frac{L}{1-L|\lambda|}(1+K\log n)$. 

\textbf{Step 3: dense regime $D_{n}>C_{0}\log n$.} Define the event
\[
\mathcal{E}_{2,n}\equiv\left\{ (1-\eta)D_{n}\le d_{i,n}\le(1+\eta)D_{n},\ \forall i\in[n]\right\} .
\]
Since $d_{i,n}\sim\mathrm{Binomial}\!\left(n-1,\frac{D_{n}}{n-1}\right)$
and $\E d_{i,n}=D_{n}$, the Chernoff bound (Exercise 2.3.5 in \citet{vershynin2018HighDimensional})
gives 
\[
\p\big(|d_{i,n}-D_{n}|>\eta D_{n}\big)\le2\exp\!\left(-c\eta^{2}D_{n}\right),
\]
where $c>0$ is an absolute constant. Since $\mathcal{E}^{c}_{2,n}\equiv\{|d_{i,n}-D_{n}|>\eta D_{n}\text{ for some }i\in[n]\}$,
by applying the union bound over all $i$, 
\[
\p(\mathcal{E}^{c}_{2,n})\le2n\exp\left(-c\eta^{2}D_{n}\right)\le2n^{1-c\eta^{2}C_{0}}\to0,
\]
because $D_{n}>C_{0}\log n$ and $C_{0}>1/\left(c\eta^{2}\right)$.
Hence $\p(\mathcal{E}_{2,n})\to1$. On the event $\mathcal{E}_{2,n}$,
define $d_{\min,n}\equiv\min_{1\le i\le n}d_{i,n}$ and $d_{\max,n}\equiv\max_{1\le i\le n}d_{i,n}$.
Then 
\[
d_{\min,n}\ge(1-\eta)D_{n}>0,\qquad d_{\max,n}\le(1+\eta)D_{n}.
\]
Fix any column $i$. Since $w_{ji,n}=\frac{A_{ji,n}}{d_{j,n}},$ we
have 
\[
\sum^{n}_{j=1}w_{ji,n}=\sum^{n}_{j=1}\frac{A_{ji,n}}{d_{j,n}}\le\frac{1}{d_{\min,n}}\sum^{n}_{j=1}A_{ji,n}=\frac{d_{i,n}}{d_{\min,n}}\le\frac{d_{\max,n}}{d_{\min,n}}\le\frac{1+\eta}{1-\eta}.
\]
Therefore, 
\begin{equation}
\|W_{n}\|_{1}\le\frac{1+\eta}{1-\eta}\qquad\text{on }\mathcal{E}_{2,n}.\label{eq:ER_W1_bound}
\end{equation}
Now use the Neumann series: $S^{+}_{n}=L\sum^{\infty}_{m=0}(L|\lambda|)^{m}W^{m}_{n}$.
On $\mathcal{E}_{2,n}$, by Eq.\eqref{eq:ER_W1_bound}, 
\begin{align*}
\mu_{\infty}=\|S^{+}_{n}\|_{1} & \le L\sum^{\infty}_{m=0}(L|\lambda|)^{m}\|W_{n}\|^{m}_{1}\le L\sum^{\infty}_{m=0}\left(L|\lambda|\frac{1+\eta}{1-\eta}\right)^{m}=\frac{L}{1-L|\lambda|(1+\eta)/(1-\eta)},
\end{align*}
where the denominator is positive by Eq.\eqref{eq:ER_eta_choice}.

\textbf{Step 4: Combining the results from the two regime.} For each
$n$, exactly one of the two deterministic alternatives holds: 
\[
D_{n}\le C_{0}\log n\qquad\text{or}\qquad D_{n}>C_{0}\log n.
\]
If the first alternative holds, then by Step 2 there exists an event
$\mathcal{E}_{1,n}$ such that $\p(\mathcal{E}_{1,n})\to1$ and $\mu_{\infty}=\|S^{+}_{n}\|_{1}\leq\frac{L}{1-L|\lambda|}(1+K\log n)$
on $\mathcal{E}_{1,n}$. If the second alternative holds, then by
Step 3 there exists an event $\mathcal{E}_{2,n}$ such that $\p(\mathcal{E}_{2,n})\to1$
and $\mu_{\infty}\leq\frac{L}{1-L|\lambda|(1+\eta)/(1-\eta)}$ holds
on $\mathcal{E}_{2,n}$. Hence, defining 
\[
\mathcal{H}_{n}=\begin{cases}
\mathcal{E}_{1,n}, & D_{n}\le C_{0}\log n,\\
\mathcal{E}_{2,n}, & D_{n}>C_{0}\log n,
\end{cases}
\]
we have $\p(\mathcal{H}_{n})\to1$, and $\mu_{\infty}\leq C_{\mu}\log n$
on $\mathcal{H}_{n}$ for some constant $C_{\mu}>0$. Therefore, on
$\mathcal{H}_{n}$
\[
\frac{\mu^{2\tilde{p}}_{2\tilde{p}}}{n^{\tilde{p}-1}}\le\frac{(C_{\mu}\log n)^{2\tilde{p}}}{n^{\tilde{p}-1}}=o(1).
\]
Moreover, taking $(u,v)=(\infty,1)$, we have 
\[
\frac{\mu^{(2\tilde{p}-1)+1/u}_{u(2\tilde{p}-1)+1}\mu^{\tilde{p}}_{v\tilde{p}}}{n^{\tilde{p}-1}}=\frac{\mu^{2\tilde{p}-1}_{\infty}\mu^{\tilde{p}}_{\tilde{p}}}{n^{\tilde{p}-1}}\le\frac{\mu^{3\tilde{p}-1}_{\infty}}{n^{\tilde{p}-1}}\leq\left(\frac{(C_{\mu}\log n)^{3\tilde{p}-1}}{n^{\tilde{p}-1}}\right)=o(1).
\]
Since $\p(\mathcal{H}_{n})\to1$, we claim that Assumption~\ref{assu:S column condition}
holds. The proof is complete.

\textbf{Proof of Proposition \ref{prop:triangle_S_column_condition}.}
The proof proceeds by controlling the maximum degree.

\textbf{Step 1: an upper bound for the degree.} Fix $i\in[n]$. Let
$G_{i,n}\equiv\sum_{j\neq i}A^{(\mathrm{ER})}_{ij,n}$ be the degree
contribution from the background ER edges. Then $G_{i,n}\sim\mathrm{Binomial}\left(n-1,\frac{D_{n}}{n-1}\right)$
and $\E G_{i,n}=D_{n}$. Next define
\[
N_{i,n}\equiv\sum_{\{j,k\}\subset[n]\setminus\{i\}}B_{\{i,j,k\},n},
\]
namely, the number of selected triangles containing node $i$. Since
there are $\binom{n-1}{2}$ unordered pairs $\{j,k\}\subset[n]\setminus\{i\}$,
we have 
\[
N_{i,n}\sim\mathrm{Binomial}\left(\binom{n-1}{2},\frac{T_{n}}{\binom{n}{3}}\right),\qquad\E N_{i,n}=\binom{n-1}{2}\frac{T_{n}}{\binom{n}{3}}=\frac{3T_{n}}{n}.
\]
We now compare the total degree $d_{i,n}$ with $G_{i,n}$ and $N_{i,n}$.
Each background ER edge can contribute at most $1$ to the degree
of node $i$. Each selected triangle containing node $i$ can contribute
at most $2$ new neighbors to node $i$. Therefore, 
\begin{equation}
d_{i,n}\equiv\sum_{j\neq i}A_{ji,n}\le G_{i,n}+2N_{i,n}.\label{eq:triangle_degree_domination}
\end{equation}

\textbf{Step 2: a high-probability logarithmic bound for the maximum
degree.} Choose 
\[
K\equiv\max\{2e^{2}C_{0},\,8\}.
\]
We claim that 
\begin{equation}
d_{\infty,n}\equiv\max_{1\le i\le n}d_{i,n}<K\log n\qquad\text{with probability approaching one.}\label{eq:triangle_max_degree_log}
\end{equation}
Fix $i\in[n]$. If $d_{i,n}\ge K\log n$, then by \eqref{eq:triangle_degree_domination},
either 
\[
G_{i,n}\ge\frac{K}{2}\log n\qquad\text{or}\qquad N_{i,n}\ge\frac{K}{4}\log n.
\]
Hence 
\begin{equation}
\p(d_{i,n}\ge K\log n)\le\p\!\left(G_{i,n}\ge\frac{K}{2}\log n\right)+\p\!\left(N_{i,n}\ge\frac{K}{4}\log n\right).\label{eq:triangle_split_tail}
\end{equation}

We first bound the ER part. Since Eq.\eqref{eq:triangle_sparse_condition}
implies $D_{n}\le C_{0}\log n$, and 
\[
\frac{K}{2}\log n\ge e^{2}C_{0}\log n\ge e^{2}D_{n}>D_{n},
\]
we have $\frac{eD_{n}}{(K/2)\log n}\le\frac{1}{e}$. Immediately,
the Chernoff inequality (Theorem 2.3.1 in \citet{vershynin2018HighDimensional})
yields 
\begin{equation}
\p\!\left(G_{i,n}\ge\frac{K}{2}\log n\right)\le\left(\frac{eD_{n}}{(K/2)\log n}\right)^{(K/2)\log n}\leq\left(\frac{1}{e}\right)^{(K/2)\log n}=n^{-K/2}.\label{eq:triangle_ER_tail}
\end{equation}

We next bound the triangle part. Since $\frac{6T_{n}}{n}\le C_{0}\log n$
from Eq.\eqref{eq:triangle_sparse_condition}, we have $\E N_{i,n}=\frac{3T_{n}}{n}\le\frac{C_{0}}{2}\log n$.
Also, 
\[
\frac{K}{4}\log n\ge\frac{e^{2}C_{0}}{2}\log n\ge e^{2}\E N_{i,n}>\E N_{i,n}.
\]
By this inequality and Chernoff's inequality to $N_{i,n}$, we have
\begin{equation}
\p\!\left(N_{i,n}\ge\frac{K}{4}\log n\right)\le\left(\frac{e\E N_{i,n}}{(K/4)\log n}\right)^{(K/4)\log n}\le\left(\frac{1}{e}\right)^{(K/4)\log n}=n^{-K/4}.\label{eq:triangle_triangle_tail}
\end{equation}

Combining Eqs.\eqref{eq:triangle_split_tail}, \eqref{eq:triangle_ER_tail},
and \eqref{eq:triangle_triangle_tail}, we obtain 
\[
\p(d_{i,n}\ge K\log n)\le n^{-K/2}+n^{-K/4}\le2n^{-K/4}.
\]
Now apply the union bound over all $i\in[n]$: 
\[
\p(d_{\infty,n}\ge K\log n)\le\sum^{n}_{i=1}\p(d_{i,n}\ge K\log n)\le2n^{\,1-K/4}\to0,
\]
because $K\ge8$. This proves Eq.\eqref{eq:triangle_max_degree_log}.

\textbf{Step 3: verify Assumption \ref{assu:S column condition}.}
By Lemma \ref{lem:undirected_row_normalized_Splus} and Eq.\eqref{eq:triangle_max_degree_log},
\[
\mu_{\infty}=\|S^{+}_{n}\|_{1}\leq\frac{L}{1-L|\lambda|}(1+K\log n)\qquad\text{with probability approaching one.}
\]
Since $\mu_{q}\le\mu_{\infty}$ for every $q\ge1$, we obtain 
\[
\frac{\mu^{2\tilde{p}}_{2\tilde{p}}}{n^{\tilde{p}-1}}\le\frac{\mu^{2\tilde{p}}_{\infty}}{n^{\tilde{p}-1}}=O_{\p}\left(\frac{(\log n)^{2\tilde{p}}}{n^{\tilde{p}-1}}\right)=o_{\p}(1)
\]
Moreover, taking $(u,v)=(\infty,1)$, we have 
\[
\frac{\mu^{(2\tilde{p}-1)+1/u}_{u(2\tilde{p}-1)+1}\mu^{\tilde{p}}_{v\tilde{p}}}{n^{\tilde{p}-1}}=\frac{\mu^{2\tilde{p}-1}_{\infty}\mu^{\tilde{p}}_{\tilde{p}}}{n^{\tilde{p}-1}}\le\frac{\mu^{3\tilde{p}-1}_{\infty}}{n^{\tilde{p}-1}}=O_{\p}\left(\frac{(\log n)^{3\tilde{p}-1}}{n^{\tilde{p}-1}}\right)=o_{\p}(1)
\]
Therefore Assumption \ref{assu:S column condition} holds. The proof
is completed.$\hfill\qedsymbol$

\textbf{Proof of Proposition \ref{prop:SBM_S_column_condition}.}
We again split the proof into sparse and dense regimes, but before
doing so, we first show that the random block sizes are uniformly
well concentrated.

\textbf{Step 1: concentration of the block sizes.} For each block
$m\in\{1,\dots,M_{n}\}$, define 
\[
N_{m,n}\equiv\sum^{n}_{i=1}1\{g_{i,n}=m\}.
\]
Then 
\[
N_{m,n}\sim\mathrm{Binomial}\left(n,\frac{1}{M_{n}}\right),\qquad\E N_{m,n}=\frac{n}{M_{n}}.
\]
Fix any $\varepsilon\in(0,1/4)$ and define the event 
\[
\mathcal{B}_{n}(\varepsilon)\equiv\left\{ \left|N_{m,n}-\frac{n}{M_{n}}\right|\le\varepsilon\frac{n}{M_{n}},\ \forall m=1,\dots,M_{n}\right\} .
\]
By the Chernoff bound (Exercise 2.3.5 in \citet{vershynin2018HighDimensional}),
\[
\p\!\left(\left|N_{m,n}-\frac{n}{M_{n}}\right|>\varepsilon\frac{n}{M_{n}}\right)\le2\exp\!\left(-c\frac{n}{M_{n}}\varepsilon^{2}\right),
\]
where $c>0$ is an absolute constant. Applying the union bound over
$m=1,\dots,M_{n}$, we obtain 
\[
\p\!\big(\mathcal{B}_{n}(\varepsilon)^{c}\big)\le2M_{n}\exp\!\left(-c\frac{n}{M_{n}}\varepsilon^{2}\right).
\]
Condition $M_{n}\log M_{n}=o(n)$ implies $\log M_{n}=o\!\left(\frac{n}{M_{n}}\right)$,
and therefore $\p\!\big(\mathcal{B}_{n}(\varepsilon)^{c}\big)\to0$.
Hence 
\begin{equation}
\p\!\big(\mathcal{B}_{n}(\varepsilon)\big)\to1.\label{eq:SBM_B_event}
\end{equation}

We note that $M_{n}\log M_{n}=o(n)$ also implies $\frac{n}{M_{n}}\to\infty$.
Indeed, since $M_{n}\ge2$, we have $\log M_{n}\ge\log2>0$, so $M_{n}=o(n)$.
Hence $n/M_{n}\to\infty$.

\textbf{Step 2: conditional mean degree on the block-size event.}
Fix $i\in[n]$, and let $m=g_{i,n}$ be the block containing node
$i$. Conditional on the labels $g_{1,n},\dots,g_{n,n}$, the degree
$d_{i,n}$ is the sum of $(N_{m,n}-1)$ independent Bernoulli($\frac{D_{\mathrm{wb},n}}{n/M_{n}-1}$)
random variables and $(n-N_{m,n})$ independent Bernoulli($\frac{D_{\mathrm{bb},n}}{n-n/M_{n}}$)
random variables. Therefore, conditional on the labels, 
\[
\E\big(d_{i,n}\mid g_{1,n},\dots,g_{n,n}\big)=\left(N_{m,n}-1\right)\frac{D_{\mathrm{wb},n}}{n/M_{n}-1}+\left(n-N_{m,n}\right)\frac{D_{\mathrm{bb},n}}{n-n/M_{n}}.
\]
For brevity, write $\mu_{i,n}\equiv\E\big(d_{i,n}\mid g_{1,n},\dots,g_{n,n}\big).$
We now bound $\mu_{i,n}$ on the event $\mathcal{B}_{n}(\varepsilon)$.
Since $n/M_{n}\to\infty$, we have $n/M_{n}\ge2$ for all sufficiently
large $n$.

First, on $\mathcal{B}_{n}(\varepsilon)$, $N_{m,n}\ge(1-\varepsilon)\frac{n}{M_{n}}$.
Hence, for all sufficiently large $n$, 
\begin{align*}
N_{m,n}-1 & \ge\left(\frac{n}{M_{n}}-1\right)-\varepsilon\frac{n}{M_{n}}\geq\left(\frac{n}{M_{n}}-1\right)-2\varepsilon\left(\frac{n}{M_{n}}-1\right)=(1-2\varepsilon)\left(\frac{n}{M_{n}}-1\right).
\end{align*}
where in the second inequality we used $\frac{n}{M_{n}}\ge2$. Similarly,
since $N_{m,n}\le(1+\varepsilon)\frac{n}{M_{n}}$, we obtain 
\begin{align*}
N_{m,n}-1 & \le(1+\varepsilon)\frac{n}{M_{n}}-1\le(1+2\varepsilon)\left(\frac{n}{M_{n}}-1\right),
\end{align*}
again for all sufficiently large $n$. Therefore, on $\mathcal{B}_{n}(\varepsilon)$,
\begin{equation}
(1-2\varepsilon)D_{\mathrm{wb},n}\le\left(N_{m,n}-1\right)\frac{D_{\mathrm{wb},n}}{n/M_{n}-1}\le(1+2\varepsilon)D_{\mathrm{wb},n}.\label{eq:SBM_within_mean_bound}
\end{equation}

Next, on $\mathcal{B}_{n}(\varepsilon)$, 
\[
n-N_{m,n}\ge n-(1+\varepsilon)\frac{n}{M_{n}}=\left(n-\frac{n}{M_{n}}\right)-\varepsilon\frac{n}{M_{n}}.
\]
Since $M_{n}\ge2$, we have $\frac{n}{M_{n}}\le n-\frac{n}{M_{n}}$.
Hence $n-N_{m,n}\ge(1-\varepsilon)\left(n-\frac{n}{M_{n}}\right)$.
Similarly, 
\[
n-N_{m,n}\le n-(1-\varepsilon)\frac{n}{M_{n}}=\left(n-\frac{n}{M_{n}}\right)+\varepsilon\frac{n}{M_{n}}\le(1+\varepsilon)\left(n-\frac{n}{M_{n}}\right).
\]
Therefore, on $\mathcal{B}_{n}(\varepsilon)$, 
\begin{equation}
(1-\varepsilon)D_{\mathrm{bb},n}\le\left(n-N_{m,n}\right)\frac{D_{\mathrm{bb},n}}{n-n/M_{n}}\le(1+\varepsilon)D_{\mathrm{bb},n}.\label{eq:SBM_between_mean_bound}
\end{equation}

Combining Eqs.\eqref{eq:SBM_within_mean_bound} and \eqref{eq:SBM_between_mean_bound},
we conclude that on $\mathcal{B}_{n}(\varepsilon)$, 
\begin{equation}
(1-2\varepsilon)\big(D_{\mathrm{wb},n}+D_{\mathrm{bb},n}\big)\le\mu_{i,n}\le(1+2\varepsilon)\big(D_{\mathrm{wb},n}+D_{\mathrm{bb},n}\big)\label{eq:SBM_mean_degree_bound}
\end{equation}
for all $i=1,\dots,n$ and all sufficiently large $n$.

\textbf{Step 3: choose tolerances for the dense regime.} Since $L|\lambda|<1$,
we may choose $\varepsilon\in(0,1/4)$ and $\tau\in(0,1)$ small enough
such that 
\begin{equation}
L|\lambda|\frac{(1+\tau)(1+2\varepsilon)}{(1-\tau)(1-2\varepsilon)}<1.\label{eq:SBM_ratio_choice}
\end{equation}
Fix such $\varepsilon$ and $\tau$. Next choose a constant 
\[
C_{0}>\max\left\{ 1,\frac{1}{c(1-2\varepsilon)\tau^{2}}\right\} ,
\]
where $c>0$ is the absolute constant from the Exercise 2.3.5 in \citet{vershynin2018HighDimensional}.
For each $n$, exactly one of the following two deterministic alternatives
holds: 
\[
\text{(i) }D_{\mathrm{wb},n}+D_{\mathrm{bb},n}\le C_{0}\log n,\text{ and }\text{(ii) }D_{\mathrm{wb},n}+D_{\mathrm{bb},n}>C_{0}\log n.
\]

\textbf{Step 4: sparse regime $D_{\mathrm{wb},n}+D_{\mathrm{bb},n}\le C_{0}\log n$.}
We first work on the event $\mathcal{B}_{n}(\varepsilon)$. Conditional
on the labels, $d_{i,n}$ is a sum of independent Bernoulli variables
with mean $\mu_{i,n}$. By Eq.\eqref{eq:SBM_mean_degree_bound}, 
\[
\mu_{i,n}\le(1+2\varepsilon)C_{0}\log n\qquad\text{on }\mathcal{B}_{n}(\varepsilon).
\]
Let 
\[
K\equiv e^{2}(1+2\varepsilon)C_{0}.
\]
Then $K\log n>\mu_{i,n}$ for all sufficiently large $n$ on $\mathcal{B}_{n}(\varepsilon)$.
Using the Chernoff inequality (Theorem 2.3.1 in \citet{vershynin2018HighDimensional}),
\[
\p(d_{i,n}\ge K\log n\mid g_{1,n},\dots,g_{n,n})\le\left(\frac{e\mu_{i,n}}{K\log n}\right)^{K\log n}\qquad\text{on }\mathcal{B}_{n}(\varepsilon).
\]
Since $\mu_{i,n}\le(1+2\varepsilon)C_{0}\log n$ and $K=e^{2}(1+2\varepsilon)C_{0}$,
\[
\frac{e\mu_{i,n}}{K\log n}\le\frac{e(1+2\varepsilon)C_{0}\log n}{e^{2}(1+2\varepsilon)C_{0}\log n}=\frac{1}{e}.
\]
Therefore, on $\mathcal{B}_{n}(\varepsilon)$, 
\[
\p(d_{i,n}\ge K\log n\mid g_{1,n},\dots,g_{n,n})\le n^{-K}.
\]
Applying the union bound over $i=1,\dots,n$, 
\[
\p\!\left(\max_{1\le i\le n}d_{i,n}\ge K\log n\,|\,g_{1,n},\dots,g_{n,n}\right)\le n^{1-K}\qquad\text{on }\mathcal{B}_{n}(\varepsilon).
\]
Taking expectation and using Eq.\eqref{eq:SBM_B_event}, we conclude
that 
\[
d_{\infty,n}\equiv\max_{1\le i\le n}d_{i,n}<K\log n\qquad\text{with probability approaching one.}
\]
Let $\mathcal{E}_{1,n}\equiv\left\{ \mu_{\infty}<\frac{L}{1-L|\lambda|}(1+K\log n)\right\} $.
It follows from Eq.\eqref{eq:generic_mu_infty} that $\p(\mathcal{E}_{1,n})\to1$
as $n\to\infty$.

\textbf{Step 5: dense regime $D_{\mathrm{wb},n}+D_{\mathrm{bb},n}>C_{0}\log n$.}
Again work on the event $\mathcal{B}_{n}(\varepsilon)$. Define the
event 
\[
\mathcal{E}_{2,n}\equiv\left\{ |d_{i,n}-\mu_{i,n}|\le\tau\mu_{i,n},\ \forall i\in[n]\right\} .
\]
Conditional on the labels $g_{1,n},\ldots,g_{n,n}$, $d_{i,n}$ is
a sum of independent Bernoulli variables with mean $\mu_{i,n}$. Hence
the Chernoff bound (Exercise 2.3.5 in \citet{vershynin2018HighDimensional})
gives 
\[
\p\big(|d_{i,n}-\mu_{i,n}|>\tau\mu_{i,n}\mid g_{1,n},\dots,g_{n,n}\big)\le2\exp\!\left(-c\tau^{2}\mu_{i,n}\right),
\]
where $c$ is an absolute constant. On $\mathcal{B}_{n}(\varepsilon)$,
Eq.\eqref{eq:SBM_mean_degree_bound} implies 
\[
\mu_{i,n}\ge(1-2\varepsilon)\big(D_{\mathrm{wb},n}+D_{\mathrm{bb},n}\big)>(1-2\varepsilon)C_{0}\log n.
\]
Therefore, 
\[
\p\big(|d_{i,n}-\mu_{i,n}|>\tau\mu_{i,n}\mid g_{1,n},\dots,g_{n,n}\big)\le2n^{-c(1-2\varepsilon)\tau^{2}C_{0}}\qquad\text{on }\mathcal{B}_{n}(\varepsilon).
\]
Applying the union bound over $i=1,\dots,n$, we obtain 
\[
\p(\mathcal{E}^{c}_{2,n}\mid g_{1,n},\dots,g_{n,n})\le2n^{\,1-c(1-2\varepsilon)\tau^{2}C_{0}}\qquad\text{on }\mathcal{B}_{n}(\varepsilon).
\]
Since $C_{0}>1/(c(1-2\varepsilon)\tau^{2})$, the right-hand side
of the last display tends to zero. Combining this with Eq.\eqref{eq:SBM_B_event},
we conclude that 
\[
\p(\mathcal{B}_{n}(\varepsilon)\cap\mathcal{E}_{2,n})\to1.
\]

Now fix any realization in the event $\mathcal{B}_{n}(\varepsilon)\cap\mathcal{E}_{2,n}$.
By Eq.\eqref{eq:SBM_mean_degree_bound} and the definition of $\mathcal{E}_{2,n}$,
\[
d_{i,n}\ge(1-\tau)\mu_{i,n}\ge(1-\tau)(1-2\varepsilon)\big(D_{\mathrm{wb},n}+D_{\mathrm{bb},n}\big),
\]
and 
\[
d_{i,n}\le(1+\tau)\mu_{i,n}\le(1+\tau)(1+2\varepsilon)\big(D_{\mathrm{wb},n}+D_{\mathrm{bb},n}\big).
\]
Therefore, 
\[
d_{\min,n}\ge(1-\tau)(1-2\varepsilon)\big(D_{\mathrm{wb},n}+D_{\mathrm{bb},n}\big),
\]
\[
d_{\max,n}\le(1+\tau)(1+2\varepsilon)\big(D_{\mathrm{wb},n}+D_{\mathrm{bb},n}\big),
\]
and hence 
\begin{equation}
\frac{d_{\max,n}}{d_{\min,n}}\le\frac{(1+\tau)(1+2\varepsilon)}{(1-\tau)(1-2\varepsilon)}.\label{eq:SBM_degree_ratio}
\end{equation}
We now bound the column-sum norm of $W_{n}$. Fix any column $i$.
Then 
\[
\sum^{n}_{j=1}w_{ji,n}=\sum^{n}_{j=1}\frac{A_{ji,n}}{d_{j,n}}\le\frac{1}{d_{\min,n}}\sum^{n}_{j=1}A_{ji,n}=\frac{d_{i,n}}{d_{\min,n}}\le\frac{d_{\max,n}}{d_{\min,n}}.
\]
Using Eq.\eqref{eq:SBM_degree_ratio}, we obtain 
\[
\|W_{n}\|_{1}\le\frac{(1+\tau)(1+2\varepsilon)}{(1-\tau)(1-2\varepsilon)}\qquad\text{on }\mathcal{B}_{n}(\varepsilon)\cap\mathcal{E}_{2,n}.
\]
Since $L|\lambda|\frac{(1+\tau)(1+2\varepsilon)}{(1-\tau)(1-2\varepsilon)}<1$,
by Eq.\eqref{eq:SBM_ratio_choice}, the Neumann series implies 
\begin{align*}
\|S^{+}_{n}\|_{1} & \le L\sum^{\infty}_{m=0}(L|\lambda|)^{m}\|W_{n}\|^{m}_{1}\le L\sum^{\infty}_{m=0}\left(L|\lambda|\frac{(1+\tau)(1+2\varepsilon)}{(1-\tau)(1-2\varepsilon)}\right)^{m}=\frac{L}{1-L|\lambda|\frac{(1+\tau)(1+2\varepsilon)}{(1-\tau)(1-2\varepsilon)}}.
\end{align*}
Hence $\mu_{\infty}=\|S^{+}_{n}\|_{1}=O(1)$ on $\mathcal{B}_{n}(\varepsilon)\cap\mathcal{E}_{2,n}$.

\textbf{Step 6: Combining the results from the two regime.} For each
$n$, exactly one of the two deterministic alternatives holds: 
\[
\text{(i) }D_{\mathrm{wb},n}+D_{\mathrm{bb},n}\le C_{0}\log n,\text{ and }\text{(ii) }D_{\mathrm{wb},n}+D_{\mathrm{bb},n}>C_{0}\log n.
\]
If the first alternative holds, then by Step 4 there exists an event
$\mathcal{E}_{1,n}$ such that $\p(\mathcal{E}_{1,n})\to1$ and $\mu_{\infty}\leq\frac{L}{1-L|\lambda|}(1+K\log n)$
on $\mathcal{E}_{1,n}$. If the second alternative holds, then by
Step 5 there exists an event $\mathcal{B}_{n}(\varepsilon)\cap\mathcal{E}_{2,n}$
such that $\p(\mathcal{B}_{n}(\varepsilon)\cap\mathcal{E}_{2,n})\to1$
and $\mu_{\infty}=O(1)$ holds on $\mathcal{B}_{n}(\varepsilon)\cap\mathcal{E}_{2,n}$.
Hence, defining 
\[
\mathcal{H}_{n}=\begin{cases}
\mathcal{E}_{1,n}, & D_{\mathrm{wb},n}+D_{\mathrm{bb},n}\le C_{0}\log n,\\
\mathcal{B}_{n}(\varepsilon)\cap\mathcal{E}_{2,n}, & D_{\mathrm{wb},n}+D_{\mathrm{bb},n}>C_{0}\log n,
\end{cases}
\]
we have $\p(\mathcal{H}_{n})\to1$, and $\mu_{\infty}\leq C_{\mu}\log n$
on $\mathcal{H}_{n}$ for some constant $C_{\mu}>0$. Therefore, on
$\mathcal{H}_{n}$
\[
\frac{\mu^{2\tilde{p}}_{2\tilde{p}}}{n^{\tilde{p}-1}}\le\frac{(C_{\mu}\log n)^{2\tilde{p}}}{n^{\tilde{p}-1}}=o(1).
\]
Moreover, taking $(u,v)=(\infty,1)$, we have 
\[
\frac{\mu^{(2\tilde{p}-1)+1/u}_{u(2\tilde{p}-1)+1}\mu^{\tilde{p}}_{v\tilde{p}}}{n^{\tilde{p}-1}}=\frac{\mu^{2\tilde{p}-1}_{\infty}\mu^{\tilde{p}}_{\tilde{p}}}{n^{\tilde{p}-1}}\le\frac{\mu^{3\tilde{p}-1}_{\infty}}{n^{\tilde{p}-1}}\leq\left(\frac{(C_{\mu}\log n)^{3\tilde{p}-1}}{n^{\tilde{p}-1}}\right)=o(1).
\]
Since $\p(\mathcal{H}_{n})\to1$, we claim that Assumption~\ref{assu:S column condition}
holds. The proof is complete. $\hfill\qedsymbol$

\textbf{Proof of Proposition \ref{prop:tobit consistency}. }The proof
proceeds in three steps.

\textbf{Step 1.} Let $p>6$. We calculate the FDM of $\{Y_{i,n}\}$,
$\{z_{i,n}(\theta)\}$, $\left\{ a_{i,n}=\log\Phi(z_{i,n}(\theta))\right\} $
and $\left\{ b_{i,n}=z_{i,n}(\theta)^{2}\right\} $. For $Y_{i,n}$,
Proposition~\ref{prop: SAR FDM} gives $\delta^{(Y)}_{p,n}(j,i,\mathcal{C}_{n})\le2\|\epsilon\|_{L^{p},\mathcal{C}_{n}}S^{+}_{ji,n}$.
Then we calculate the $L^{p}$-FDM of $\{z_{i,n}(\theta)\}$. Let
$z_{j,n,i}(\theta)$, $Y_{j,n,i}$, and $Y_{n,i}$ be the coupled
versions of $z_{j,n}(\theta)$, $Y_{j,n}$, and $Y_{n}$ with $\epsilon_{i,n}$
replaced by its i.i.d. copy $\epsilon^{*}_{i,n}$. We have
\begin{align*}
 & \delta^{(z)}_{p,n}(j,i,\mathcal{C}_{n})=\left\Vert z_{j,n}(\theta)-z_{j,n,i}(\theta)\right\Vert _{L^{p},\c_{n}}\\
= & \left\Vert \frac{Y_{j,n}-\lambda\sum^{n}_{k=1}w_{jk,n}Y_{k,n}-X^{\prime}_{j,n}\beta}{\sigma}-\frac{Y_{j,n,i}-\lambda\sum^{n}_{k=1}w_{jk,n}Y_{k,n,i}-X^{\prime}_{j,n}\beta}{\sigma}\right\Vert _{L^{p},\c_{n}}\\
= & \frac{1}{\sigma}\left\Vert Y_{j,n}-\lambda\sum^{n}_{k=1}w_{jk,n}Y_{k,n}-Y_{j,n,i}+\lambda\sum^{n}_{k=1}w_{jk,n}Y_{k,n,i}\right\Vert _{L^{p},\c_{n}}\\
\leq & \frac{1}{\sigma}\left\Vert Y_{j,n}-Y_{j,n,i}\right\Vert _{L^{p},\c_{n}}+\frac{\left|\lambda\right|}{\sigma}\sum^{n}_{k=1}\left|w_{jk,n}\right|\left\Vert Y_{k,n}-Y_{k,n,i}\right\Vert _{L^{p},\c_{n}}\\
= & \frac{1}{\sigma}\delta^{(Y)}_{p,n}(j,i,\mathcal{C}_{n})+\frac{\left|\lambda\right|}{\sigma}\sum^{n}_{k=1}\left|w_{jk,n}\right|\delta^{(Y)}_{p,n}(k,i,\mathcal{C}_{n})\\
\le & \frac{2}{\sigma}\left\Vert \epsilon\right\Vert _{L^{p},\c_{n}}\left(S^{+}_{ji,n}+\left|\lambda\right|\sum^{n}_{k=1}\left|w_{jk,n}\right|S^{+}_{ki,n}\right)\\
= & \frac{2}{\sigma}\left\Vert \epsilon\right\Vert _{L^{p},\c_{n}}\left(S^{+}_{ji,n}+\left(\left|\lambda\right|\left|W_{n}\right|S^{+}_{n}\right)_{ji}\right)\\
= & \frac{2}{\sigma}\left\Vert \epsilon\right\Vert _{L^{p},\c_{n}}\left(S^{+}_{ji,n}+\frac{\left|\lambda\right|}{\left|\lambda_{0}\right|}\left(S^{+}_{n}-I_{n}\right)_{ji}\right)\le\frac{2}{\sigma}\left\Vert \epsilon\right\Vert _{L^{p},\c_{n}}\left(1+\frac{\left|\lambda\right|}{\left|\lambda_{0}\right|}\right)S^{+}_{ji,n},
\end{align*}
where the last equality follows from $(I_{n}-\left|\lambda_{0}\right|\left|W_{n}\right|)S^{+}_{n}=I_{n}$.
By \citet{xu2015maximum}, $\left|\log\Phi(x_{1})-\log\Phi(x_{2})\right|\le C_{1}(|x_{1}|+|x_{2}|+1)(|x_{1}-x_{2}|)$.
We also have $|x^{2}_{1}-x^{2}_{2}|\le(|x_{1}|+|x_{2}|)(|x_{1}-x_{2}|)$.
By Proposition~\ref{prop:unbdd lip 1}, there exist constants $C_{1}$
and $C_{2}$ such that $\delta^{(a)}_{p/2,n}(j,i,\mathcal{C}_{n})\le C_{1}S^{+}_{ji,n}$
and $\delta^{(b)}_{p/2,n}(j,i,\mathcal{C}_{n})\le C_{2}S^{+}_{ji,n}$,
where $\delta^{(a)}_{p/2,n}(j,i,\mathcal{C}_{n})$ and $\delta^{(b)}_{p/2,n}(j,i,\mathcal{C}_{n})$
denote the FDM of $\left\{ a_{i,n}=\log\Phi(z_{i,n}(\theta))\right\} $
and $\left\{ b_{i,n}=z_{i,n}(\theta)^{2}\right\} $ respectively.

\textbf{Step 2.} Next, we construct a smoothed log likelihood function.
We use a continuous function $\kappa_{h}(y)$ to approximate $1(Y_{i,n}>0)$.
Take
\[
\kappa_{h}(y)=\begin{cases}
0, & y\le0,\\
y/h, & 0<y<h,\\
1, & y\ge h.
\end{cases}
\]
Then $0\le\kappa_{h}\le1$ and the Lipschitz constant of $\kappa_{h}$
is $\mathrm{Lip}(\kappa_{h})=h^{-1}$. We have that $\delta^{(\kappa_{h}(Y))}_{p,n}(i,k,\mathcal{C}_{n})\le h^{-1}\delta^{(Y)}_{p,n}(i,k,\mathcal{C}_{n})$.
Define $G_{n,h}(Y_{n})=\text{diag}(\kappa_{h}(Y_{1,n}),\dots,\kappa_{h}(Y_{n,n}))$.
Replace $1(Y_{i,n}>0)$ in the original log-likelihood by $\kappa_{h}(Y_{i,n})$,
obtaining a smoothed version of the likelihood function $\ell_{n,h}(\theta)=R_{n,h}(\theta)+J_{n,h}(\theta),$
where
\[
R_{n,h}(\theta)\equiv\sum^{n}_{i=1}(1-\kappa_{h}(Y_{i,n}))\log\Phi(z_{i,n}(\theta))-\frac{1}{2}\log(2\pi\sigma^{2})\sum^{n}_{i=1}\kappa_{h}(Y_{i,n})-\frac{1}{2}\sum^{n}_{i=1}\kappa_{h}(Y_{i,n})z_{i,n}(\theta)^{2},
\]
and
\[
J_{n,h}(\theta)\equiv\log\det(I_{n}-\lambda G_{n,h}(Y_{n})W_{n}G_{n,h}(Y_{n})).
\]
Let $\{h_{n}\}$ be a positive sequence such that $h_{n}=\sqrt{\mu_{2}}/n^{1/4}=o_{\p}(1)$,
then $\frac{\mu_{2}}{\sqrt{n}h_{n}}=\sqrt{\mu_{2}}/n^{1/4}=o_{\p}(1)$.
We will show that $\frac{1}{n}R_{n,h_{n}}(\theta)=\frac{1}{n}\E_{\c_{n}}R_{n,h_{n}}(\theta)+o_{\p}(1)$.
Let 
\[
R_{i,n,h}(\theta)=(1-\kappa_{h}(Y_{i,n}))\log\Phi(z_{i,n}(\theta))-\frac{1}{2}\log(2\pi\sigma^{2})\kappa_{h}(Y_{i,n})-\frac{1}{2}\kappa_{h}(Y_{i,n})z_{i,n}(\theta)^{2},
\]
then $R_{n,h}(\theta)=\sum^{n}_{i=1}R_{i,n,h}(\theta)$. By Propositions~\ref{prop:+ FDM}
and \ref{prop:FDM prod 1}, there exists a constant $C_{3}$ such
that the $L^{p/3}$-FDM of $\left\{ R_{i,n,h}(\theta)\right\} $ satisfies
$\delta^{(R_{h})}_{p/3,n}(j,i,\mathcal{C}_{n})\le C_{3}h^{-1}S^{+}_{ji,n}$.
Since $p>6$, we have $p/3>2$, and therefore $\delta^{(R_{h})}_{2,n}(j,i,\c_{n})\le\delta^{(R_{h})}_{p/3,n}(j,i,\c_{n})\le C_{3}h^{-1}S^{+}_{ji,n}$.
Hence, by Theorem~\ref{thm: Rosenthal},
\begin{align}
 & \Biggl\|\frac{1}{n}\sum^{n}_{i=1}[R_{i,n,h}(\theta)-\E_{\c_{n}}R_{i,n,h}(\theta)]\Biggr\|_{L^{2},\mathcal{C}_{n}}\le\left\{ \frac{1}{n^{2}}\sum^{n}_{i=1}\left[\sum^{n}_{j=1}\delta^{(R_{h})}_{2,n}(j,i,\mathcal{C}_{n})\right]^{2}\right\} ^{1/2}\nonumber \\
\le & \Biggl\{\frac{1}{n^{2}}\sum^{n}_{i=1}\left(C_{3}h^{-1}_{n}\left\Vert S^{+}_{\cdot i,n}\right\Vert _{1}\right)^{2}\Biggr\}^{1/2}=O_{\p}\left(\frac{\mu_{2}}{\sqrt{n}h_{n}}\right)=o_{\p}(1).\label{eq:R-ER}
\end{align}

Next, we deal with $J_{n,h_{n}}(\theta)$. Let $A_{n,h}=\lambda G_{n,h}(Y_{n})W_{n}G_{n,h}(Y_{n}).$
Since $\left\Vert A_{n,h}\right\Vert _{\infty}<1$, we have that $J_{n,h}(\theta)=-\sum^{\infty}_{\ell=1}\frac{1}{\ell}\tr(A^{\ell}_{n,h})$.
We further define $A_{n,h,k}\equiv\lambda G_{n,h}(Y_{n,k})W_{n}G_{n,h}(Y_{n,k})$,
$g_{i,n}\equiv\kappa_{h}(Y_{i,n})$ and $g_{i,n,k}\equiv\kappa_{h}(Y_{i,n,k})$,
where $Y_{n,k}\equiv(Y_{1,n,k},\ldots,Y_{n,n,k})^{\prime}$ denote
the coupled version of $Y_{n}$ after $e_{k,n}$ with replaced with
its i.i.d. version $e^{*}_{k,n}$ conditional on $\c_{n}$ and $G_{n,h}(Y_{n,k})=\text{diag}(\kappa_{h}(Y_{1,n,k}),\dots,\kappa_{h}(Y_{n,n,k}))$.
Then
\[
\tr(A^{\ell}_{n,h})=\lambda^{\ell}\sum_{i_{1},\dots,i_{\ell}}w_{i_{1}i_{2},n}\cdots w_{i_{\ell}i_{1},n}\prod^{\ell}_{m=1}g^{2}_{i_{m},n}.
\]
For any $a_{m},b_{m}\in[0,1]$, $\Bigl|\prod^{\ell}_{m=1}a_{m}-\prod^{\ell}_{m=1}b_{m}\Bigr|\le\sum^{\ell}_{m=1}|a_{m}-b_{m}|.$
Therefore,
\[
\begin{aligned} & \frac{1}{n}\bigl|\tr(A^{\ell}_{n,h})-\tr((A_{n,h,k})^{\ell})\bigr|\\
\le & \frac{\left|\lambda\right|^{\ell}}{n}\sum^{\ell}_{m=1}\sum_{i_{1},\dots,i_{\ell}}\left|w_{i_{1}i_{2},n}\cdots w_{i_{\ell}i_{1},n}\right|\left|g^{2}_{i_{m},n}-g^{2}_{i_{m},n,k}\right|\\
= & \frac{\left|\lambda\right|^{\ell}}{n}\sum^{\ell}_{m=1}\sum^{n}_{r=1}\left|g^{2}_{r,n}-g^{2}_{r,n,k}\right|\sum_{i_{1},\dots,i_{\ell}:i_{m}=r}\left|w_{i_{1}i_{2},n}\cdots w_{i_{\ell}i_{1},n}\right|\\
= & \frac{\left|\lambda\right|^{\ell}}{n}\sum^{\ell}_{m=1}\sum^{n}_{r=1}\left|g^{2}_{r,n}-g^{2}_{r,n,k}\right|\bigl(|W_{n}|^{\ell}\bigr)_{rr}\\
\le & 2\frac{\left|\lambda\right|^{\ell}}{n}\sum^{\ell}_{m=1}\sum^{n}_{r=1}\left|g_{r,n}-g_{r,n,k}\right|\bigl(|W_{n}|^{\ell}\bigr)_{rr}\leq2\frac{\left|\lambda\right|^{\ell}}{n}\sum^{\ell}_{m=1}\sum^{n}_{r=1}\left|g_{r,n}-g_{r,n,k}\right|\left\Vert W_{n}\right\Vert ^{\ell}_{\infty}\\
= & 2\frac{\ell\left|\lambda\right|^{\ell}}{n}\sum^{n}_{r=1}\left|g_{r,n}-g_{r,n,k}\right|\le2\frac{\ell\left|\lambda\right|^{\ell}}{nh}\sum^{n}_{r=1}\left|Y_{r,n}-Y_{r,n,k}\right|.
\end{aligned}
\]
Let $P_{i}J_{n,h}(\theta)=\E_{\c_{n}}(J_{n,h}(\theta)|\mathcal{F}_{i,n})-\E_{\c_{n}}(J_{n,h}(\theta)|\mathcal{F}_{i-1,n})$,
where $\mathcal{F}_{i,n}$ is defined in Section~\ref{sec:Property}.
We further calculate 
\begin{align}
 & \left\Vert \frac{1}{n}J_{n,h_{n}}(\theta)-\frac{1}{n}\E_{\c_{n}}J_{n,h_{n}}(\theta)\right\Vert _{L^{p},\c_{n}}=\frac{1}{n}\left\Vert \sum^{n}_{i=1}P_{i}J_{n,h_{n}}(\theta)\right\Vert _{L^{p},\c_{n}}=\frac{1}{n}\left\Vert \sum^{n}_{i=1}P_{i}\left(\sum^{\infty}_{\ell=1}\frac{1}{\ell}\tr(A^{\ell}_{n,h_{n}})\right)\right\Vert _{L^{p},\c_{n}}\nonumber \\
\le & \frac{1}{n}\sum^{\infty}_{\ell=1}\frac{1}{\ell}\left\Vert \sum^{n}_{i=1}P_{i}\tr(A^{\ell}_{n,h_{n}})\right\Vert _{L^{p},\c_{n}}\leq\frac{1}{n}\sum^{\infty}_{\ell=1}\frac{1}{\ell}\sqrt{p-1}\left(\sum^{n}_{i=1}\left\Vert P_{i}\tr(A^{\ell}_{n,h_{n}})\right\Vert ^{2}_{L^{p},\c_{n}}\right)^{\frac{1}{2}}\nonumber \\
\leq & \frac{1}{n}\sum^{\infty}_{\ell=1}\frac{1}{\ell}\sqrt{p-1}\left[\sum^{n}_{i=1}\left(2\frac{\ell\left|\lambda\right|^{\ell}}{h_{n}}\sum^{n}_{r=1}\left\Vert Y_{r,n}-Y_{r,n,i}\right\Vert _{L^{p},\c_{n}}\right)^{2}\right]^{\frac{1}{2}}\nonumber \\
\le & \frac{2}{nh_{n}}\frac{1}{1-\left|\lambda\right|}\sqrt{p-1}\left[\sum^{n}_{i=1}\left(\sum^{n}_{r=1}\left\Vert Y_{r,n}-Y_{r,n,i}\right\Vert _{L^{p},\c_{n}}\right)^{2}\right]^{\frac{1}{2}}\nonumber \\
\le & \frac{2}{nh_{n}}\frac{1}{1-\left|\lambda\right|}\sqrt{p-1}\left\Vert \epsilon\right\Vert _{L^{p},\c_{n}}\left[\sum^{n}_{i=1}\left\Vert S^{+}_{\cdot i,n}\right\Vert ^{2}_{1}\right]^{\frac{1}{2}}\leq\frac{2}{1-\left|\lambda\right|}\sqrt{p-1}\left\Vert \epsilon\right\Vert _{L^{p},\c_{n}}\frac{\mu_{2}}{\sqrt{n}h_{n}}=o_{\p}(1),\label{eq:J-EJ}
\end{align}
where the second inequality follows from Lemma~\ref{lem: Burkholder}
and the third inequality follows from Lemma~\ref{lem:general pred<FDM}.
Combining Eqs.(\ref{eq:R-ER}) and (\ref{eq:J-EJ}) yields that 
\begin{align*}
 & \frac{1}{n}\left\Vert \ell_{n,h_{n}}(\theta)-\E_{\c_{n}}\ell_{n,h_{n}}(\theta)\right\Vert _{L^{2},\c_{n}}\\
\leq & \frac{1}{n}\left\Vert R_{n,h}(\theta)-\E_{\c_{n}}R_{n,h}(\theta)\right\Vert _{L^{2},\c_{n}}+\frac{1}{n}\left\Vert J_{n,h_{n}}(\theta)-\E_{\c_{n}}J_{n,h_{n}}(\theta)\right\Vert _{L^{2},\c_{n}}=o_{\p}(1).
\end{align*}
It follows from \citet[Lemma 6.1]{chernozhukov2018Double} that $\frac{1}{n}[\ell_{n,h_{n}}(\theta)-\E_{\c_{n}}\ell_{n,h_{n}}(\theta)]=o_{\p}(1)$.

\textbf{Step 3. }We show that $\frac{1}{n}|\ell_{n}(\theta)-\ell_{n,h_{n}}(\theta)|=o_{\p}(1)$.
Since $|1(Y_{i,n}>0)-\kappa_{h_{n}}(Y_{i,n})|\le1(0<Y_{i,n}<h_{n}),$
and $Y_{i,n}=\max(0,Y^{*}_{i,n})$, we have $1(0<Y_{i,n}<h_{n})=1(0<Y^{*}_{i,n}<h_{n}).$
By Assumption \ref{assu:consis}(v), $\p_{\c_{n}}(0<Y_{i,n}<h_{n})\le C_{f}h_{n}.$
Let $B_{i,n}(\theta)=\left|\frac{1}{2}\log(2\pi\sigma^{2})\right|+|\log\Phi(z_{i,n}(\theta))|+\frac{1}{2}z_{i,n}(\theta)^{2}$.
We have 
\begin{align*}
 & \frac{1}{n}\left\Vert R_{n}(\theta)-R_{n,h_{n}}(\theta)\right\Vert _{L^{p/2},\c_{n}}\le\frac{1}{n}\sum^{n}_{i=1}\left\Vert B_{i,n}(\theta)1(0<Y_{i,n}<h_{n})\right\Vert _{L^{p/2},\c_{n}}\\
\le & \frac{1}{n}\sum^{n}_{i=1}\left\Vert B_{i,n}(\theta)\right\Vert _{L^{p},\c_{n}}\left\Vert 1(0<Y_{i,n}<h_{n})\right\Vert _{L^{p},\c_{n}}=h^{1/p}_{n}O_{\p}(1)=o_{\p}(1),
\end{align*}
where the $O_{\p}(1)$ can be obtained using similar techniques from
\citet{xu2015maximum}. For $J_{n}(\theta)=\log\det(I_{n}-\lambda G_{n}(Y_{n})W_{n}G_{n}(Y_{n}))$,
we have that 
\begin{align*}
 & \frac{1}{n}\left\Vert J_{n}(\theta)-J_{n,h_{n}}(\theta)\right\Vert _{L^{p},\c_{n}}\\
= & \frac{1}{n}\left\Vert \sum^{\infty}_{\ell=1}\frac{1}{\ell}\tr(A^{\ell}_{n})-\sum^{\infty}_{\ell=1}\frac{1}{\ell}\tr(A^{\ell}_{n,h_{n}})\right\Vert _{L^{p},\c_{n}}\le\frac{1}{n}\sum^{\infty}_{\ell=1}\frac{1}{\ell}\left\Vert \tr(A^{\ell}_{n})-\tr(A^{\ell}_{n,h_{n}})\right\Vert _{L^{p},\c_{n}}\\
\le & \frac{2}{n}\sum^{\infty}_{\ell=1}\left|\lambda\right|^{\ell}\sum^{n}_{i=1}\left\Vert 1(Y_{i,n}>0)-\kappa_{h_{n}}(Y_{i,n})\right\Vert _{L^{p},\c_{n}}\le\frac{2}{n}\frac{1}{1-\left|\lambda\right|}\sum^{n}_{i=1}\left\Vert 1(0<Y_{i,n}<h_{n})\right\Vert _{L^{p},\c_{n}}\\
\le & O_{\p}(1)\cdot h^{1/p}_{n}=o_{\p}(1).
\end{align*}
Thus 
\begin{align*}
 & \frac{1}{n}\left\Vert \ell_{n,h_{n}}(\theta)-\ell_{n}(\theta)\right\Vert _{L^{p/2},\c_{n}}\\
\leq & \frac{1}{n}\left\Vert R_{n}(\theta)-R_{n,h_{n}}(\theta)\right\Vert _{L^{p/2},\c_{n}}+\frac{1}{n}\left\Vert J_{n}(\theta)-J_{n,h_{n}}(\theta)\right\Vert _{L^{p},\c_{n}}=o_{\p}(1)
\end{align*}
and by \citet[Lemma 6.1]{chernozhukov2018Double}, it holds that $\frac{1}{n}|\ell_{n}(\theta)-\ell_{n,h_{n}}(\theta)|=o_{\p}(1)$.
Finally, we have
\[
\begin{aligned} & \frac{1}{n}[\ell_{n}(\theta)-\E_{\c_{n}}\ell_{n}(\theta)]\\
= & \frac{1}{n}[\ell_{n}(\theta)-\ell_{n,h_{n}}(\theta)]+\frac{1}{n}[\ell_{n,h_{n}}(\theta)-\E_{\c_{n}}\ell_{n,h_{n}}(\theta)]+\frac{1}{n}\E_{\c_{n}}[\ell_{n,h_{n}}(\theta)-\ell_{n}(\theta)]=o_{\p}(1).
\end{aligned}
\]
The proof is completed.$\hfill\qedsymbol$

\section{Proofs for Section \ref{sec: Transformation}}

\textbf{Proof of Proposition \ref{prop: Lips}. }By Eq.\eqref{eq:lip condi}
and $B_{j,n}\left(y,y^{\bullet}\right)\leq C$, we have 
\begin{align*}
 & \delta^{(Z)}_{p,n}\left(j,i,\c_{n}\right)=\left\Vert H_{j,n}\left(Y_{j,n}\right)-H_{j,n}\left(Y_{j,n,i}\right)\right\Vert _{L^{p},\c_{n}}\leq\left\Vert B_{j,n}(Y_{j,n},Y_{j,n,i})\left|Y_{j,n}-Y_{j,n,i}\right|\right\Vert _{L^{p},\c_{n}}\\
\leq & C\left\Vert Y_{j,n}-Y_{j,n,i}\right\Vert _{L^{p},\c_{n}}=C\delta^{(Y)}_{p,n}\left(j,i,\c_{n}\right)
\end{align*}
a.s.$\hfill\qedsymbol$

\textbf{Proof of Proposition \ref{prop:unbdd lip 1}.} By Eq.\eqref{eq:lip condi}
and $B_{j,n}\left(y,y^{\bullet}\right)\leq C_{1}(\left|y\right|^{a}+\left|y^{\bullet}\right|^{a}+1)$,
we have 
\begin{align*}
 & \delta^{(Z)}_{p,n}\left(j,i,\c_{n}\right)=\left\Vert H_{j,n}\left(Y_{j,n}\right)-H_{j,n}\left(Y_{j,n,i}\right)\right\Vert _{L^{p},\c_{n}}\\
\leq & C_{1}\left\Vert \left(\left|Y_{j,n}\right|^{a}+\left|Y_{j,n,i}\right|^{a}+1\right)\cdot\left|Y_{j,n}-Y_{j,n,i}\right|\right\Vert _{L^{p},\c_{n}}\\
\leq & C_{1}\left\Vert \left|Y_{j,n}\right|^{a}+\left|Y_{j,n,i}\right|^{a}+1\right\Vert _{L^{r},\c_{n}}\left\Vert Y_{j,n}-Y_{j,n,i}\right\Vert _{L^{q},\c_{n}}\leq C_{1}(2\left\Vert Y\right\Vert ^{a}_{L^{ar},\c_{n}}+1)\delta^{(Y)}_{q,n}\left(j,i,\c_{n}\right),
\end{align*}
where the second inequality follows from conditional generalized Hölder's
inequality (because $p^{-1}=q^{-1}+r^{-1}$), and the third inequality
follows from conditional Minkowski's inequality. $\hfill\qedsymbol$

\textbf{Proof of Proposition \ref{prop:unbdd lip 2}.} Denote $Z_{j,n,i}\equiv H_{j,n}\left(Y_{j,n,i}\right)$,
$B\equiv\left|Y_{j,n}\right|^{a}+\left|Y_{j,n,i}\right|^{a}+1$, $\rho\equiv\left|Y_{j,n}-Y_{j,n,i}\right|$,
$r=q/\left(a+1\right)>p$. By conditional Lyapunov's inequality and
conditional Minkowski's inequality, we have 
\[
\left\Vert B\right\Vert _{L^{p/\left(p-1\right)},\c_{n}}\leq\left\Vert B\right\Vert _{L^{q/a},\c_{n}}\leq\left\Vert Y^{a}_{j,n}\right\Vert _{L^{q/a},\c_{n}}+\left\Vert \left|Y^{*}_{j,n}\right|^{a}\right\Vert _{L^{q/a},\c_{n}}+1\leq2\left\Vert Y\right\Vert ^{a}_{L^{q},\c_{n}}+1<\infty
\]
a.s., and 
\[
\left\Vert \rho\right\Vert _{L^{q},\c_{n}}\leq2\left\Vert Y_{j,n}\right\Vert _{L^{q},\c_{n}}<2\left\Vert Y\right\Vert _{L^{q},\c_{n}}<\infty\text{ a.s.}
\]
Using the above two results, by conditional generalized Hölder's inequality
($\frac{1}{r}=\frac{a+1}{q}=\frac{a}{q}+\frac{1}{q}$), we have 
\[
\left\Vert B\rho\right\Vert _{L^{r},\c_{n}}\leq\left\Vert B\right\Vert _{L^{q/a},\c_{n}}\left\Vert \rho\right\Vert _{L^{q},\c_{n}}\leq\left(2\left\Vert Y\right\Vert ^{a}_{L^{q},\c_{n}}+1\right)2\left\Vert Y\right\Vert _{L^{q},\c_{n}}<\infty\text{ a.s.}
\]
Thus, by Lemma S.7 in the Supplementary Material of \citet{wu2023application},
for all $i$, $j$, and $n\geq1$, it holds that 
\begin{align*}
 & \delta^{(Z)}_{p,n}\left(j,i,\c_{n}\right)=\left\Vert Z_{j,n}-Z_{j,n,i}\right\Vert _{L^{p},\c_{n}}\leq C_{1}\left\Vert B\rho\right\Vert _{L^{p},\c_{n}}\\
\leq & 2C_{1}\left(\left\Vert B\right\Vert ^{r-p}_{L^{p/\left(p-1\right)},\c_{n}}\left\Vert \rho\right\Vert ^{r-p}_{L^{p},\c_{n}}\left\Vert B\rho\right\Vert ^{(p-1)r}_{L^{r},\c_{n}}\right)^{1/\left(pr-p\right)}\\
\leq & C_{2}\left(\c_{n}\right)\left(\left\Vert Y_{j,n}-Y_{j,n,i}\right\Vert _{L^{p},\c_{n}}\right)^{\left(q-ap-p\right)/\left(pq-ap-p\right)}\\
= & C_{2}\left(\c_{n}\right)\left[\delta^{(Y)}_{p,n}\left(j,i,\c_{n}\right)\right]^{\left(q-ap-p\right)/\left(pq-ap-p\right)},
\end{align*}
where $C_{2}\left(\c_{n}\right)<\infty$ a.s. $\hfill\qedsymbol$

\textbf{Proof of Proposition \ref{prop:1 >0}.} For any $\epsilon>0$,
let $B=\left\{ \left|Y_{j,n}\right|<\epsilon,\left|Y_{j,n,i}\right|<\epsilon\right\} $.
Since 
\[
\left|1\left(x_{1}>0\right)-1\left(x_{2}>0\right)\right|\leq\frac{\left|x_{1}-x_{2}\right|}{\epsilon}1\left(\left|x_{1}\right|\geq\epsilon\ \text{or}\ \left|x_{2}\right|\geq\epsilon\right)+1\left(\left|x_{1}\right|<\epsilon\ \text{and}\ \left|x_{2}\right|<\epsilon\right),
\]
\footnote{See the proof of Proposition 2 in \citet{xu2015maximum}.}we
have 
\begin{align*}
 & \delta^{(Z)}_{p,n}\left(j,i,\c_{n}\right)=\left\Vert 1\left(Y_{j,n}>0\right)-1\left(Y_{j,n,i}>0\right)\right\Vert _{L^{p},\c_{n}}\\
\leq & \left\{ \frac{1}{\epsilon^{p}}\int_{B^{c}}\left|Y_{j,n}-Y_{j,n,i}\right|^{p}\mathrm{d}\p_{\c_{n}}+\p\left(B|\c_{n}\right)\right\} ^{1/p}\\
\le & \frac{\left\Vert Y_{j,n}-Y_{j,n,i}\right\Vert _{L^{p},\c_{n}}}{\epsilon}+\p\left(\left|Y_{j,n}\right|<\epsilon|\c_{n}\right)^{1/p}\leq\frac{\delta^{(Y)}_{p,n}\left(j,i,\c_{n}\right)}{\epsilon}+C_{2}\epsilon^{1/p}
\end{align*}
for some constant $C_{2}>0$, where the second inequality follows
from the fact that $\left(a^{p}+b^{p}\right)^{1/p}\leq a+b$ for arbitrary
non-negative numbers $a$, $b$ and $p\geq1$, and the last one follows
from $\sup_{n\geq1,1\leq j\leq n}\sup_{y}f_{j,n}(y\mid\c_{n})<C_{1}$.
By letting $\epsilon=\left[\delta^{(Y)}_{p,n}\left(j,i,\c_{n}\right)\right]^{p/\left(p+1\right)}$,
we have $\delta^{(Z)}_{p,n}\left(j,i,\c_{n}\right)\leq\left(1+C_{2}\right)\left[\delta^{(Y)}_{p,n}\left(j,i,\c_{n}\right)\right]^{1/\left(p+1\right)}$.
$\hfill\qedsymbol$

\textbf{Proof of Proposition \ref{prop:FDM prod 1}.} We have 
\begin{align*}
 & \delta^{(YZ)}_{p,n}\left(j,i,\c_{n}\right)=\left\Vert Y_{j,n}Z_{j,n}-Y_{j,n,i}Z_{j,n,i}\right\Vert _{L^{p},\c_{n}}\\
= & \left\Vert Y_{j,n}Z_{j,n}-Y_{j,n,i}Z_{j,n}+Y_{j,n,i}Z_{j,n}-Y_{j,n,i}Z_{j,n,i}\right\Vert _{L^{p},\c_{n}}\\
\leq & \left\Vert Y_{j,n}Z_{j,n}-Y_{j,n,i}Z_{j,n}\right\Vert _{L^{p},\c_{n}}+\left\Vert Y_{j,n,i}Z_{j,n}-Y_{j,n,i}Z_{j,n,i}\right\Vert _{L^{p},\c_{n}}\\
= & \left\Vert \left(Y_{j,n}-Y_{j,n,i}\right)Z_{j,n}\right\Vert _{L^{p},\c_{n}}+\left\Vert Y_{j,n,i}\left(Z_{j,n}-Z_{j,n,i}\right)\right\Vert _{L^{p},\c_{n}}\\
\leq & \left\Vert Z_{j,n}\right\Vert _{L^{r_{1}},\c_{n}}\left\Vert Y_{j,n}-Y_{j,n,i}\right\Vert _{L^{q_{1}},\c_{n}}+\left\Vert Y_{j,n}\right\Vert _{L^{r_{2}},\c_{n}}\left\Vert Z_{j,n}-Z_{j,n,i}\right\Vert _{L^{q_{2}},\c_{n}}\\
\leq & \left\Vert Z\right\Vert _{L^{r_{1}},\c_{n}}\delta^{(Y)}_{q_{1},n}\left(j,i,\c_{n}\right)+\left\Vert Y\right\Vert _{L^{r_{2}},\c_{n}}\delta^{(Z)}_{q_{2},n}\left(j,i,\c_{n}\right),
\end{align*}
where the first inequality follows from conditional Minkowski inequality,
and the last inequality follows from conditional Hölder's inequality.
$\hfill\qedsymbol$

\textbf{Proof of Proposition \ref{prop:FDM prod 2}. }Let $B=\left|Z_{j,n}\right|$,
$\rho=\left|Y_{j,n}-Y_{j,n,i}\right|$, $r=\frac{q}{2}>p$, by Lyapunov's
inequality and Minkowski's inequality, we have 
\[
\left\Vert B\right\Vert _{L^{p/\left(p-1\right)},\c_{n}}\leq\left\Vert B\right\Vert _{L^{q},\c_{n}}=\left\Vert Z_{j,n}\right\Vert _{L^{q},\c_{n}}\leq\left\Vert Z\right\Vert _{L^{q},\c_{n}}<\infty\text{ a.s.,}
\]
and 
\[
\left\Vert \rho\right\Vert _{L^{q},\c_{n}}=\left\Vert Y_{j,n}-Y_{j,n,i}\right\Vert _{L^{q},\c_{n}}\leq2\left\Vert Y_{j,n}\right\Vert _{L^{q},\c_{n}}\leq2\left\Vert Y\right\Vert _{L^{q},\c_{n}}<\infty\text{ a.s.}
\]
So, by the generalized Hölder's inequality, 
\[
\left\Vert B\rho\right\Vert _{L^{r},\c_{n}}\leq\left\Vert B\right\Vert _{L^{q},\c_{n}}\left\Vert \rho\right\Vert _{L^{q},\c_{n}}\leq2\left\Vert Y\right\Vert _{L^{q},\c_{n}}\left\Vert Z\right\Vert _{L^{q},\c_{n}}<\infty\text{ a.s.}
\]
Hence, by Lemma S.7 in the Supplementary Material of \citet{wu2023application},
\begin{align*}
 & \left\Vert \left|Y_{j,n}-Y_{j,n,i}\right|\cdot\left|Z_{j,n}\right|\right\Vert _{L^{p},\c_{n}}\leq2\left(\left\Vert B\right\Vert ^{r-p}_{L^{p/\left(p-1\right)},\c_{n}}\left\Vert \rho\right\Vert ^{r-p}_{L^{p},\c_{n}}\left\Vert B\rho\right\Vert ^{(p-1)r}_{L^{r},\c_{n}}\right)^{1/\left(pr-p\right)}\\
\leq & C_{1}\left(\c_{n}\right)\left(\left\Vert Y_{j,n}-Y_{j,n,i}\right\Vert _{L^{p},\c_{n}}\right)^{\left(q-2p\right)/\left(pq-2p\right)}
\end{align*}
where $C_{1}\left(\c_{n}\right)<\infty$ a.s. Similarly, we have 
\[
\left\Vert \left|Y_{j,n,i}\right|\cdot\left|Z_{j,n}-Z_{j,n,i}\right|\right\Vert _{L^{p},\c_{n}}\leq C_{2}\left(\c_{n}\right)\left(\left\Vert Z_{j,n}-Z_{j,n,i}\right\Vert _{L^{p},\c_{n}}\right)^{\left(q-2p\right)/\left(pq-2p\right)}
\]
where $C_{2}\left(\c_{n}\right)<\infty$ a.s. By the above two inequalities,
we have 
\begin{align*}
 & \delta^{(YZ)}_{p,n}\left(j,i,\c_{n}\right)=\left\Vert Y_{j,n}Z_{j,n}-Y_{j,n,i}Z_{j,n,i}\right\Vert _{L^{p},\c_{n}}\\
= & \left\Vert Y_{j,n}Z_{j,n}-Y_{j,n,i}Z_{j,n}+Y_{j,n,i}Z_{j,n}-Y_{j,n,i}Z_{j,n,i}\right\Vert _{L^{p},\c_{n}}\\
\leq & \left\Vert \left|Y_{j,n}-Y_{j,n,i}\right|\cdot\left|Z_{j,n}\right|\right\Vert _{L^{p},\c_{n}}+\left\Vert \left|Y_{j,n,i}\right|\cdot\left|Z_{j,n}-Z_{j,n,i}\right|\right\Vert _{L^{p},\c_{n}}\\
\leq & C_{1}\left(\c_{n}\right)\left(\left\Vert Y_{j,n}-Y_{j,n,i}\right\Vert _{L^{p},\c_{n}}\right)^{\left(q-2p\right)/\left(pq-2p\right)}+C_{2}\left(\c_{n}\right)\left(\left\Vert Z_{j,n}-Z_{j,n,i}\right\Vert _{L^{p},\c_{n}}\right)^{\left(q-2p\right)/\left(pq-2p\right)}\\
= & C_{1}\left(\c_{n}\right)\left[\delta^{(Y)}_{p,n}\left(j,i,\c_{n}\right)\right]^{\left(q-2p\right)/\left(pq-2p\right)}+C_{2}\left(\c_{n}\right)\left[\delta^{(Z)}_{p,n}\left(j,i,\c_{n}\right)\right]^{\left(q-2p\right)/\left(pq-2p\right)}
\end{align*}
a.s. $\hfill\qedsymbol$ 
\end{document}